\documentclass[pdflatex,sn-basic]{sn-jnl}

\usepackage{amsmath}

\newcommand{\beq}{\begin{equation}}
\newcommand{\eeq}{\end{equation}}
\newcommand{\beqn}{\begin{eqnarray}}
\newcommand{\eeqn}{\end{eqnarray}}

\jyear{2022}%

\begin{document}

\title[Neutrino transport in mergers]{Neutrino transport in general relativistic neutron star merger simulations}


\author*[1]{\fnm{Francois} \sur{Foucart}}\email{francois.foucart@unh.edu}

\affil*[1]{\orgdiv{Department of Physics and Astronomy}, \orgname{University of New Hampshire}, \orgaddress{\street{9 Library Way}, \city{Durham}, \postcode{03824}, \state{NH}, \country{USA}}}


\abstract{Numerical simulations of neutron star--neutron star and neutron star--black hole binaries play an important role in our ability to model gravitational wave and electromagnetic signals powered by these systems. These simulations have to take into account a wide range of physical processes including general relativity, magnetohydrodynamics, and neutrino radiation transport. The latter is particularly important in order to understand the properties of the matter ejected by many mergers, the optical/infrared signals powered by nuclear reactions in the ejecta, and the contribution of that ejecta to astrophysical nucleosynthesis. However, accurate evolutions of the neutrino transport equations that include all relevant physical processes remain beyond our current reach. In this review, I will discuss the current state of neutrino modeling in general relativistic simulations of neutron star mergers and of their post-merger remnants. I will focus on the three main types of algorithms used in simulations so far: leakage, moments, and Monte-Carlo scheme. I will review the advantages and limitations of each scheme, as well as the various neutrino-matter interactions that should be included in simulations. We will see that the quality of the treatment of neutrinos in merger simulations has greatly increased over the last decade, but also that many potentially important interactions remain difficult to take into account in simulations (pair annihilation, oscillations, inelastic scattering).}


\maketitle

\pagestyle{myheadings}
\markright{F. Foucart} 
{\small
\setcounter{tocdepth}{3}
\tableofcontents
}

\section{Introduction}\label{sec:intro}

Over the last decade, the study of merging compact objects has made tremendous progress. Recently observed astrophysical events provide us with some of the most reliable information currently at our disposal regarding the population of stellar mass black holes in the nearby Universe. Rarer events that include neutron stars also inform us about the mass distribution of neutron stars, the equation of state of dense matter, and the origin of heavy elements formed through rapid neutron capture nucleosynthesis (r-process). Our ability to study these systems has largely grown in tandem with the sensitivity of the LIGO and Virgo gravitational wave detectors. Gravitational wave observatories have now detected dozens of binary black hole (BBH) mergers, as well as two likely binary neutron star (BNS) mergers and at least two likely neutron star-black hole (NSBH) mergers (see Sect.~\ref{sec:obs} for a more detailed discussion of these events). An overview of these events can be found in the three GWTC catalogues \citep{LIGOScientific:2018mvr,LIGOScientific:2021usb,LIGOScientific:2021djp}. 

While BNS and NSBH mergers are not as commonly observed as BBH mergers, they do have important advantages for nuclear astrophysics. The presence of a neutron star means that these systems can potentially be used to constrain the equation of state of cold, neutron rich dense matter \citep{GW170817-NSRadius} -- a crucial source of information about many-nucleon interactions and, potentially, the high-density states of quantum chromodynamics. Additionally, some mergers and post-merger remnants eject material that undergoes r-process nucleosynthesis. The radioactive decay of the ashes of the r-process can then power optical/infrared emission days to weeks after the merger: a kilonova \citep{1976ApJ...210..549L,Li:1998bw,2010MNRAS.406.2650M,Roberts2011,Kasen:2013xka}. The production site(s) of r-process elements remain(s) very uncertain today, and the observation of neutron star mergers and associated kilonovae may help us solve the long-standing problem of their astrophysical origin. Additionally, some post-merger remnants likely produce collimated relativistic outflows (jets) that are currently believed to be the source of short-hard gamma-ray bursts (SGRBs) \citep{Eichler:1989ve,Nakar:2007yr,Fong2013}. The exact process powering SGRBs is however not well understood, and further observations of neutron star mergers could help us ellucidate how these high-energy events occur in practice. Finally, joint observations of neutron star mergers using both gravitational and electromagnetic waves may also provide additional information about the properties of the merging compact objects, the position of the merging binary, and even the value of the Hubble constant \citep{2005ApJ...629...15H,2010ApJ...725..496N,LIGOScientific:2017adf,Hotokezaka:2018dfi}.

Neutron star mergers involve a wide range of nonlinear physical processes, preventing us from providing quantitative theoretical predictions for the result of a merger using purely analytical methods. As a result, numerical simulations are an important tool in current attempts to model the gravitational wave and electromagnetic signals powered by compact binary mergers. Gravity, fluid dynamics, magnetic fields and neutrinos all play major roles during and after neutron star mergers, with out-of-equlibrium nuclear reactions also becoming important on longer time scales ($\sim$ seconds). In theory, merger simulations thus need to solve Boltzmann's equations of radiation transport coupled to the relativistic equations of magnetohydrodynamics and Einstein's equation of general relativity. However, no simulation can do this with the desired level of realism at this point. Two major roadblocks to this modeling efforts are our inability to properly resolve magnetohydrodynamical instabilities during merger (and thus the dynamo process that may follow the growth of magnetic fields due to these instabilities) \citep{Kiuchi2015}, as well as the difficulty of properly solving Boltzmann's equation of radiation transport for the evolution of neutrinos \citep{Foucart:2018gis}. In this review, we focus on the second problem. The role of magnetic fields in merger simulations is discussed in more detail, for example, in \cite{Baiotti:2016qnr,Paschalidis2017,Burns:2019byj}.

Neutrinos play a number of roles in neutron star mergers, with particularly noticeable impacts on the production of r-process elements and the properties of kilonovae. However, properly accounting for neutrino-matter interactions in neutron star mergers remains a difficult problem because, within a merger remnant, neutrinos transition from being in equilibrium with the fluid (in dense hot regions) to mostly free-streaming through the ejected material (far away). In the intermediate regions, neutrino-matter interactions play an important role in the evolution of the temperature and composition of the fluid, but neutrinos cannot be assumed to be in equilibrium with the fluid. Numerical methods that properly capture both regimes are technically challenging and/or computationally expensive. As a result, most merger simulations use approximate neutrino transport algorithms that introduce potentially significant and often hard to quantify errors in our predictions for the nuclei produced during r-process nucleosynthesis and for the properties of kilonovae. 

The main objective of this review is to provide an overview of the various algorithms currently used in general relativistic simulations of neutron star mergers and of their post-merger remnants. These can be broadly classified into three groups: leakage methods, which do not explicitly transport neutrinos; moment schemes, which evolve a truncated expansion of the transport equations in momentum space with methods highly similar to those used to evolve the equations of relativistic magnetohydrodynamics; and Monte-Carlo methods, which sample the distribution of neutrinos with packets (or superparticles) propagating through numerical simulations. These are discussed in detail in Sect.~\ref{sec:algorithms}. Section~\ref{sec:background} aims to provide some scientific background about merging neutron stars, while Sect.~\ref{sec:nu} provides an overview of neutrino physics in neutron star mergers, and of the important neutrino-matter interactions that are currently included or neglected in simulations. Finally, Sect.~\ref{sec:results} discusses what existing simulations can tell us about the ways in which our choice of algorithm impacts our numerical results. We note that the objective here is not to review all results in the study of neutron star mergers with neutrinos, but rather to focus on the numerical methods used to perform general relativistic radiation transport. We will thus focus on comparisons of different numerical methods, rather that provide an extensive review of existing simulations that make use of neutrino transport.

\emph{Conventions}: In this manuscript, latin letters are used for the indices of spatial 3-dimensional vectors/tensors, while greek letters are used for the indices of 4-dimensional vectors / tensors. Sections discussing numerical methods will often use units such that $h=c=G=1$, but we explicitly keep physical constants in our expressions when discussing interaction rates.

\section{Scientific background}
\label{sec:background}

\subsection{Overview of neutron star mergers physics}
\label{sec:overview}

Before delving deeper into the topic of radiation transport in neutron star mergers, it is worth reviewing how we currently understand the evolution of these systems, as well as when different physical processes are expected to play an important role. When discussing neutron star merger simulations, we are typically concerned with the evolution of a binary from tens of milliseconds before merger to a few seconds after merger, i.e., from the moment standard post-Newtonian methods can no longer accurately model the gravitational wave signal to the moment when the accretion disk formed during a merger has lost most of its mass. In the late inspiral ($O(10)$ orbits before merger), the tidal distortion of a neutron star by its binary companion has a potentially measurable impact on the gravitational wave signal, which can be used to put constraints on the equation of state of neutron stars \citep{Flanagan2008,GW170817-NSRadius}. The main role of numerical simulations in that regime is to help test and calibrate analytical waveform models used in the analysis of gravitational wave events (e.g., \citealt{Bernuzzi2012b,Hinderer:2016a,Akcay:2018yyh} for BNS mergers and \cite{Thompson:2020nei,Matas:2020wab} for NSBH mergers). General relativity, fluid dynamics, and the choice of equation of state are important at that stage, but magnetic fields only impact relatively weak pre-merger electromagnetic signals and neutrinos have practically no impact on the evolution of the system.

\begin{figure}[ht]
\centering
\includegraphics[width=0.4\textwidth]{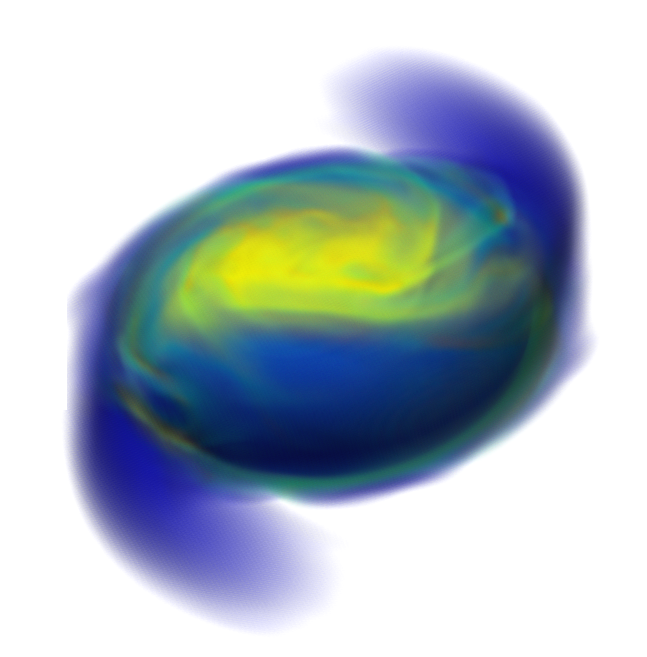}
\includegraphics[width=0.4\textwidth]{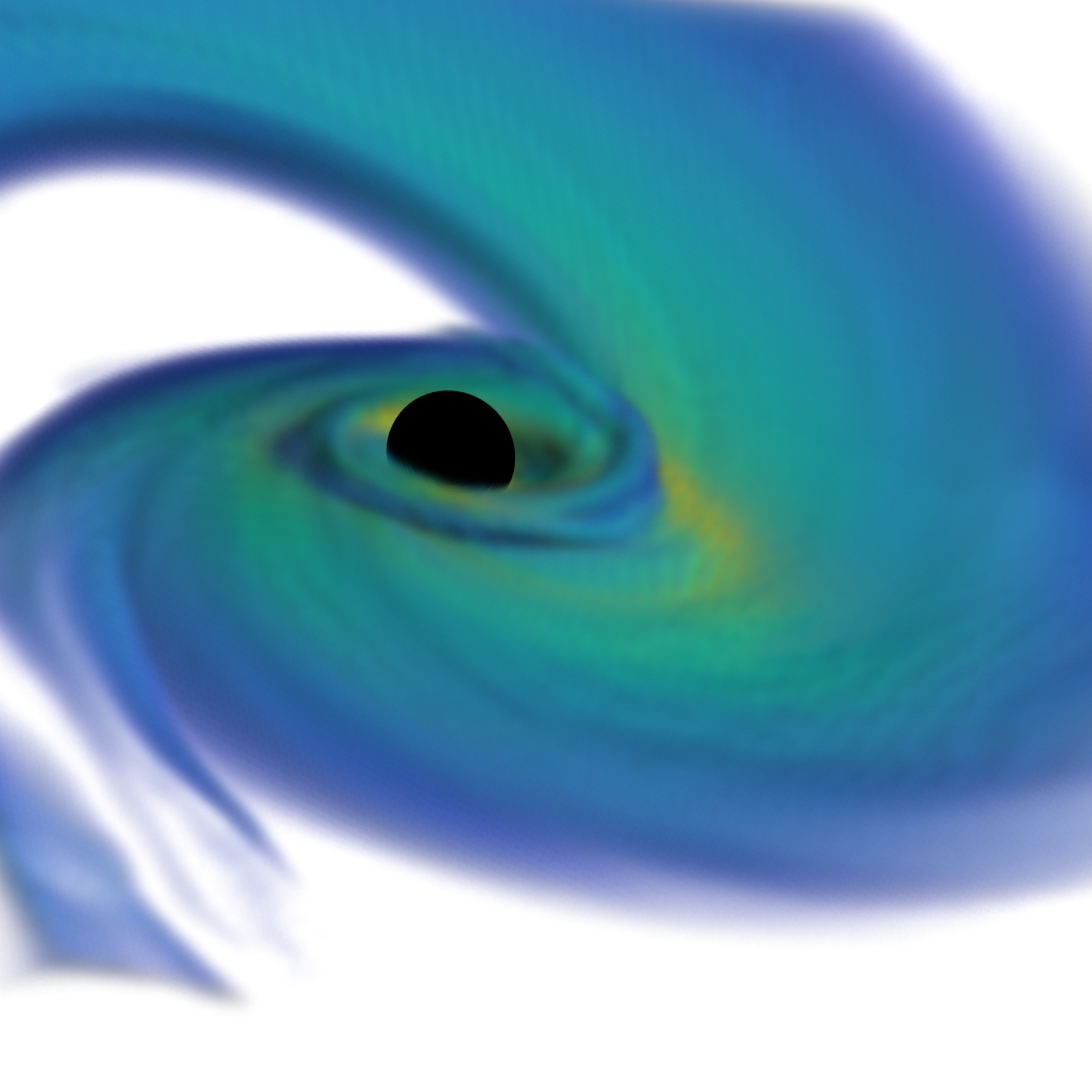}
\caption{Merger of a disrupting NSBH binary (Right) and of a low-mass NSNS binary (Left). In disrupting NSBH systems, most of the matter is rapidly accreted onto the black hole, while the rest forms an accretion disk and extended tidal tail. Low-mass NSNS binaries form a massive neutron star remnant surrounded by a bound disk, with a smaller amount of material ejected in the tidal tail. The right panel is reproduced from \cite{Foucart:2016vxd}; the left panel visualizes a simulation from \cite{Foucart:2015gaa}.}
\label{fig:merger}
\end{figure}

For NSBH binaries, the same remains true during the merger itself, i.e., the few milliseconds during which the neutron star is either tidally disrupted by its black hole companion, or absorbed whole by the black hole. The outcome of the merger is determined by the masses and spins of the compact objects, the equation of state of dense matter \citep{1976ApJ...210..549L,Pannarale:2010vs,Foucart2012}, and the eccentricity of the orbit \citep{East:2015yea}. Numerical simulations of low-eccentricity binaries have shown that only low mass and/or high spin black holes disrupt their neutron star companions ($M_{\rm BH}\lesssim 5\,M_\odot$ for non-spinning compact objects and circular orbits), a prerequisite to the production of any post-merger electromagnetic signal. If the neutron star is tidally disrupted, a few percents of a solar mass of very neutron rich, cold matter is typically ejected, and tenths of a solar mass remain in a bound accretion disk and/or tidal tail around the black hole (see e.g., \citealt{Foucart:2020ats,Kyutoku:2021icp} for recent reviews, and Fig.~\ref{fig:merger}). In eccentric binaries, neutron stars are typically easier to disrupt, and eject more mass in their tidal tails. 

For BNS systems, on the other hand, other physical processes become important once the neutron stars collide. First, the shear region that is naturally created between the merging neutron stars is unstable to the Kelvin--Helmoltz instability, leading to the rapid growth of small scale turbulence \citep{Kiuchi2015}. Magnetic fields are quickly amplified to $B\sim 10^{16}\,{\rm G}$ as a result, and start to play an important role in the evolution of the system. Whether a dynamo process can generate a large scale magnetic field from this turbulent state is an important open questions that simulations have not so far been able to answer. The collision of the two neutron stars also creates hot regions where neutrino emission and absorption can no longer be ignored. BNS mergers eject relatively small amounts of cold tidal ejecta ($\lesssim 0.01\,M_\odot$), as well as hotter material coming from the regions where the cores of the neutron stars collide. We will see that neutrinos play an important role in the evolution of that hot ejecta. Depending on the equation of state and on the mass of the system, the remnant may immediately collapse to a black hole (on milliseconds time scales), remain temporarily supported by rotation and/or thermal pressure, or form a long-lived neutron star (as on Fig.~\ref{fig:merger}). In all cases, that remnant is surrounded by a hot accretion disk -- with more asymmetric systems producing more massive disks (see e.g., \citealt{Baiotti:2016qnr,Burns:2019byj,Radice:2020ddv} for recent reviews).

After merger, neutrino emission is the main source of cooling for the accretion disk and remnant neutron star (if there is one), and neutrino-matter interactions drive changes in the composition of the disk material and of the outflows. 
Initially, the efficiency of neutrinos in cooling the disk lies in between the radiatively efficient (thin disks) and radiatively inefficient (thick disks) regimes observed in AGNs. NSBH and BNS simulations including radiation transport show a disk aspect ratio $H/R \sim (0.2-0.3)$ (with $H$ the scale height of the disk and $R$ its radius)~\citep{FoucartM1:2015,Fujibayashi:2017puw}.
Hydrodynamical shocks and/or fluid instabilities and then turbulence driven by the magnetorotational instability (MRI) lead to angular momentum transport and heating in the disk, and drive accretion onto the compact object. If a large scale poloidal magnetic field threads the disk, magnetically driven outflows are likely to unbind $\sim 20\%$ of the mass of the disk \citep{Siegel:2017nub,Fernandez:2018kax} -- but this is not a given considering uncertainties about the large scale structure of the magnetic field in post-merger accretion disks. Indeed, while it is possible to grow such a large scale field after merger \citep{Christie:2019lim}, this takes too long to efficiently contribute to the production of winds. A large scale magnetic field generated during merger appears to be required for these winds to exist. 

After $O(100\,{\rm ms})$, the density of the disk decreases enough that neutrino cooling becomes inneficient \citep{Fernandez:2013tya,De:2020jdt}, while the MRI remains active. The disk becomes \emph{advection dominated}. It puffs up to $H/R\sim 1$, and viscous spreading of the disk leads to the ejection of $5\%-25\%$ of the disk mass (viscous outflows) \citep{Fernandez:2013tya}. Neutrino-matter interactions directly impact the properties of magnetically driven outflows, and indirectly impact the properties of viscous outflows (due to neutrino-matter interactions during the early evolution of the disk, before weak-interaction freeze-out).

\begin{figure}[ht]
\centering
\includegraphics[width=0.4\textwidth]{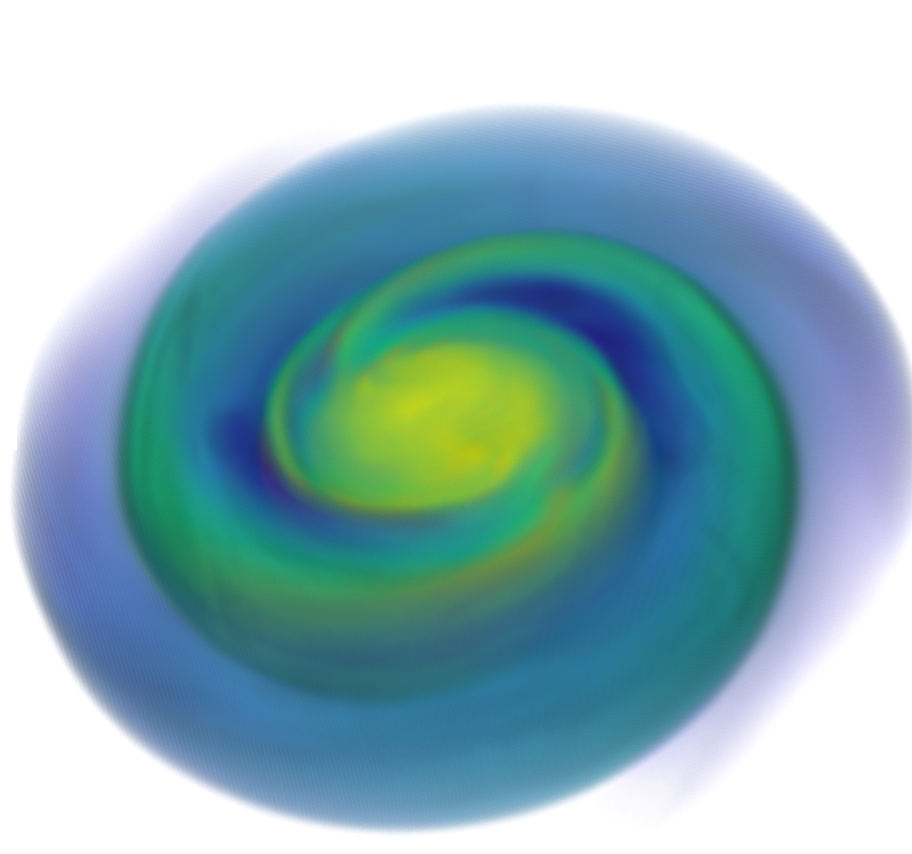}\hspace{1cm}
\includegraphics[width=0.4\textwidth]{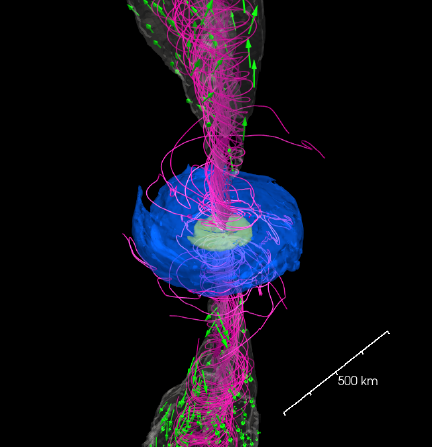}
\caption{Post-merger remnant a few milliseconds after a BNS merger (Left), and $0.3\,{\rm s}$ after a NSBH merger (Right). The BNS system forms a massive, differentially rotating neutron star surrounded by a low-mass accretion disk, with shocked spiral arms visible in the disk. The NSBH system forms an extended accretion disk around the remnant black hole, with collimated magnetic fields in the polar region. The right panel is reproduced from \cite{Hayashi:2021oxy}; the left panel visualizes a simulation from \cite{Foucart:2015gaa}.}
\label{fig:postmerger}
\end{figure}

The post-merger evolution is also impacted by the presence and life time of a massive neutron star remnant. A hot neutron star remnant is a bright source of neutrinos that can accelerate changes to the composition of matter outflows in the polar regions. How efficiently matter can accrete onto the neutron star remains uncertain. Axisymmetric simulations treating the neutron star surface as a hard boundary predict the eventual ejection of most of the remnant disk \citep{Metzger:2014ila}; whether this would remain true for more realistic boundary conditions is unclear, but it is at least likely that a larger fraction of the disk is eventually unbound for neutron star remnants than for black hole remnants. The neutron star remnants themselves are initially differentially rotating, and simulations generally find rotation profiles that are stable to the MRI in most of the star (the angular velocity increases with radius). Some other angular momentum transport mechanism is thus required to bring these remnants to uniform rotation, e.g., convection and/or the Spruit--Taylor dynamo \citep{Margalit:2022rde}. The exact impact of the interaction between the neutron star remnant, its external magnetic field, and the surrounding accretion disk on the evolution of the system remains very uncertain. Examples of post-merger remnants are shown in Fig.~\ref{fig:postmerger}

\subsection{Observables and existing observations}
\label{sec:obs}

The main signals observed so far in neutron star mergers include gravitational wave emission during the late inspiral of the binary towards mergers, SGRBs (and their multi-wavelength afterglows) likely due to relativistic jets powered by the post-merger remnant, and kilonovae. For a system with component masses $m_1,m_2$, the gravitational waves provide us with a very accurate measurement of the chirp mass $M_c = (m_1m_2)^{0.6}/(m_1+m_2)^{0.2}$, as well as, for sufficiently loud signals, less accurate information about the mass ratio (and thus the component masses), the spins of the compact objects, the equation of state of neutron stars (through their tidal deformability), as well as the distance, orientation, and sky localization of the source (especially for multi-detector observations). We will not discuss the gravitational wave signal in much more detail here, as it is not meaningfully impacted by neutrinos. Outflows generated during and after the merger (see previous section) will be the main source of post-merger electromagnetic signals. Relativistic collimated outflows power SGRBs detectable by observers located along  the spin axis of the remnant. As the jet material becomes less relativistic, SGRBs are followed by longer wavelength afterglows detectable by off-axis observers \citep{Fong:2015}. The gamma-ray emission is very short lived ($\lesssim 2\,{\rm s}$ for a typical SGRB), but radio afterglows can still be observed a year after the merger \citep{Mooley:2018clx}. The exact mechanism powering the relativistic jet remains unknown. The most commonly discussed model requires the formation of a large scale poloidal magnetic field threading a black hole remnant, with energy extraction from the black hole's rotation though a Blandford-Znajek-like process \citep{1977MNRAS.179..433B}. Some SGRB models are however powered by neutrino-antineutrino pair annihilations in the polar regions. Explaining the most energetic SGRBs through this mechanism is difficult given what is currently known of the neutrino luminosity of post-merger remnants and the efficiency of the pair annihilation process \citep{just:16}, yet even in a magnetically-powered SGRB, energy deposition due to neutrino pair annihilation or baryon loading of the polar regions due to neutrino-driven winds could impact the formation of a jet \citep{Fujibayashi:2017xsz}.

The properties of kilonovae and the role of neutron star mergers in astrophysical nucleosynthesis are likely to be much more significantly impacted by neutrinos than gravitational waves or even SGRBs. Absorption and emission of electron-type neutrinos ($\nu_e$) and antineutrinos ($\bar \nu_e$) modifies the relative number of neutrons and protons in the fluid. This is usually expressed through the lepton fraction
\beq
Y_l = \frac{n_{e^-}-n_{e^+}+n_{\nu_e}-n_{\bar\nu_e}}{n_p + n_n} 
\eeq
with $n_{e^\pm},n_n,n_p,n_{\nu_e},n_{\bar\nu_e}$ the number density of electrons, positrons, neutrons, protons, $\nu_e$ and $\bar\nu_e$ respectively. Many simulations use the net electron fraction $Y_e$ instead of the lepton fraction, and assume that charge neutrality requires $n_{e^-}-n_{e^+}=n_p$ \footnote{Note that this assumes that muon and tau leptons have vanishing net lepton numbers, i.e. that we have an equal number of particles and antiparticles for heavy leptons.}, so that
\beq
Y_e = \frac{n_p}{n_p+n_n}.
\eeq
The electron fraction is a crucial determinant of the outcome of r-process nucleosynthesis in merger outflows. Low $Y_e$ outflows (roughly $Y_e\lesssim 0.25$) produce heavier r-process elements, while higher $Y_e$ outflows produce lighter r-process elements \citep{Lippuner2015}.
In particular, for the conditions typically observed in merger outflows, there is not much production of elements above the ``2nd peak'' of the r-process (at atomic number $A\sim 130$) for high-$Y_e$ outflows, and an under-production of elements below the 2nd peak for neutron-rich (low $Y_e$) outflows.
Cold outflows that do not interact much with neutrinos are typically neutron-rich, but hotter outflows can end up with $Y_e\sim 0.4-0.5$ due to neutrino-matter interactions. Cooling from neutrino emission and heating from neutrino absorption are also important to the thermodynamics of the remnant and of the outflows, and neutrino absorption in the disk corona and close to the neutron star surface can lead to the production of neutrino-driven winds \citep{Dessart:2008zd}. It is thus clear that neutrino-matter interactions should be properly understood if we aim to model the role of neutron star mergers in the production of r-process elements. 

The impact of neutrinos on kilonovae is less direct but no less important. Most of the r-process occurs within a few seconds of the merger, after which the outflows are mainly composed of radioactively unstable heavy nuclei. Radioactive decays of these nuclei will continue to release energy over much longer timescales. Initially, the outflows are  opaque to most photons, and decay products are thermalized -- except for neutrinos, which immediately escape the outflows. As the density of the outflows decrease, however, they will eventually become optically thin to optical/infrared photons. When this transition happens depends on the composition of the outflows. Lanthanides and actinides, which are among the heavier r-process elements that are only produced by neutron-rich outflows, have much higher opacities than other nuclei produced during the r-process. As a result, neutron-rich outflows become optically thin later than higher $Y_e$ outflows ($\sim 10$ days vs $\sim 1$ day), and the corresponding kilonova signal is redder (peaks in the infrared, instead of in the optical). Overall, the duration, color, and magnitude of a kilonova tell us about the mass of the outflows, their composition, and their velocity \citep{2013ApJ...775...18B}. For a given binary merger, it will also depend on the relative orientation of the binary and the observer, as different types of outflows have different geometry.

Other electromagnetic counterparts to neutron star mergers have been proposed, with no confirmed observations so far. This include bursts of radiation before merger \citep{Tsang:2013}, continuous emission from magnetosphere interactions \citep{PalenzuelaEtAl:2013}, coherent emission from magnetosphere interactions \citep{Most:2022ayk}, and months to decades-long synchrotron radio emission from the mildly relativistic ejecta as it interacts with the interstellar medium \citep{Hotokezaka:2016clu}. Neutrinos have no impact on the first three, however, and only a minor impact on the third (as neutrino-matter interactions may slightly change the mass / velocity of the outflows). More detailed discussions of the range of electromagnetic transients that may follow a merger can be found, e.g., in \cite{Fernandez:2015use,Burns:2019byj}

Electromagnetic emission from neutron star mergers has likely been observed for decades now in the form of SGRBs, and a first kilonova may have been observed in the afterglow of GRB130603B as early as 2013 \citep{Tanvir:2013,Berger:2013}. However, our current understanding of the engine powering SGRBs is not sufficient to provide us with much information about the parameters of the binary system that created the burst -- or even to differentiate between a BNS and NSBH merger. Gravitational wave observations provide more direct information about the properties of the compact objects. So far, two systems have been observed with component masses most easily explained by the merger of two neutron stars: GW170817 \citep{LIGOScientific:2017vwq} and GW1902425 \citep{Abbott:2020uma}. The former is a relatively low mass system, whose observation was followed by a weak SGRB (most likely observed off-axis), radio emission most likely associated with a relativistic jet, and a clear kilonova signal most easily explain by a combination of at least two outflow components -- one that led to strong r-process nucleosynthesis, and one that did not. The exact process that produced these outflows remain a subject of research today. GW190425 has a higher total mass ($3.4\,M_\odot$). There was no observed electromagnetic counterpart to that signal, a relatively unsurprising result considering the large uncertainty in the location of the source and the high likelihood that such a system did not eject a significant amount of matter \citep{Barbieri:2020ebt,Raaijmakers:2021slr,Dudi:2021abi,Camilletti:2022jms}. At least two NSBH mergers were observed in 2020 \citep{LIGOScientific:2021qlt}, with more candidates also available in the latest gravitational wave catalogue \citep{LIGOScientific:2021djp}. None of these systems was however expected to lead to the disruption of their neutron star, and thus their lack of electromagnetic counterpart was unsurprising.

Overall, we note that the analysis of current and future observations of neutron star mergers would benefit from accurate models of kilonova signals, as well as from an improved understanding of the engine behind gamma-ray bursts. In that respect, it is particularly important to understand the role of neutrinos in setting the composition of the outflows powering kilonovae, and possibly their impact on the production of relativistic jets. In the rest of this review, we will mainly focus on these issues, and on the methods available to evolve neutrinos in merger simulations.

\section{Neutrinos in mergers}
\label{sec:nu}

\subsection{Definitions}

When solving the general relativistic equations of radiation transport, we would ideally evolve Boltzmann's equation, or the quantum kinetics equations (QKE, when accounting for neutrino oscillations). Classically, we evolve the distribution function of neutrinos $f_\nu(t,x^i,p_j)$, defined such that 
\beq
N = \int_V d^3x \frac{d^3p}{h^3} f_\nu(t,x^i,p_j)
\eeq
is the number of neutrinos within a 6D volume of phase space $V$. Here, $x^i$ are the spatial coordinates and $p_j$ the spatial components of the 4-momentum one-form $p_\mu$, while $h$ is Planck's constant. 

When using the classical equations of radiation transport, we usually neglect neutrino masses and assume $p^\mu p_\mu = 0$. Boltzmann's equation is then
\beq
p^\alpha \left[\frac{\partial f_\nu}{\partial x^\alpha}-\Gamma^i_{\alpha\gamma} p^\gamma \frac{\partial f_\nu}{\partial p^i}\right] = \nu \left[\frac{df_\nu}{d\tau}\right]_{\rm collisions}
\eeq
with $\tau$ the proper time in the fluid frame, $\nu$ the neutrino energy in the fluid frame, and $\Gamma^\alpha_{\beta\gamma}$ the Christoffel symbols. The left-hand side simply implies that neutrinos follow null geodesics, while the right-hand side includes all neutrino-matter and neutrino-neutrino interactions, and thus hides most of the complexity in these equations. We note that we should evolve a separate $f_\nu$ for each type of neutrinos ($\nu_e,\nu_\mu,\nu_\tau$) and antineutrinos ($\bar\nu_e,\bar\nu_\mu,\bar\nu_\tau$); and that these distributions functions may be coupled through the collision terms. As neutrinos are fermions, we have $0\leq f_\nu \leq 1$. 

The spacial coordinate volume $d^3x = dx dy dz$ and momentum volume $d^3p = dp_x dp_y dp_z$ are not invariant under coordinate transformations, but $d^3x p^t \sqrt{-g}$ and $d^3p (p^t \sqrt{-g})^{-1}$ are, with $g$ the determinant of the spacetime metric $g_{\mu\nu}$. Thus $d^3x d^3p$ is invariant under coordinate transformations. The stress-energy tensor of neutrinos at $(t,x^i)$ is
\beq
T^{\alpha\beta}(t,x^i) = \int \frac{d^3p}{h^3 p^t \sqrt{-g}} p^\alpha p^\beta f_\nu(t,x^i,p_j).  
\label{eq:Tab}
\eeq

In general relativistic merger simulations, we often use the 3+1 decomposition of the metric
\beq
ds^2 = g_{\mu\nu} dx^\mu dx^\nu = -\alpha^2 dt^2 + \gamma_{ij} (dx^i + \beta^i dt) (dx^j +\beta^j dt),
\eeq
with $\alpha$ the lapse, $\beta^i$ the shift, and $\gamma_{ij}$ the 3-metric on a slice of constant time $t$. The unit normal one-form to such a slice is then $n_\mu=(-\alpha,0,0,0)$, and the 4-vector $n^\mu=g^{\mu\nu}n_\nu$ can be interpreted as the 4-velocity of an observer moving along that normal -- which we will call \emph{normal observer} from now on. From there, we can deduce that $\epsilon=-p^\mu n_\mu=\alpha p^t$ is the energy of a neutrino of 4-momentum $p^\mu$ as measured by a normal observer. More generally, the energy of a neutrino measured by an observer with 4-velocity $u^\mu$ is $\nu = -p^\mu u_\mu$. Here, we will generally reserve the symbol $\epsilon$ for the energy measured by normal observers, and $\nu$ for the energy measured in the fluid rest frame, i.e., when $u^\mu$ is the 4-velocity of the fluid.

\subsubsection{Equilibrium distribution}

We will often make use of the equilibrium distribution of neutrinos. For neutrinos in equilibrium with a fluid at temperature $T$ moving with 4-velocity $u^\mu$, that is the Fermi-Dirac distribution
\beq
f^{\rm eq} = \frac{1}{1+\exp{\left[\frac{\nu-\mu}{k_B T}\right]}}
\label{eq:feq}
\eeq
with $\mu$ the chemical potential of neutrinos, and $k_B$ Boltzmann's constant. We note that in an orthonormal frame $(\hat t,\hat x^i)$ the energy density of neutrinos is
\beq
E_\nu = T_{\hat t \hat t} = \int \frac{d^3p}{h^3} \hat\epsilon f_\nu
\eeq
with $\hat \epsilon=p^{\hat t}$ the energy of neutrinos as measured by a stationary observer in the orthonormal frame. We thus see that we recover the expected results for the equilibrium energy of a fermion gas in the fluid frame,
\beq
E_{\rm eq} = \int \frac{d^3p}{h^3} \frac{\nu}{1+\exp{\left[\frac{\nu-\mu}{k_B T}\right]}} = 4\pi \int \frac{d\nu}{(hc)^3} \frac{\nu^3}{1+\exp{\left[\frac{\nu-\mu}{k_B T}\right]}},
\eeq
where in the last expression we used the special relativistic result $\nu = \|p\|c$. This is more easily expressed in terms of the Fermi integrals $F_n$, which we will use extensively in this section:
\beq
F_n(\eta) = \int_0^\infty dx \frac{x^n}{1+\exp{(x-\eta)}}.
\eeq
From this definition, we see that
\beq
E_{\rm eq}(T,\mu) = 4\pi \frac{(k_BT)^4}{(hc)^3} F_3\left(\frac{\mu}{k_BT}\right).
\eeq
Similarly, the equilibrium number density of neutrinos is
\beq
N_{\rm eq}(T,\mu) = 4\pi \frac{(k_BT)^3}{(hc)^3} F_2\left(\frac{\mu}{k_BT}\right)
\eeq
and the average energy of neutrinos in equilibrium with the fluid
\beq
\langle \nu_{\rm eq}\rangle =  \frac{F_3\left(\frac{\mu}{k_BT}\right)}{F_2\left(\frac{\mu}{k_BT}\right)} k_B T
\eeq
(which asymptotes to $3.15k_BT$ at low densities, when $\|\mu\| \ll k_B T$).

\subsection{Commonly considered reactions}
\label{sec:reac}

Let us now discuss the various neutrino-matter interactions that are commonly considered in neutron star merger simulations. Our objective here is not to provide detailed derivations of all interaction rates, but rather to review the reactions that may be taken into consideration and to get reasonable estimates of the scaling of reaction rates with the fluid properties. This will allow us to estimate when different reactions become important to the evolution of the system. Accordingly, for the sake of brevity, the cross-sections and reaction rates presented here sometimes make stronger approximations than what is done in merger simulations. However, for each reaction we provide references to more detailed discussions of these cross-sections. We will also make use of our discussion of the $p+e^- \leftrightarrow n+\nu_e$ and $e^+e^-\leftrightarrow \nu\bar\nu$ reactions to illustrate a number of issues that arise when attempting to include collision terms in the radiation transport equations, and thus discuss these reactions in more detail than the others. Given the significant overlap between reactions important to neutron star merger simulations and reactions important to core-collapse supernova simulations, a number of expressions in this section are slight modifications of the interaction rates presented in the review of neutrino reactions in core-collapse supernovae of  \cite{Burrows:2004vq}, though for numerical estimates of interaction rates we focus on the conditions most commonly found in neutron star mergers and post-merger remnants.

\subsubsection{Charged-current reactions}

The reactions with the strongest impact on the observable properties of neutron star mergers involve absorption and emission of $\nu_e$ and $\bar \nu_e$. Indeed, these reactions are often (but not always) the main source of cooling in the system, and they are the only reactions that lead to changes in the electron fraction $Y_e$ of the fluid. In the hot, dense remnant of a BNS or NSBH merger, this mostly occurs through the reactions
\beq
p + e^- \leftrightarrow n + \nu_e;\,\, n+ e^+ \leftrightarrow p + \bar \nu_e
\eeq
which are typically included at least approximately in all merger simulations that attempt to account for neutrino-matter interactions. 

Self-consistently calculating the forward and backward reaction rates can be difficult. Final state blocking means that these reactions depend on the distribution functions of $p,n,e^+,e^-,\nu_e,\bar \nu_e$. While we can typically assume equilibrium distributions at the fluid temperature and composition for $n,p,e^+,e^-$ in neutron star mergers, at least in regions where neutrino-matter interactions are important, the neutrinos may be far out of equilibrium -- and many approximate schemes used in simulations today do not contain enough information about the neutrino distribution function to fully account for the value of $f_\nu$ in all reactions. 

To illustrate these issues, and some of the ways in which they are handled in existing simulations, let us consider the cross-section per baryon for the reaction $n+\nu_e \rightarrow p+e^-$, the dominant absorption process in merger outflows, derived by~\cite{1985ApJS...58..771B}. Following the notation of \cite{Burrows:2004vq}, we get
\beq
\sigma_{\nu_e n} = 1.38 \sigma_0 \left(\frac{\nu_{\nu_e}+\Delta_{np}}{m_e c^2}\right)^2 
\left[1-\left(\frac{m_e c^2}{\nu_{\nu_e}+\Delta_{np}}\right)^2\right]^{1/2} W_M
\eeq
with
\beq
\sigma_0 = 1.705\times 10^{-44} \,{\rm cm^{2}},
\eeq
$\nu_{\nu_e}$ the fluid frame neutrino energy, $\Delta_{np} = (m_n-m_p)c^2 = 1.293\,{\rm MeV}$ the difference in rest mass energy between neutrons and protons, $m_e$ the mass of an electron, and $W_M$ a small correction for weak magnetism and recoil ($2.5\%$ for $20\,{\rm MeV}$ neutrinos) \citep{PhysRevD.29.1918}. Neutrinos in BNS and NSBH mergers have typical energies $\nu \gtrsim 10\,{\rm MeV}$, significantly larger than the rest mass energy of an electron. Thus, to a reasonably good approximation (for the purpose of our qualitative discussion here at least),
\beq
\sigma_{\nu_e n} \approx 1.38 \sigma_0 \left(\frac{\nu_{\nu_e}}{m_e c^2}\right)^2.
\label{eq:nnue}
\eeq
This dependence of neutrino cross-sections on the square of the neutrino energies is found in many reactions relevant to neutron star mergers, and is going to be a significant source of uncertainty in our simulations, as many approximate transport algorithms do not provide detailed information about the neutrino spectrum.

The opacity for the absorption of $\nu_e$ on $n$ is then 
\beq
\kappa_a \approx \int \frac{2d^3p_n}{h^3} f_n(E) (1-f_p(E)) \sigma_{\nu_e n} \approx n_n \sigma_{\nu_e n}
\eeq
with $f_n,f_p$ the distribution functions of neutrons and protons, and $E\approx p^2/2m$ the kinetic energy of the baryons (ignoring the difference in mass between protons and neutrons and momentum transfer onto the proton). In the last expression, which ignores the final state blocking factor of the protons, $n_n$ is the neutron number density. That expression would be very inaccurate in the densest region of a star (where $f_p$ cannot be neglected), but is quite accurate in the lower-density regions where neutrinos decouple from the fluid. 

To gain a more intuitive understanding of the rate of these interactions, let us assume that the typical length scale within a neutron star is $\sim 1\,{\rm km}$. We can see from this expression that for a $20\,{\rm MeV}$ neutrino, we expect $\kappa_a = 1\,{\rm km^{-1}}$ for $n_n\sim 10^{-3}\,{\rm fm^{-3}}$, i.e., for a neutron mass density of $\sim 10^{12}\,{\rm g/cm^3}$. As the center of a neutron star has density $\rho_c \sim 10^{15}\,{\rm g/cm^3}$, we see that neutrinos inside the neutron star have a mean free path much shorter than the size of the star, and decouple from the matter as they move through the crust of the neutron star. 

Similar scalings apply to the $p+\bar\nu_e \rightarrow n + e^+$ reaction, as
 \beqn
\sigma_{\bar\nu_e p} &=& 1.38 \sigma_0 \left(\frac{\nu_{\bar\nu_e}-\Delta_{np}}{m_e c^2}\right)^2 
\left[1-\left(\frac{m_e c^2}{\nu_{\bar\nu_e}-\Delta_{np}}\right)^2\right]^{1/2} W_{\bar M}\\
&\approx& 1.38 \sigma_0 \left(\frac{\nu_{\bar\nu_e}}{m_e c^2}\right)^2
\eeqn
and $\kappa_a \approx n_p \sigma_{\bar\nu_e p}$ for the absorption of $\bar \nu_e$ on protons, under the same assumptions as for absorption onto neutrons. The correction $W_{\bar M}$ is more significant than $W_M$ ($\sim 15\%$ at $20\,{\rm MeV}$) \citep{PhysRevD.29.1918,PhysRevD.65.043001}, though still not large enough to impact our order of magnitude estimates. As $n_p<n_n$ in most regions of a neutron star merger remnant, the absorption opacity for $\bar\nu_e$ is smaller than for $\nu_e$. 

It is also possible to include in simulations the impact of $\nu_e$ and/or $\bar\nu_e$ absorption on atomic nuclei. This is typically more important in the core-collapse context than in mergers, as in mergers most of the matter is in the form of free nucleons in regions where neutrino-matter interactions are significant. Additionally, simulations do not keep track of the abundances of individual nuclei, and equations of state for the fluid do not always contain that information, complicating any estimate of the absorption cross-section for this process. Cross-sections for the absorption of $\nu_e$ onto nuclei can be found in \cite{1985ApJS...58..771B}. In high-density, low-temperature, neutron-rich regions inside of merging neutron stars, the modified URCA processes \citep{Yakovlev:2000jp,Alford:2021ogv}
\beq
N+n\rightarrow N+p+e^-+\bar \nu_e;\,\,N+p+e^-\rightarrow N+n+\nu
\eeq
(with $N$ a spectator nucleon) may also play a role in the evolution of the system through the creation of an effective bulk viscosity in the post-merger remnant \citep{Alford:2017rxf}.

In the expressions derived so far for neutrino absorption, we have generally ignored final state blocking factors. These can however be approximately calculated if we rely on the fact that the fluid particles are in statistical equilibrium at a given temperature. Final state blocking factors for neutrino emission are slightly more complex to take into account. For neutrinos of a given energy and momentum, the neutrino emission rate will generally be of the form $\eta = \eta^* (1-f_\nu)$, where the $(1-f_\nu)$ term captures Pauli blocking for neutrinos in the final state. This is not a form that is practical to use in simulations, as we would like the emission rate and opacities to depend solely on the properties of the fluid, without any dependency on $f_\nu$. \cite{Burrows:2004vq} show that a convenient redefinition of the emissivity and absorption opacity can solve this problem. If we directly use $\eta^*$ as our emission rate (without neutrino blocking factor), and define $\kappa_a^* = \kappa_a / (1-f_\nu^{\rm eq})$ as our absorption opacity (with 
$f_{\nu}^{\rm eq}$ taken from Eq.~\eqref{eq:feq}), then the collision term for charged-current reactions in Boltzmann's equation can be written in the two equivalent ways
\beq
\left[\frac{df_{\nu}}{d\tau}\right]_{\rm collisions} = \eta_\nu - c\kappa_a f_\nu = \eta_\nu^* - c\kappa_a^* f_\nu,
\eeq
with $\eta$ the emissivity per unit of solid angle and neutrino energy. Importantly, in the first expression $\eta_\nu$ depends of $f_\nu$, but in the second $\eta_\nu^*$ does not. Accordingly, most simulations use $\eta^*$ and $\kappa_a^*$ to parametrize neutrino-matter interactions. In our discussion of numerical algorithms for neutrino transport, \emph{emissivity} and \emph{absorption opacity} will generally refer to these corrected values.

We also note that when all reactions are accounted for, $\eta^* =c\kappa_a^* f_\nu^{\rm eq}$ (Kirchoff's law).
This allows us to calculate only one of $(\eta^*,\kappa^*_a)$, then set the other to make sure that the equilibrium energy density of neutrinos has the desired physical value. This is particularly useful in dense, hot regions, where neutrinos quickly reach equilibrium with the fluid. In that regime, the exact emission and absorption rate can be more difficult to calculate (due to blocking factors), but they are also fairly unimportant: what matters is that neutrinos quickly reach their equilibrium density, and then diffuse through the dense regions. This is guaranteed when using Kirchoff's law, even if $\eta^*$ and $\kappa_a^*$ are not extremely accurate.

The total emission rate of neutrinos due to a given reaction can be calculated by integrating $\eta^*$ over both solid angle and neutrino energy. In terms of the absorption opacity, we get
\beq
Q = 4\pi \int \frac{d\nu}{(hc)^3} \frac{\nu^3}{1+\exp{\left[\frac{\nu-\mu_\nu}{k_B T}\right]}} \kappa_a^*(\nu) c.
\eeq
For comparison with results for other reactions, we can estimate this emission rate for $\nu_e$, ignoring the final state blocking factor of protons in the inverse reaction and using $W_M\sim 1$. We then get for the emission of electron neutrinos due to electron capture on protons (energy per unit volume)
\beq
Q_{pe^-} \approx 1.38 (4\pi) \sigma_0 c n_n \left(\frac{k_B T}{m_e c^2}\right)^2
\left(\frac{k_B T}{hc}\right)^3 F_5(\eta_{\nu_e}^{\rm eq}) (k_B T)
\eeq
with $\eta = \mu/(k_B T)$. Similarly, the number of neutrinos emitted per unit volume is simply
\beq
N_{pe^-} \approx 1.38 (4\pi) \sigma_0 c n_n \left(\frac{k_B T}{m_e c^2}\right)^2
\left(\frac{k_B T}{hc}\right)^3 F_4(\eta_{\nu_e}^{\rm eq}),
\eeq
and the average energy of emitted neutrinos
\beq
\langle \nu \rangle = \frac{ F_5(\eta_{\nu_e}^{\rm eq})}{ F_4(\eta_{\nu_e}^{\rm eq})} k_B T.
\eeq
For $\|\eta_\nu\|\ll1$, $\langle \nu \rangle\sim 5.1 k_B T$. We note that this is higher than the average energy of neutrinos in equilibrium with the fluid. This will generally be true whenever neutrinos are allowed to directly escape from an emission region instead of thermalizing with the fluid first. A more explicit expression for $Q_{pe^-}$ is
\beq
Q_{pe^-} \approx (3.4\times 10^{30}{\rm erg\,s^{-1}cm^{-3}}) \left[\frac{k_BT}{\rm MeV}\right]^6 \frac{n_n}{10^{36}\,{\rm cm^{-3}}} \frac{F_5(\eta_{\nu_e}^{\rm eq})}{F_5(0)}.
\eeq
We see that the emission rate of neutrinos has a strong dependence in the fluid temperature, with $Q\propto T^6$, and a linear dependence in the fluid density (ignoring the Fermi integral term). The emission rate of $\bar\nu_e$ can be computed in the exact same manner,
\beq
Q_{ne^+} \approx (3.4\times 10^{30}{\rm erg\,s^{-1}cm^{-3}}) \left[\frac{k_BT}{\rm MeV}\right]^6 \frac{n_p}{10^{36}\,{\rm cm^{-3}}} \frac{F_5(-\eta_{\nu_e}^{\rm eq})}{F_5(0)}.
\eeq
In this expression, we made use of the fact that $\eta_{\bar \nu_e}^{\rm eq} = - \eta_{\nu_e}^{\rm eq}$. The dependence of these emission rates on $n_n$ and $n_p$ may seem counterintuitive, as $Q_{pe^-}$ involves absorption of electrons on protons, yet is proportional to $n_n$. This is however a natural result of using Kirchoff's law; the complete dependence of  $Q_{pe^-}$ in the density of all fluid particles is practically hidden in the Fermi integral term $F_5(\eta_{\nu_e}^{\rm eq})$, and the assumption of statistical equilibrium in the fluid. In particular, as $F_n(\eta)$ monotonically increase with $\eta$, and neutrino emission in post-merger remnants comes from regions of the fluid where $\eta_{\nu_e}<0$ (more neutron-rich than in equilibrium), we generally get $Q_{ne^+}>Q_{pe^-}$ even though $n_p<n_n$.

\subsubsection{Pair processes}
\label{sec:thpairs}

After charged current reactions, the most commonly considered processes for the emission and asborption of neutrinos are the pair processes
\beq
e^+ e^- \leftrightarrow \nu \bar\nu;\,\, \gamma\gamma \leftrightarrow \nu \bar\nu;\,\,
N + N \leftrightarrow N + N + \nu \bar \nu
\eeq
i.e., electron-positron annihilation, plasmon decay, and nucleon-nucleon Bremsstrahlung. Here, each pair can be $\nu_e\bar\nu_e$, $\nu_\mu\bar\nu_\mu$, or $\nu_\tau\bar\nu_\tau$. Pair processes will be the dominant source of neutrino emission for muon and tau neutrinos, as charged-current reactions involving muons and taus are significantly less common than charged current reactions involving electrons in the merger context (the mass of a muon is $105\,{\rm MeV}$, while most of the post-merger remnant has temperature $T\lesssim 50\,{\rm MeV}$, and the neutrinospheres and optically thin regions are at even lower temperatures). Pair processes are however harder to accurately include in simulations due to their nonlinear dependencies in the neutrino distribution functions. The reaction rates for the $\nu\bar\nu$ pair productions (forward reactions) depend on the distribution function of both neutrinos and antineutrinos through blocking factors, which are typically difficult to estimate accurately with existing transport algorithms. Worse, the reaction rates for pair annihilations (inverse reactions) are directly propoprtional to the product of the distribution functions of neutrinos and antineutrinos. 

Let us consider for example the reactions $\nu\bar\nu \rightarrow e^+ e^-$, for neutrinos of energy significantly higher than $m_e c^2$ (as appropriate in neutron star mergers). We can slightly adapt the results of \cite{1999ApJ...517..859S}, based on the Newtonian rate calculations of \cite{1986ApJ...309..653C,1987ApJ...314L...7G}, to find the rate of momentum deposition per unit volume 
\beq
Q^\alpha_{\nu\bar \nu} = \int\int \frac{d^3p_{(\nu)}}{\sqrt{-g}p^t_{(\nu)}h^3}\frac{d^3p_{(\bar\nu)}}{\sqrt{-g}p^t_{(\bar\nu)}h^3} f_{(\nu)} f_{(\bar \nu)} (p^\alpha_{(\nu)} +p^\alpha_{(\bar\nu)})
\frac{DG_F^2}{3\pi}\left(-p_{(\nu)}^\beta p_{\beta(\bar\nu)}\right)^2
\label{eq:Qpairs}
\eeq
with $G_F=5.29\times 10^{-44}\,{\rm cm^2\, MeV^{-2}}$ and $D=2.34$ for electron type neutrinos, while $D=0.50$ for muon or tau neutrinos. We have here chosen to rewrite the results of \cite{1999ApJ...517..859S} into a manifestly covariant expression more appropriate for general relativistic simulations. From this expression, we can see that the probability that a given neutrino is annihilated will depend on both the momentum of that neutrino and the distribution function of its antiparticle. 

To limit the computational cost of this calculation, it is often convenient to make some assumptions regarding the distribution function of neutrinos, e.g. ignoring neutrino blocking factors (for the forward reactions), assuming equilibrium distributions of neutrinos (for either direction), or, in moment schemes, using approximate moments of the distribution functions (for the backward reactions). The most common strategy in existing merger simulations has been to compute the forward reaction rates assuming equilibrium distributions of neutrinos or ignoring blocking factors, either for all neutrinos or only for the muon and tau neutrinos. The inverse reaction rates are then computed using Kirchhoff's law, even though that law is not necessarily valid for pair processes \citep{2015ApJS..219...24O}. These approximations are generally reasonable for heavy lepton neutrinos close to the neutrinosphere, i.e., where most of the neutrinos that leave the remnant are emitted, because as long as charged-current reactions including muons and taus are negligible, the distribution functions of $\nu_\mu,\bar\nu_\mu,\nu_\tau,\bar \nu_\tau$ are all identical and close to equilibrium. They are however very unreliable for electron-type neutrinos and for calculations of the rate of $\nu\bar\nu$ annihilation in regions where the neutrinos are not in equilibrium with the fluid (e.g. in the polar regions). We will consider some of the ways in which the latter process has been studied in our discussion of specific radiation transport algorithm and simulations.

Now that we have established the difficulty of properly treating pair processes, let us estimate their importance in the merger context. We begin again with $e^+e^-$ annihilation. Ignoring neutrino blocking factors, \cite{Burrows:2004vq} integrate the reactions rate of \cite{PhysRevD.6.941} to find a total emissivity in $\nu\bar\nu$ pair of
\beq
Q_{e^+ e^-} = Q_0 \left[\frac{k_BT}{\rm MeV}\right]^9 \frac{F_4(\eta_e)F_3(-\eta_e) + F_3(\eta_e) F_4(-\eta_e)}{2F_4(0)F_3(0)}
\eeq
with $Q_0=9.76\times 10^{24} {\rm ergs\,cm^{-3}\,s^{-1}}$ for $\nu_e\bar\nu_e$, and $Q_0=4.17\times 10^{24} {\rm ergs\,cm^{-3}\,s^{-1}}$ for all other neutrinos combined. The average energy of the emitted neutrinos in the fluid frame is
\beq
\langle \nu \rangle=\frac{1}{2} \left(\frac{F_4(\eta_e)}{F_3(\eta_e)}+\frac{F_4(-\eta_e)}{F_3(-\eta_e)}\right) T
\eeq
(e.g., $\langle \nu \rangle \approx 4.1T$ when $\eta_e=0$). Approximate expressions for the energy spectrum of the neutrinos are also found in \cite{Burrows:2004vq}, following the work of \cite{1985ApJS...58..771B}. If we compare this result to $Q_{pe^-}$ and $Q_{ne^+}$ from the previous section, we see that for $\nu_e \bar\nu_e$, pair processes will only dominate over charge current reactions in very hot and/or low density regions of the fluid, where neutrinos will either rapidly reach their equilibrium distribution or rapidly cool the fluid. In such cases, getting exact reaction rates is not overly important as long as we obtain the correct equilibrium distribution and have sufficiently high emission rates. Even in dense regions of the fluid, neglecting $\nu_e\bar\nu_e$ production is thus not a particularly strong approximation (but neglecting pair annihilation in low-density regions might be, as we will see).

What about the heavy-lepton neutrinos? The equilibrium energy density of neutrinos is
\beq
E_\nu \approx (6\times 10^{25} {\rm ergs\,cm^{-3}}) \left[\frac{k_BT}{\rm MeV}\right]^4 
 \frac{F_3(\eta_{\nu})}{F_3(0)}.
\eeq
At $T=1\,{\rm MeV}$, the timescale for neutrinos to reach that equilibrium density solely through $e^+e^-$ emission is thus $O(10\,{\rm s})$, but at $T=10\,{\rm MeV}$, it is $O(0.1\,{\rm ms})$, i.e., much shorter than the dynamical timescale of a neutron star merger. In hot regions, heavy-lepton neutrinos (muons and taus) will thus reach their expected equilibrium density, and the neutrino luminosity of $\nu_\mu\bar\nu_\mu \nu_\tau \bar\nu_\tau$ will be set by the diffusion timescale of neutrinos through the hot, dense remnant. For heavy-lepton neutrinos, ignoring pair processes (and missing the associated cooling of the remnant) would be significantly worse than incuding approximate reaction rates, as long as those rates properly recover the equilibrium energy of neutrinos in dense regions.

Let us now briefly consider other pair processes. For nucleon-nucleon Bremsstrahlung, \cite{Burrows:2004vq} (building on results by \cite{PhysRevD.38.2338,Hannestad:1997gc}) find the total neutrino emissivity per species to be
\beq
Q_{nb} \approx  (1.5\times 10^{26} {\rm ergs\,cm^{-3}\,s^{-1}}) \left(\frac{n_n}{10^{36}\,{\rm cm^{-3}}}\right)^2 \left[\frac{k_BT}{\rm MeV}\right]^{5.5}.
\eeq
We see that Bremsstrahlung will dominate over $e^+e^-$ annihilation in denser, colder regions. In the densest region of a post-merger accretion disk (typically $n_n \sim 10^{35-37}\,{\rm cm^{-3}}$, $T\sim (1-10)\,{\rm MeV}$), we see that the process dominating the production of heavy-lepton neutrinos may thus vary, and we can neglect neither Bremsstrahlung nor $e^+e^-$ pair production/annihilation.

Approximate formula for the total energy emission from plasmon decays and for the average energy of the neutrinos emitted through that process can be found in \cite{Ruffert1996}. They are equivalent to
\beq
Q_{pl} = Q_{0,pl} \left[\frac{k_BT}{\rm MeV}\right]^{9} \gamma^6 e^{-\gamma} (1+\gamma) \left(2+\frac{\gamma^2}{1+\gamma}\right) B
\eeq
with $Q_{0,pl} =(6\times 10^{23} {\rm ergs\,cm^{-3}\,s^{-1}})$ per species for electron-type neutrinos and 
 $Q_{0,pl} =(10^{21} {\rm ergs\,cm^{-3}\,s^{-1}})$ per species for other neutrinos. The blocking factor $B=\langle(1-f_\nu) \rangle\langle(1-f_{\bar\nu}) \rangle$ can only be evaluated assuming a specific neutrino distribution function, while $\gamma \approx 0.056 \sqrt{(\pi^2+3\eta_e^2)/3}$ is a parameter with strong sensitivity to the electron degeneracy parameter $\eta_e$. We see that we need relatively fine-tuned conditions for plasmon decay to dominate over pair annihilation, especially for heavy-lepton neutrinos -- as  a result, this reaction is often ignored in merger simulations.

From these estimates of the emissivity of pair processes, we can understand one additional difficulty in the use of these processes in simulations. Both $e^+e^-$ creation/annihilation and plasmon decays have $Q\propto T^9$, with no explicit dependence in the fluid density (at least in regions where blocking factors are negligible). This can prove problematic in merger simulations, where numerical errors can lead to the creation of hot low-density regions whose properties are not necessarily well modeled by equations of state built to capture the properties of dense matter. As a result, some simulations ignore pair processes below an ad-hoc density threshold.

\subsubsection{Neutrino scattering}

Scattering of neutrinos on protons, neutrons, nuclei and electrons plays an important role in setting the diffusion timescale of neutrinos through the densest regions of merger remnants. The total cross-sections per baryon for the nearly elastic scattering of neutrinos onto protons and neutrons are \citep{1976Ap&SS..41..221Y,1985ApJS...58..771B,Burrows:2004vq}
\beq
\sigma_{s,p} \approx 1.1\sigma_0 \left[\frac{\nu}{1\,{\rm MeV}}\right]^2;\,\,
\sigma_{s,n} \approx 1.3\sigma_0 \left[\frac{\nu}{1\,{\rm MeV}}\right]^2
\eeq
and the scattering opacity for these two processes combined is thus
\beq
\kappa_s \approx (1.1n_p + 1.3n_n) \sigma_0 \left[\frac{\nu}{1\,{\rm MeV}}\right]^2.
\eeq
We see that these quasi-elastic scatterings are about as likely as charged-current absorption for $\nu_e$ and $\bar\nu_e$, and will often be a dominant contribution to the total opacity of the fluid for heavy-lepton neutrinos. We also note that the differential cross-sections are
\beq
\frac{d\sigma_{n,p}}{d\Omega} = \frac{\sigma_{n,p}}{4\pi} \left(1 + \delta_{n,p}\mu\right)
\eeq
with $\mu=\cos\theta$ and $\theta$ the scattering angle. As $\delta_p\sim -0.2$ and $\delta_n\sim -0.1$ at the most relevant neutrino energies \citep{Burrows:2004vq}, back-scattering is favored. However, most merger simulations assume isotropic elastic scatterings in the fluid frame ($\delta_{n,p}=0$), an approximation whose impact on simulation results has not been tested so far.

Similar calculations can be made for scattering on atomic nuclei. For example, elastric scattering on $\alpha$ particles has a total cross-section per nucleus of \citep{1976Ap&SS..41..221Y,Burrows:2004vq}
\beq
\sigma_{s,\alpha} \approx 0.8 \sigma_0 \left[\frac{\nu}{1\,{\rm MeV}}\right]^2.
\eeq
We note that in the neutron star merger context, we typically have $n_\alpha \ll n_n$ in regions where neutrino scattering is important, and similar results apply to heavier nuclei. As many equations of state used in merger simulations do not provide detailed information about the abundances of individual atomic nuclei, the contribution of nuclei to the total scattering opacity is often only approximately taken into account (e.g. considering only $\alpha$ particles, or $\alpha$ particles and some `representative' nucleus of fixed proton number $Z$ and atomic number $A$), or completely ignored. 

Including inelastric scattering of neutrinos on electrons is a more difficult problem, and as a result inelastic scattering has not so far been taken into account in merger simulations. To understand these issues, we can look at the methods used to treat inelastic scattering in core-collapse supernovae \citep{1985ApJS...58..771B,Burrows:2004vq}. The relevant part of Boltzmann's equation can be written
\beq
\left[\frac{d f_\nu}{d \tau}\right] = (1-f_\nu) \int \frac{d^3p'}{h^3} f'_\nu R^{\rm in}(\nu,\nu',\cos\theta) - f_\nu \int \frac{d^3p'}{h^3} (1-f'_\nu) R^{\rm out}(\nu,\nu',\cos\theta).
\eeq
Note that $f_\nu$ is the distribution function for neutrinos with energy $\nu$ and momentum $p_\mu$, while $f_\nu'$ is the distribution function for neutrinos with energy $\nu'$ and momentum $p_\mu'$. $R^{\rm in,out}$ are the scattering kernels to scatter into/out of the energy bin $\nu$ from/to $\nu'$. Even ignoring the blocking factors $(1-f_\nu)$ and $(1-f_\nu')$, the collision terms clearly depend on the distribution function of neutrinos, and couple the values of $f_\nu$ at all neutrino momenta. One possible approximation is to use a truncated expansion of the kernels in $\cos\theta$:
\beq
R^{\rm in,out}(\nu,\nu',\cos\theta) \approx \frac{1}{2} \Phi_0^{\rm in,out}(\nu,\nu') + \frac{3}{2} \Phi_1^{\rm in,out}(\nu,\nu') \cos\theta
\eeq
with $\Phi_{0,1}$ known functions of the incoming and outgoing neutrino energies. The integrals over $p_i'$ are then similarly truncated using moments of the distribution function $f_\nu'$. While this makes the evolution of $f_\nu$ slightly more tractable numerically, we still end up with numerically stiff terms coupling every pair of neutrino energies, which makes these reactions expensive to include in simulations. 

The scattering kernels have complex dependencies in the incoming and outgoing neutrino energies, and the temperature of the fluid (which sets the electron distribution function). As a very rough order of magnitude estimate, and assuming that the neutrinos have energies larger than or comparable to the electrons, we have
\beq
R^{\rm in,out} \sim \sigma_0 c  \left[\frac{\epsilon_\nu\epsilon_e}{ (m_ec^2)^2}\right]
\eeq
with $\epsilon_\nu$ the typical energy of neutrinos at the current point, and $\epsilon_e$ the typical energy of electrons. This leads to an effective opacity (i.e., the inverse of the mean free path of neutrinos with respect to scattering on electrons)
\beq
\kappa_{s} \sim \sigma_0 \left[\frac{\epsilon_\nu\epsilon_e}{ (m_ec^2)^2}\right]\left[\frac{\epsilon_\nu}{hc}\right]^3  \sim \sigma_0 \left[\frac{\epsilon_\nu^4\epsilon_e}{({\rm MeV})^5}\right] \left(10^{30}\,{\rm cm^{-3}}\right).
\eeq
We thus see that at the densities at which neutrino-matter interactions are most important in neutron star mergers, inelastic scattering on electrons has a significantly lower opacity than elastic scattering on nucleons or charged-current reactions, but not necessarily smaller than absorption opacities for pair processes. Accordingly, its direct impact on $\nu_e$ and $\bar\nu_e$ is likely subdominant, but it could be important to the thermalization of heavy-lepton neutrinos. 

Finally, we note that scattering on nucleons is not perfectly elastic. The typical exchange of energy between neutrinos and the fluid is much smaller for nucleon scatterings than for electron scatterings, but as seen above, the cross-sections for nucleon scatterings are larger in regions where neutrino-matter interactions are important. In core-collapse supernovae, \cite{Wang:2020udq} showed that the smaller energy transfer during each scattering can be used to treat inelastic scattering on nucleons as a diffusion process in energy space, leading to much cheaper calculations than when using scattering kernels: there is no need to couple all energy bins through numerically stiff interaction terms. In \cite{Wang:2020udq}, the impact of neutrino-nucleon scattering on the thermalization of heavy-lepton nucleons was also shown to be comparable to the impact of neutrino-electron scattering. Accounting for inelastic scattering on nucleons could thus provide an avenue to partially account for the thermalization effect of scattering events without an implementation of inelastic scattering on electrons.

\subsubsection{Discussion}

From the previous sections, we see that the reactions currently used in our most advanced merger simulations can, if properly included in a transport algorithm, capture the dominant processes for emission, absorption, diffusion, and thermalization of $\nu_e$ and $\bar \nu_e$ in most of a post-merger remnant. Without even getting into the complications of approximate transport methods, however, we see that the situation is already more complex for other species of neutrinos. Emission of heavy-lepton neutrinos is dominated by pair processes which are poorly modeled as soon as neutrinos are out of equilibrium with the fluid. Thermalization of these neutrinos is likely impacted by inelastic scattering, which current simulations do not take into account. Finally, pair annihilation of all types of neutrinos in low-density regions is difficult to include, but possibly important to jet formation. It is thus worth noting that uncertainties in transport schemes are not the only potential sources of errors in our modeling of neutrinos today; the choice of physical processes included in the simulations, and the accuracy to which they are modeled, remains an area where significant improvements are possible.

\subsection{Quantum kinetics and neutrino oscillations}

So far, we have considered neutrinos as particles in well-defined flavor states (electron, muon, tau). However, we know that this is only an approximation. Even in vacuum, the fact that the mass eigenstates of neutrinos are different from their flavor eigenstates leads to oscillations between flavors. Vacuum oscillations occur on length scales too long to impact the evolution of a post-merger remnant, though if neutrinos from a neutron star merger were ever to be observed, oscillations between the source and the Earth would certainly be significant. There are however other processes that lead to flavor transformation with more relevance to the merger problem. Generally, any process that transform electron type neutrinos into heavy-lepton neutrinos (or vice-versa) close enough to the merger remnant that neutrino-matter interactions are still impacting the composition of the outflows has the potential to change the properties of kilonovae and the outcome of nucleosynthesis in neutron star mergers. 

One way to study neutrino oscillations is through the quantum kinetic equations (QKE). In that formalism, neutrinos are described by the $3\times 3$ density matrix $\rho(t,x^i,p^\mu)$. The diagonal terms of this matrix can be understood as equivalent to the distribution functions $f_{\nu_e}$, $f_{\nu_\mu}$, $f_{\nu_\tau}$, while the off-diagonal terms encode quantum coherence between flavors. A second matrix $\bar \rho$ contains information about antineutrinos. The density matrix evolves according to \citep{Vlasenko:2013fja}
\beq
\frac{D\rho}{d\tau} = -i \left[ H,\rho\right] + C[\rho]
\eeq
where the left-hand side is a total time derivative in phase-space, and the two terms on the right-hand side are responsible for, respectively, oscillations and collisions. The Hamiltonian $H$ can be decomposed as 
\beq
H = H_{\rm vac} + H_{\rm mat} + H_{\nu\nu},
\eeq
with $H_{\rm vac}$ responsible for vacuum oscillations, $H_{\rm mat}$ for interactions between neutrinos and the matter potential, and $H_{\nu\nu}$ for neutrino self-interactions. 

At least two types of oscillations have been found to be potentially improtant in the merger context. The Matter-Neutrino Resonance (MNR) occurs when the matter potential is equal to the neutrino self-interaction potential, and can impact the luminosity of $\nu_e$ and $\bar\nu_e$ within a few radii of the post-merger remnant \citep{2014JPhG...41d4004C,Zhu:2016mwa}. The flast-flavor instability (FFI), on the other hand, is due solely to the neutrino self-potential, and seems to occur in regions where the sign of the net lepton flux (number flux of $\nu_e$ minus number flux of $\bar\nu_e$) changes between different directions of propagation of the neutrinos \citep{PhysRevD.84.053013,Wu:2017drk,Grohs:2022fyq}.
The FFI occurs on very short timescales ($\sim$ns, i.e., cm length scales), and is likely active in many regions close to the post-merger remnant \citep{Grohs:2022fyq}. How much flavor transformation occurs as a result of the FFI remains uncertain, but recent studies using simplified prescriptions for where the FFI occurs and how much flavor transformation happens as a result have shown that it could plausibly lead to significant changes in the composition of matter outflows \citep{Li:2021vqj,Fernandez:2022yyv}. As quantum kinetics is not at this point studied as part of general relativistic radiation transport algorithms coupled to merger simulations, but rather evaluated using either simple approximations or specialized zoomed-in simulations, we do not discuss it in more detail here. We do however emphasize that these oscillations could very well have an impact on the composition of the matter outflows produced in mergers. The fact that they are not included directly within simulations is due largely to the additional technical difficulty of evolving the quantum kinetic equations and to the very short timescales involved in the FFI, rather than to a certainty that oscillations are not important to astrophysical results. Obtaining better models for the role of oscillations in merger simulations is certainly an important open problem in merger simulations today.

\section{Radiation transport algorithms}
\label{sec:algorithms}

Having discussed the reactions that we would like to take into account in neutron star merger simulations, we can now turn to a discussion of the various methods used so far to treat neutrino transport and neutrino-matter interactions. These can be broadly classified into quasi-local leakage schemes, approximate transport schemes based on the moment formalism, and Monte-Carlo evolution of Boltzmann's equation. Multiple simulations have also considered mixed leakage-moment schemes, while algorithms mixing Monte-Carlo methods with a moment scheme have been considered but not successfully used in merger simulations. For most of this section, we attempt to keep the discussion focused to the methods used for \emph{general relativistic radiation hydrodynamics} simulations, either in the context of neutron star merger simulations or for the evolution of their post-merger remnant. We will however discuss along the way a number of techniques that were first developed for Newtonian simulations or for simulations using quasi-Newtonian potentials that have either been ported to general relativistic simulations, or are likely to be used in that context in the near future. This is particularly true for the more advanced leakage schemes, which were first developed in non-relativistic codes but are currently being integrated in general relativistic simulations. We also note that neutrino radiation transport algorithms used in simulations of core-collapse supernovae are often more advanced than any of the algorithms used in merger simulations~(e.g.\cite{1993ApJ...405..637M,Liebendoerfer:2007dz,Takiwaki:2013cqa,Kuroda:2015bta,OConnor:2015rwy,Roberts:2016lzn,Bruenn:2018wpz,Skinner:2018iti}). In fact, many algorithms used in merger simulations today are directly inspired from work done in the core-collapse community. Accordingly, while we do not attempt to review the algorithms used in the core-collapse context, we will occcasionally refer to methods developed for core-collapse simulations if they have been used in the merger context. More advanced methods have also been proposed, but not yet applied to the merger problem; e.g. methods for a fully covariant evolution of the radiative transport equations appropriate for a direct discretization of Boltzmann's equation have recently been studied in \cite{2020ApJ...888...94D}, Lattice-Boltzmann methods have been implemented and used on test problems in \cite{Weih:2020qyh}, and the MOCMC (Method of Characteristics Moment Closure) method has shown that it is possible to combine particle and moment formalisms to improve on the convergence properties of a pure Monte-Carlo radiation transport code \citep{Ryan_2020}.

In this review, we focus particularly on moment methods, as they have been used in the majority of the most advanced radiation hydrodynamics simulations of mergers to-date. Most general relativistic simulations using leakage schemes use methods that are at best order-of-magnitude accurate, while very few simulations have been performed with the recently developed Monte-Carlo algorithms. Accordingly, moment schemes remain at the moment our best source of information about the role of neutrinos in neutron star mergers.

We note that while most simulations consider $3$ species of neutrinos and $3$ species of antineutrinos, it is fairly common for simulations to assume that the distribution function of all heavy-lepton neutrinos $\nu_\mu,\nu_\tau,\bar\nu_\mu, \bar\nu_\tau$ are identical, and thus to replace the evolution of those $4$ species by the evolution of a single species $\nu_x$ that represent them all; we will use the notation $\nu_x$ to represent all heavy-lepton neutrinos here as well.

\subsection{Leakage Schemes}

\subsubsection{Overview}

Leakage algorithms are the simplest methods used to treat neutrinos in neutron star merger and post-merger simulations. In their most basic form, they can capture the cooling of the post-merger remnant at the order-of-magnitude level, but not the evolution of the composition of the outflows. More advanced leakage schemes have however been developed for post-merger simulations, and mixed leakage-moment schemes have been used in general relativistic merger simulations. Those advanced schemes can at least approximately capture absorption within the outflows of neutrinos emitted by a post-merger accretion disk or by a neutron star remnant. 

The leakage schemes used in merger simulations today were first developed by \cite{Ruffert1996,Rosswog:2003rv}. They generally rely on a local computation of the neutrino energy and number emission rates per unit volume, $Q_{\nu,\rm free}$ and $R_{\nu,\rm free}$, for each species of neutrinos. In addition, leakage schemes compute an estimate of the optical depth between a given grid cell and the outer boundary of the computational domain, in order to estimate the diffusion timescale of trapped neutrinos through the remnant. 

The energy emission rate $Q_{\nu,\rm free}$ is calculated as described in Sect.~\ref{sec:reac}. In the merger context, simulations have usually considered the total (energy-integrated) emission rate, but energy-dependent leakage schemes have been developed for post-merger simulations \citep{Perego:2015agy}. In the former case, 
$R_{\nu,\rm free} = Q_{\nu,\rm free}/\langle \epsilon_\nu\rangle$, 
with $\langle \epsilon_\nu\rangle$ the average energy of emitted neutrinos. In the latter case, $R_{\nu,\rm free}$ for each energy bin is just $Q_{\nu,\rm free}/\epsilon_\nu$, with $\epsilon_\nu$ the energy at the center of the bin. In optically thin regions, this is sufficient to calculate the cooling rate and composition changes of the fluid. In optically thick regions, however, the rates at which neutrinos carry away energy and lepton number are much lower than the free emission rates. In those regions, we expect neutrinos to quickly reach their equilibrium distribution function $f_{\nu}^{\rm eq}$, and to slowly diffuse out of the remnant over a time scale $t_{\rm diff}$. The rate at which neutrinos carry energy away from a given cell is then approximately given, for neutrinos of a given energy $\nu$, by
\beq
Q_{\rm diff} = \frac{E_\nu^{\rm eq}}{t_{\rm diff}},
\eeq
with $E_\nu^{\rm eq}$ the equilibrium energy density of neutrinos. Most leakage schemes used in merger simulations implement the diffusion time scale prescription of \cite{Rosswog:2003rv}
\beq
t_{\rm diff} = \frac{\alpha_{\rm diff}\tau_\nu^2}{\kappa_{\rm tot}c}
\eeq
where $\kappa_{\rm tot}$ is the total opacity at the current point (including all absorption and scattering processes considered in the simulation),
and $\tau_\nu$ is the estimated optical depth between that point and the domain boundary. The parameter $\alpha_{\rm diff}$ is calibrated to the result of transport simulations; \cite{Rosswog:2003rv} use $\alpha_{\rm diff}=3$, but this choice is not unique (e.g. \cite{OConnor:2009iuz} argue for an increase of $\alpha_{\rm diff}$ by a factor of two). The optical depth is defined as
\beq
\tau_\nu = \min_\Gamma \left(\int_{\Gamma} ds \kappa_{\rm tot}\right),
\eeq
with the minimum taken over all possible paths $\Gamma$ starting from the current point and ending at the boundary of the computational domain. The optical depth $\tau_\nu$ is energy dependent, but the reactions that dominate the calculation of $\kappa_{\rm tot}$ all have $\kappa \propto \nu^2$ (charged-current reactions and elastic scatterings). Calculating $\tau_\nu$ at a single energy and then assuming $\tau_\nu \propto \nu^2$ is thus a reasonable approximation. We discuss different methods to estimate $\tau_\nu$ later in this section. More complex estimates of $t_{\rm diff}$ are also possible; for example, in the Improved leakage-equilibration-absorption scheme (ILEAS) of \cite{Ardevol-Pulpillo:2018btx}, separate diffusion timescales are calculated for the number and energy emission rate from the local gradient of the number density and energy density. Such a local calculation also has the advantage of allowing for emission rates that match the expected diffusion limit in optically thick regions, which is not possible using the simpler dimensional analysis of earlier schemes.

In an energy integrated leakage scheme, one then needs to integrate $Q_{\rm diff}(\nu)$ and $R_{\rm diff}(\nu)$ separately over $\nu$:
\beq
Q_{\rm diff} = \int d\nu \frac{\nu^3}{(hc)^3} \frac{f_\nu^{\rm eq}}{t_{\rm diff}}
\eeq
or, if we assume $t_{\rm diff}\propto \nu^2$,
\beq
Q_{\rm diff} = \frac{(k_B T)^2}{(hc)^2} \frac{(1\,{\rm MeV})}{hc}\frac{1\,{\rm MeV}}{t_{\rm diff}[1\,{\rm MeV}]} F_1(\eta_\nu)
\eeq
and similarly for the number diffusion rate
\beq
R_{\rm diff} = \frac{(k_B T)}{(hc)} \frac{(1\,{\rm MeV})^2}{(hc)^2}
\frac{1}{t_{\rm diff}(1\,{\rm MeV})}
F_0(\eta_\nu).
\eeq
The average energy of escaping neutrinos in the diffusion regime is then
\beq
\langle \epsilon_\nu\rangle_{\rm diff} = \frac{F_1(\eta_\nu)}{F_0(\eta_\nu)} k_B T.
\eeq
We note that the average energy of diffusing neutrinos is significantly lower than the average energy of a thermal spectrum, reflecting the fact that low energy neutrinos diffuse faster than high energy neutrinos.

The actual rate at which neutrinos leave a given region of the fluid is then given by an interpolation between the estimates valid at low and high optical depth, effectively considering that neutrino transport is limited by the lowest of those two rates. \cite{Ruffert1996} uses
\beq
Q_\nu = \frac{Q_{\nu,\rm free}Q_{\nu,\rm diff}}{Q_{\nu,\rm free}+Q_{\nu,\rm diff}};\,\,
R_\nu = \frac{R_{\nu,\rm free}R_{\nu,\rm diff}}{R_{\nu,\rm free}+R_{\nu,\rm diff}}.
\eeq
Alternatively, \cite{Sekiguchi:2010fh} considers an exponential transition between the two regimes
\beq
Q_\nu = Q_{\nu,\rm free} e^{-3\tau_\nu/2} + Q_{\nu,\rm diff} \left(1-e^{-3\tau_\nu/2}\right).
\eeq
These results can then be coupled to the evolution of the fluid equations using conservation of energy-momentum and lepton number
\beqn
\nabla_\alpha T^{\alpha\beta}_{\rm fluid} &=& \sum_{\nu_i} \left(-Q_{\nu_i} u^\beta - (g^{\alpha\beta}+u^\alpha u^\beta)\nabla_\alpha P_{\nu_i}\right)
\label{eq:leakE}\\
\nabla_\alpha (n_p u^\alpha) &=& R_{\bar\nu_e}-R_{\nu_e}.
\label{eq:leakY}
\eeqn
The neutrino pressure term can for example be computed assuming a relativistic gas of neutrinos in equilibrium with the fluid, i.e., $P_{\nu_i} = E_{\nu_i}^{\rm eq}/3$ \citep{OConnor:2009iuz}, potentially suppressed in regions where neutrinos are not trapped. Alternatively, we will see below that more advanced leakage schemes have been coupled to evolution equations for the trapped neutrinos -- in which case the neutrino pressure is calculated directly from the estimated energy density of trapped neutrinos.

\subsubsection{Leakage in general relativistic merger simulations}

In general relativistic merger simulations, the first published leakage scheme was developed by \cite{Sekiguchi:2010fh}. This algorithm is a mixed moment-leakage scheme with significantly more complexity that the simple scheme described above. The algorithm
assumes evolution equations
\beqn
\nabla_\alpha T^{\alpha\beta}_{\rm fl+tr} &=& -Q_{\rm leak} u^\beta;\,\, \nabla_\alpha T^{\alpha\beta}_{\rm st} = Q_{\rm leak} u^\beta;\\
\nabla_\alpha \left(Y_l u^\alpha\right) &=& (R_{\bar\nu_e,\rm leak}-R_{\nu_e,\rm leak})u^\alpha
\eeqn
for the stress-energy tensor $T^{\alpha\beta}_{\rm fl+tr}$ of the fluid and trapped neutrinos combined, $T^{\alpha\beta}_{\rm st}$ of streaming neutrinos, and for the lepton fraction $Y_l$. The streaming neutrinos are evolved using a moment scheme (see Sect.~\ref{sec:mom}), which allowed later iterations of this algorithm to realitively easily take into account reabsorption of the streaming neutrinos in low density regions. We can see from the evolution equations that this algorithm has the advantage of guaranteeing exact conservation of energy-momentum. Additionally, the algorithm evolves the fractions $Y_e$, $Y_{\nu_e}$, $Y_{\bar \nu_e}$ and $Y_{\nu_x}$ of electrons and neutrinos, assuming that these fractions reach their equilibrium value (at given temperature, density and lepton fraction) in regions where neutrino-matter interactions are fast compared to the numerical time step. This allows for relatively simple estimate of the contribution of neutrinos to the fluid pressure, assuming that the neutrinos are a relativistic gas.

An algorithm closer to the original methods of \cite{Ruffert1996,Rosswog:2003rv} was first used in general relativistic simulations by \cite{Deaton:2013sla}. In that work, the only contribution of neutrinos to the evolution of the system is the source terms of Eqs.~\eqref{eq:leakE}--\eqref{eq:leakY}. 
In \cite{Deaton:2013sla}, the minimum optical depth was calculated by considering lines along the coordinate directions $\hat x,\hat y,\hat z$ of a cartesian grid, as well as along the diagonals of a cube in the same coordinates, a method similar to that previously used by \cite{Ruffert:1998qg}.  This algorithm however requires global communications between all points of the numerical grid whenever $\tau_\nu$ is computed, and creates preferred directions along the axis of the cartesian coordinates. An improved method to calculate $\tau_\nu$ was later proposed by \cite{Neilsen:2014hha}, and is now the most commonly adopted algorithm in numerical relativity simulations. Their method relies on finding the path of shortest optical depth linking neighboring cell centers on a grid. We can indeed discretize our equation for $\tau_\nu$ at point $\bar x$ as
\beq
\tau_\nu(\bar x) = \min_n \left(\tau_\nu(\bar x_n) + \frac{\kappa_{\rm tot}(\bar x)+\kappa_{\rm tot}(\bar x_n)}{2} \Delta s_n \right)
\eeq
with the minimum taken over all neighboring points $\bar x_n$. Here, $\Delta s_n$ is the distance between $\bar x$ and $\bar x_n$. Given an initial guess $\tau_\nu^0$ for the optical depth at each point, we can solve this equation iteratively, using
\beq
\tau_\nu^{k+1}(\bar x) = \min_n \left(\tau_\nu^k(\bar x_n) + \frac{\kappa_{\rm tot}(\bar x)+\kappa_{\rm tot}(\bar x_n)}{2} \Delta s_n \right)
\eeq
until $\|\tau_\nu^{k+1}-\tau_\nu^{k}\|<\epsilon$ at all points for some small constant $\epsilon$. Because the optical depth evolves slowly over time, a single iteration initialized with the value of $\tau_\nu$ at the previous time step is generally sufficient to maintain a good estimate of $\tau_\nu$ everywhere, except when computing $\tau_\nu$ for the first time. This method is now widely used in neutron star merger simulations \citep{Foucart:2014nda,Radice:2018pdn,Mosta:2020hlh,Cipolletta:2020kgq,Most:2021ktk}. A conceptually similar algorithm that does not rely on the existence of an underlying cartesian grid has also been developed in \cite{Perego:2014qda}, allowing for the easy use of this method in grid-less simulations (e.g. SPH), while an improved numerical methods to solve for $\tau_\nu$ by solving the eikonal equation has been proposed by \cite{Palenzuela:2022kqk}.

\subsubsection{Leakage limitations and improved leakage schemes}

The accuracy of a leakage scheme can be very problem dependent. The free parameters in most leakage schemes are calibrated to spherically symmetric transport problems, and tend to perform best in that context -- while neutron star mergers and their post-merger remnants are very asymmetric. Even ignoring symmetry issues, however, the standard scheme discussed above has a number of important limitations. We have already mentioned the fact that the simplest leakage schemes do not accurately capture the local diffusion rate of neutrinos in the high optical depth limit; in this section we consider a few additional notable issues.

First, in regions where the total optical depth is high but the absorption and inelastic scattering optical depths are low, the assumption that neutrinos reach their equilibrium distribution function can be inaccurate. This is particularly problematic for heavy-lepton neutrinos, which typically have much lower absorption opacity than scattering opacity. One way to solve this issue is to keep track of the energy density of neutrinos, rather than assuming an equilibrium energy density. In Newtonian simulations of post-merger disks, the Advanced Spectal Leakage (ASL) scheme of \cite{Perego:2015agy}, for example, calculates the energy and number density of trapped neutrinos assuming that the distribution function of trapped neutrinos $f_\nu^{\rm tr}$ satisfies the equation
\beq
\frac{df_\nu^{\rm tr}}{dt} =  \dot f_{\nu,\rm prod} - \dot f_{\nu,\rm diff}
\eeq
with
\beqn 
\dot f_{\nu,\rm prod} &=& \frac{f_\nu^{\rm eq}-f_\nu^{\rm tr}}{\max{(t_{\nu,\rm prod},\Delta t)}}
\exp{(-\frac{t_{\nu,\rm diff}}{t_{\nu,\rm prod}})};\\
\dot f_{\nu,\rm diff} &=& \frac{f_\nu^{\rm tr}}{\max{(t_{\nu,\rm diff},\Delta t)}}
\exp{(-\frac{t_{\nu,\rm prod}}{t_{\nu,\rm diff}})}
\eeqn
for the estimated production time scale $t_{\nu,\rm prod}$ and diffusion time scale $t_{\nu,\rm diff}$. Here, $\Delta t$ is the time step of the evolution. We see that in this scheme, if the time scale to produce neutrinos is long, the trapped neutrinos do not reach equilibrium with the fluid. \cite{Perego:2015agy} further assume
\beq
f_\nu^{\rm tr} = \gamma f_\nu^{\rm eq} \left(1-e^{-\tau_{\rm en}}\right)
\eeq
with $\tau_{\rm en}$ the optical depth ignoring elasctic scatterings, and $\gamma$ a function of position only. This allows for the use of the single unknown $\gamma$ to represent the function $f_\nu^{\rm tr}$.
The ASL scheme has been adapted for use in merger simulations within a smoothed particle hydrodynamics (SPH) code \citep{Gizzi:2021ssk}, which in conjunction with the development of a first general relativistic SPH code \citep{Rosswog:2022gxz} should allow for the use of ASL in general relativsitic merger simulations in the near future.
We note that the general relativistic algorithm of \cite{Sekiguchi:2010fh} also keeps track of the energy density of trapped neutrinos when, in the notation of \cite{Perego:2015agy}, $t_{\rm diff} \leq t_{\rm prod}$. However, as \cite{Sekiguchi:2010fh} calculates the neutrino diffusion rate from the equilibrium density of neutrinos rather than from the energy density evolved within the simulation, the scheme only partially correct for the difference between those two estimates of the energy. The ILEAS scheme \citep{Ardevol-Pulpillo:2018btx} similarly tracks trapped neutrinos through an ``equilibration-advection'' step where neutrinos in optically thick regions are assumed to be in equilibrium with the fluid, and ``trapped'' neutrinos are otherwise advected through the computational grid to guarantee exact conservation of lepton numbers as well as to approximately account for the impact of trapped neutrinos on the properties of the fluid.

A second common issue with leakage schemes is that the simple interpolation between free emission and diffusion is naturally going to be inexact in semi-transparent regions. Corrections to the emission rate can be calibrated to specific systems (e.g., \citealt{Perego:2015agy,Gizzi:2021ssk} for accretion disks and mergers), but the region where most of the neutrinos leaving the system are emitted is, by definition, particularly difficult to model accurately in a leakage scheme.

A third issue is that the energy of neutrinos diffusing through the system is not constant; inelastic scatterings, as well as absorption and re-emission, are both taken into account in the energy diffusion rate, and these processes tend to bring the neutrinos closer to thermal equilibrium with the fluid. On the other hand, the integrated values of $Q_{\rm diff}$ and $R_{\rm diff}$ computed above assume that neutrinos random walk through the fluid at constant energy. Diffusing neutrinos have lower average energies than the emitted neutrinos, but only because the opacity of the system to high-energy neutrinos is larger; the scheme does not account for thermalization of the neutrinos. The ASL scheme attempts to partially account for this thermalization by suppresing $R_\nu$ by a factor of $e^{-\tau_{\rm en}/10}$, then renormalizing the energy-intergrated $R_\nu$ to keep the total number of neutrinos constant while reducing their average energy.

The last important issue is that, in its simplest form, a leakage scheme does not account for the energy, momentum, and lepton number deposited in the fluid by reabsorption of neutrinos emitted in different regions of the system. Neutrinos either leave or do not leave the simulation domain, but the interactions between neutrinos and the fluid after emission do not feedback onto the evolution of the fluid. This is particularly problematic when studying the composition of matter outflows, which is known to be significantly impacted by absorption of $\nu_e$ \citep{Wanajo:2014}. In general relativistic merger simulations, multiple leakage schemes have attempted to approximately include these absorption effects. In \cite{Mosta:2020hlh,Cipolletta:2020kgq}, absorption is taken into account by calculating the emission and absorption of neutrinos along radial rays in the post merger remnant. Along each ray, the neutrino emissivity is estimated from the energy and number density emission rates predicted by a leakage scheme. The heating rate and change in composition of the fluid can then be estimated from the neutrino luminosity integrated over each ray, and from a local estimate of $\kappa_a$. \cite{Radice:2018pdn} instead consider approximate transport along each ray, evolving the energy density of the free-streaming neutrinos (zeroth moment, see Sect.~\ref{sec:mom}). In all of these works, absorption of free streaming neutrinos is only included in optically thin regions, with an exponential cutoff $\propto e^{-\tau_\nu}$ applied to the absorption rate in optically thick regions. In the ILEAS scheme \citep{Ardevol-Pulpillo:2018btx}, neutrino emission (as predicted by the leakage) is similarly followed along ``rays'' and potentially reabsorbed. However, no specific geometry is assumed for these rays; instead, neutrinos follow the gradient of the optical depth. This leads to many more rays that may jointly contribute to absorption in any given cell, thus complicating the algorithm, but also to a propagation of neutrinos that should better match the geometry of the system. 

We note that whenever an attempt is made at taking into account neutrino absorption, it is important to calculate the absorption opacities for the energy of the free-streaming neutrinos, rather than for the energy of neutrinos locally in equilibrium with the fluid. Typically, neutrinos in merger simulations have $\epsilon_\nu \sim (10-20)\,{\rm MeV}$, while in the outflows $T\sim 1\,{\rm MeV}$. As $\kappa_a \propto \nu^2$, assuming thermal equilibrium of the neutrinos with the fluid can lead to large underestimates of the absorption rate \citep{Foucart:2016rxm,Radice:2018pdn}. This is an issue not only for leakage schemes, but also for energy-integrated moment schemes, as we will see.

  In Newtonian (or pseudo-Newtonian) post-merger simulations, simple lightbulb prescriptions have instead been used to model the spatial distribution of emitted neutrinos \citep{Fernandez:2013tya,Metzger:2014ila}. In a lightbulb model, the total luminosity of neutrinos is calculated by integrating over the entire simulation, then that luminosity is assumed to come from a specific region -- in the case of post-merger systems, an annulus around the densest region of the remnant accretion disk and/or the surface of the remnant neutron star (if present). Alternatively, simulations using the ASL scheme have combined that scheme with a more advanced approximate transport algorithm (see \citealt{Perego:2014fma,Gizzi:2019awu,Gizzi:2021ssk}). In that algorithm, neutrinos emitted in optically thick regions ($\tau_\nu>2/3$) are assumed to diffuse to the neutrinopshere ($\tau_\nu=2/3$ surface) along the gradient of $\tau_\nu$ (i.e., along the path of smallest optical depth), before free-streaming from all points of the neutrinosphere. Neutrinos emitted in optically thin regions just free-stream from their point of emission. The number density of neutrinos outside of the neutrinosphere can then be combined with calculations of the absorption and scattering optical depth to determine energy and momentum deposition in optically thin regions. As the calculation of the number density of neutrinos requires first the determination of a map linking each point at $\tau_\nu>2/3$ to a point on the neutrinosphere, and then for each point at $\tau_\nu<2/3$ an integral over the entire neutrinosphere and all optically thin regions, this algorithm is significantly more expensive than other leakage schemes, even in Newtonian physics where free-streaming neutrinos can be assumed to propagate along straight lines.

\subsubsection{Discussion}

Overall, we thus see that leakage schemes provide us with a simple, cost-effective method to approximately incorporate neutrino cooling in merger simulations. Using a basic leakage scheme typically has no noticeable effect on the cost of a merger simulations, but the results are only order-of-magnitude accurate and do not account for the important role of neutrino absorption in matter outflows. More advanced leakage methods can be developed through coupling to radiation transport algorithms, or by attempting to predict where the leaking neutrinos will go as they leave the system. These more advanced algorithms will naturally be more costly, but they have been shown to provide a better match to the result of full radiation transport simulations, at least on test problems. Estimating the error of such an algorithm without comparison to a more advanced simulation is however always difficult.

\subsection{Moments-based radiation transport}
\label{sec:mom}

To go beyond leakage schemes and attempt to transport neutrinos along geodesics through a numerical simulation, multiple general relativistic merger codes have now implemented ``tuncated moments'' schemes, i.e., algorithms evolving moments of the distribution function of neutrinos. The formalism for general relativistic moments schemes was derived by \cite{1981MNRAS.194..439T}, while the development of methods for the coupled evolution of the fluid and moment equations largely build on work performed for photon transport in Newtonian simulations \citep{1992ApJS...80..819S,Audit:2002hr,2011JQSRT.112.1323V,2013ApJS..206...21S}. For general relativistic transport, it saw early uses in spherical symmetry \citep{Rezzolla:1994se}, before being adapted to the methods used in 3D neutron star merger simulations \citep{shibata:11}. Our understanding of the moment equations in relativistic systems also significantly benefited from work done in the context of photon transport in accretion disks \citep{Sdowski:2012cx}. The moment formalism leads to evolution equations that are typically more complex and costly to evolve than the leakage scheme, but automatically include emission, transport, and reabsorption of neutrinos. We will see that moment schemes have been very successful in providing us with an improved understanding of the role of neutrinos in neutron star mergers, but also that the truncated expansion of the distribution function used in these schemes will force us to choose approximate analytical closure for unknown higher moments of $f_\nu$, leading to evolution equations that do not converge to the true solution of Boltzmann's equation, with hard-to-quantify errors in the results.

We organize our discussion of moment schemes as follow. In Sect.~\ref{sec:M1form}, we discuss the basic ideas behind the moment formalism and the derivation of the moment equations. In Sect.~\ref{sec:M1eq}, we review evolution equations for the moments in the frame of a numerical simulation as well as prescriptions to account for energy and momentum transfer between neutrinos and the fluid. In Sect.~\ref{sec:M1Neq}, we discuss lepton number exchange and the option for energy-integrated moment schemes to evolve the neutrino number density to guarantee exact lepton number conservation. In Sect.~\ref{sec:M1s}, we provide more details on the calculation of the coupling terms between neutrinos and matter, with a focus on issues that arise when the energy spectrum of neutrinos is not known. In Sect.~\ref{sec:M1clos} we discuss the approximate analytical prescriptions used to close the moment equations, while Sect.~\ref{sec:M1num} focuses on the numerical implementation of the moment formalism in general relativistic merger codes.
In Sect.~\ref{sec:M1scat} we discuss specific issues with the use of moment schemes in high-density regions, and ways to recover the proper diffusion rate of neutrinos through those regions. Finally, Sect.~\ref{sec:M1pairs} discusses approximate implementations of neutrino-antineutrino pair annihilation in low-density regions.

In this section, we choose units such that $h=c=1$. 

\subsubsection{Truncated moments formalism}
\label{sec:M1form}

To understand the basic idea behind the moment formalism, let us write the neutrino 4-momentum as
\beq
p^\mu = \epsilon (\hat t^\mu + l^\mu)
\eeq
with $\hat t_\mu$ a timelike unit vector, and $l^\mu$ a spacelike unit vector orthogonal to $\hat t_\mu$. We see that $\epsilon$ is the neutrino energy for an observer with 4-velocity $\hat t^\mu$, and that by construction $p^\mu p_\mu=0$.

The $n^{\rm th}$ moment of the distribution function $f_\nu$ according to our observer with 4-velocity $\hat t^\mu$ is defined as
\beq
M^{\alpha_1...\alpha_n}(t,x^i) = \epsilon^{3} \int d\Omega f_\nu(t,x^i,\epsilon,\Omega) (\hat t^{\alpha_1}+l^{\alpha_1})...(\hat t^{\alpha_n}+l^{\alpha_n}).
\eeq
with $\int d\Omega$ an integral over solid angle on the unit sphere in momentum space.
For example, the $1^{\rm st}$ moment is
\beq
M^\alpha(t,x^i,\epsilon) = \epsilon^3 \int d\Omega f_\nu(t,x^i,\epsilon,\Omega) (\hat t^{\alpha}+l^{\alpha}),
\eeq
while the energy density of neutrinos (0$^{\rm th}$ moment) for that same observer is
\beq
E(t,x^i,\epsilon) = \epsilon^3  \int d\Omega f_\nu(t,x^i,\epsilon,\Omega) = -(M^\alpha \hat t_\alpha).
\eeq
We thus see that only the spatial components of $M^\alpha$ are independent of the $0^{\rm th}$ moment; these are usually denoted as the flux density
\beq
F^\alpha = \epsilon^3  \int d\Omega f_\nu(t,x^i,\epsilon,\Omega) l^\alpha
\eeq
with, by construction, $F^\alpha \hat t_\alpha=0$, i.e., $F^\alpha$ is a purely spatial vector according to our chosen observer. Combining these results, we get
\beq
M^\alpha = E \hat t^\alpha + F^\alpha.
\eeq
Similarly, for the second moment, $M^{\alpha\beta}\hat t_\alpha = -M^\beta$ and  $M^{\alpha\beta}\hat t_\beta = -M^\alpha$, and so the only components of the second moment that cannot be directly reconstructed from $E$ and $F^\alpha$ are the spatial components of the pressure tensor
\beq
P^{\alpha\beta} = \epsilon^3  \int d\Omega f_\nu(t,x^i,\epsilon,\Omega) l^\alpha l^\beta
\eeq
with $P^{\alpha\beta}\hat t_\alpha= P^{\alpha\beta}\hat t_\beta=0$. With those definitions, we can write the second moment as
\beq
M^{\alpha\beta} = E \hat t^\alpha \hat t^\beta + F^\alpha \hat t^\beta + F^\beta \hat t^\alpha + P^{\alpha\beta}.
\eeq

We note that these moments are well-defined tensors, but they do generally require the choice of a specific frame in which neutrino energies are measured. The one exception is the energy integrated second moment of $f_\nu$, which can be written as
\beq
M_{\rm tot}^{\alpha\beta} = \int_0^\infty d\epsilon \epsilon \int d\Omega f_\nu p^\alpha p^\beta. 
\eeq
As $\epsilon d\epsilon d\Omega = d^3p / (p^{\hat t} \sqrt{-g})$ is the invariant integration volume in momentum space, written in an orthonormal frame with timelike coordinate $\hat t^\alpha$, that expression is independent of the coordinates in which we measure $\epsilon$. In fact, comparing with Eq.~\eqref{eq:Tab}, we see that $M^{\alpha\beta}_{\rm tot}$ is the stress-energy tensor for the chosen species of neutrinos. 

The first moment equation can be derived by taking the covariant divergence of the first moment with neutrino energies measured in the fluid rest frame (i.e., taking $\hat t^\mu = u^\mu$), and combining it with Boltzmann's equation. We get \citep{1981MNRAS.194..439T} 
\beq
\nabla_\alpha M^\alpha - \frac{\partial}{\partial\nu} \left(\nu M^{\alpha\beta} \nabla_\alpha u_\beta\right)
+ M^{\alpha\beta} \nabla_\alpha u_\beta= -S^\alpha u_\alpha
\eeq
with $\nu$ the neutrino energy in the fluid rest frame. The right-hand side is defined from
\beq
\left[\frac{df_\nu}{d\tau}\right]_{\rm collisions} = S(t,x^i,\nu,\Omega,f_\nu),
\eeq
with $S^\alpha$ the first moment of $S$ in the fluid rest frame, i.e.
\beq
S^\alpha(t,x^i,\nu) = \nu^3 \int d\Omega S(t,x^i,\nu,\Omega,f_\nu) (u^{\alpha}+l^{\alpha}).
\eeq
In merger simulations using the moment formalism, the source term has so far been limited to isotropic emission, absorption, and isotropic elastic scattering of neutrinos. Then
\beq
S^\alpha = \eta u^\alpha - \kappa_a J u^\alpha - (\kappa_a + \kappa_s) H^\alpha
\eeq
with $\eta$ the emissivity, $\kappa_{a,s}$ the absorption and scattering opacities, and $J,H^\alpha$ the energy density and momentum flux in the fluid frame, i.e.
\beq
M^\alpha = J u^\alpha + H^\alpha
\eeq
for our choice of $\hat t^\mu = u^\mu$. 
The second moment equation is \citep{1981MNRAS.194..439T}
\beq
\nabla_\beta M^{\alpha\beta} - \frac{\partial}{\partial\nu} \left(\nu M^{\alpha\beta\gamma} \nabla_\gamma u_\beta\right) = S^\alpha.
\label{eq:M1u}
\eeq
and we typically decompose $M^{\alpha\beta}$ in the fluid frame as
\beq
M^{\alpha\beta} = J u^\alpha u^\beta + H^\alpha u^\beta + u^\alpha H^\beta + L^{\alpha\beta}
\eeq
with $L^{\alpha\beta}$ the pressure tensor in that reference frame.

For an observer using an orthonormal tetrad in the fluid rest frame, we can interpret the second moment equation as an evolution equation for $J$ and $H^{\hat i}$. Indeed,
\beqn
\nabla_\alpha M^{\alpha \hat t} &=& \left(\partial_{\hat t} J + \partial_{\hat i} H^{\hat i}\right)\\
\nabla_\alpha M^{\alpha \hat i} &=& \left(\partial_{\hat t} H^{\hat i} + \partial_{\hat j} L^{\hat i\hat j}\right).
\eeqn
However, the evolution equations are not closed: they depend on the next two moments of $f_\nu$, $L^{\alpha\beta}$ and $M^{\alpha\beta\gamma}$. To obtain a well defined system of equations, one thus needs a \emph{closure} relation. That closure should provide higher moments of $f_\nu$ that are not evolved in our equations, as a function of the evolved variables. This is practically similar to the need for an equation of state $P(\rho,T,Y_e)$ in the evolution of the fluid equations, except that in a fluid $P$ can be calculated assuming statistical equilibrium of the fluid particles. For out-of-equilibrium particles, we do not have an analytical expression for the closure; our choice of closure will thus introduce an error in our solution.

We note that everything in our derivation so far has assumed that the moments are functions of the neutrino energy in the fluid frame. These functions can be discretized in energy space to obtain a \emph{spectral} or energy-dependent moment scheme. Such a scheme has been used for the evolution of post-merger remnants in Newtonian simulations \citep{2015MNRAS.448..541J} and in general relativistic simulations of core-collapse supernovae \citep{Kuroda:2015bta,Roberts:2016lzn}, but not so far in general relativistic merger simulations. As our focus here is mainly on the latter, we will mainly discuss the cheaper alternative: evolving energy-intergrated moments of $f_\nu$ using a \emph{grey} moment scheme. Grey moment schemes require different closure relations. On the one hand, we can see from Eq.~\eqref{eq:M1u} that, under the physical assumption that $f_\nu$ drops to zero faster than $\nu^{-1}$ as $\nu\rightarrow \infty$, the term involving $M^{\alpha\beta\gamma}$ in that equation does not contribute to the energy-intergrated moment equation. On the other hand, given the strong dependence of $\kappa_{a,s}$ on the energy of the neutrinos, the caculation of an energy-integrated $S^\alpha$ requires a choice of neutrino spectrum -- a new closure relation that will significantly impact the assumed cross-sections of neutrino-matter interactions. We will come back to this choice later in this review.

\subsubsection{Moment equations in the simulation frame}
\label{sec:M1eq}

One of the main advantage of the moment formalisms is that the evolution equations for the moments can be put into a ``conservative'' form very similar to that commonly used for the evolution of the fluid equations (see below). To do so, however, it is useful to move away from moments computed in the fluid rest frame. Consider instead the decomposition of the second moment
\beq
M^{\alpha\beta} = E n^\alpha n^\beta + F^\alpha n^\beta + n^\alpha F^\beta + P^{\alpha\beta}
\eeq
with $n^\alpha$ the unit normal to a constant-time slice in the simulation, and $F^\alpha n_\alpha = P^{\alpha\beta}n_\alpha = P^{\alpha\beta}n_\beta = 0$. We can then interpret $E,F^\alpha,P^{\alpha\beta}$ as the energy density, flux density, and pressure tensor of the neutrinos of a given species as measured by a normal observer. Remembering that the energy integrated second moment is independent of the reference frame in which we measure the neutrino energy, this expression makes it easy to calculate $E,F^\alpha,P^{\alpha\beta}$ as functions of $J,H^\alpha,L^{\alpha\beta}$ (and vice-versa) using projections of the second moment. Let us define the Lorentz factor $W$ amd the fluid 3-velocity $v^\mu$ such that $u^\mu = W (n^\mu + v^\mu)$ and $v^\mu n_\mu=0$, and the 3-metric $\gamma_{\alpha\beta} = g_{\alpha\beta} + n_\alpha n_\beta$. We then get
\beqn
E &=& M^{\alpha\beta}n_\alpha n_\beta = J W^2 - 2W (H^\alpha n_\alpha)  + L^{\alpha\beta} n_\alpha n_\beta\\
F_i &=& -M^{\alpha\beta}n_\alpha \gamma_{\beta i} = JW u_i + W H_i - u_i (H^\alpha n_\alpha) -  L^{\alpha\beta}
n_\alpha \gamma_{\beta i}\\
P_{ij} &=& M^{\alpha\beta}\gamma_{\alpha i}\gamma_{\beta j} = J u_i u_j + H_i u_j + H_j u_i + L_{ij}.
\eeqn
Alternatively, using $h_{\alpha\beta}=g_{\alpha\beta} + u_\alpha u_\beta$, we get
\beqn
J &=& EW^2 - 2W^2 (F^i v_i) + W^2 (P^{ij} v_i v_j)\\
H_\alpha &=& (E W - W F^k v_k) (n_\alpha - W u_\alpha) + W F_\alpha + W (F^k v_k) u_\alpha \nonumber\\
&&- g_{\alpha i} W v_j P^{ij} - u_\alpha W^2 (P^{ij} v_i v_j)\\ 
L_{\gamma\delta} &=& \left( E n^\alpha n^\beta + F^\alpha n^\beta + n^\alpha F^\beta + P^{\alpha\beta} \right)
h_{\alpha\gamma} h_{\beta\delta}.
\eeqn
We will not need this last expression, and thus do not provide a more explicit expansion. We note that $n_\mu = (-\alpha,0,0,0)$, with $\alpha$ the lapse function, which simplifies many expressions when calculating $E,F^\alpha,P^{\alpha\beta}$ from $J,H^\alpha,L^{\alpha\beta}$. We also note that $F^\alpha,P^{\alpha\beta}$, $v^\alpha$ are all purely spatial tensors, e.g. $F^t=0$. Indices for these tensors can be raised and lowered with the 4-metric using all indices, or with the 3-metric using only spatial indices (e.g. $v_i = \gamma_{ij}v^j$).

The second moment equation can now be recast as evolution equations for the energy integrated moments weighted by $\sqrt{\gamma}$, with $\gamma$ the determinant of the 3-metric $\gamma_{ij}$. Defining $\tilde X=X\sqrt{\gamma}$ for any tensor $X$, and considering now energy-integrated moments (i.e., all moments are integrated from $\nu=0$ to $\nu=\infty$), we get the equations for a grey two-moment scheme in the simulation frame:
\beqn
\partial_t \tilde E + \partial_i \left( \alpha \tilde F^i - \beta^i \tilde E\right)  &=& \alpha \tilde P^{ij} K_{ij} 
-\tilde F^j \partial_j \alpha - \alpha \tilde S^\mu n_\mu\\
\partial_t \tilde F_j + \partial_i \left( \alpha \tilde P^i_j - \beta^i \tilde F_j \right) &=& -\tilde E \partial_j \alpha
+ \tilde F_k \partial_j \beta^k + \frac{\alpha}{2} \tilde P^{ik} \partial_j \gamma_{ik} + \alpha \tilde S^\mu \gamma_{j\mu}
\eeqn
with $K_{ij}$ the extrinsic curvature of the underlying spacetime. These equations are particularly convenient to use in general relativistic simulations using finite-difference or finite-volume conservative methods for the evolution of the fluid equations. Indeed, the moment equations are now nearly identical to the ideal fluid equations without neutrino-matter interactions (in a fluid, $\tilde P_{ij} = \sqrt{\gamma} P \gamma_{ij}$, with $P$ the fluid pressure). The only new terms are the source terms involving $\tilde S^\mu$. For well chosen closures (see below), these equations also form a well-posed system of hypebolic, causal equations, an important property in order to obtain stable numerical evolutions \citep{Pons:2000br}. We note that we have a separate system of equations for each species of neutrinos and antineutrinos.

The back-reaction of the neutrinos onto the fluid is easily computed: energy-momentum conservation requires that the source terms transfering energy and momentum from the fluid to the neutrinos exactly cancel the source terms transfering energy and momentum from the neutrinos to the fluid. Our evolution equations are, for each neutrino species $\nu_i$,
\beq
\nabla_\alpha T^{\alpha\beta}_{(\nu_i)} = S^\beta_{(\nu_i)}
\eeq
and thus
\beq
\nabla_\alpha T^{\alpha\beta} = \nabla_\alpha T^{\alpha\beta}_{\rm fluid} +\sum_i \nabla_\alpha T^{\alpha\beta}_{(\nu_i)} = 0 \leftrightarrow \nabla_\alpha T^{\alpha\beta}_{\rm fluid} = -\sum_i S^\beta_{(\nu_i)},
\eeq
with the sum being over all species of neutrinos and antineutrinos.
General relativisitic merger simulations relying on the evolution of the moments of $f_\nu$ have evolved either $(\tilde E,\tilde F_i)$ while providing closures for $\tilde P_{ij}$ and the energy spectrum of the neutrinos; or $\tilde E$ with a closure for $\tilde F_i$, $\tilde P_{ij}$, and the energy spectrum.

At this point, it is worth commenting on the use of grey schemes in general relativistic merger simulations so far. As we will see over the course of this section, the formalism for an energy-dependent moment scheme in general relativity has been fully developed, and has been used e.g. in core-collapse simulations \citep{Kuroda:2015bta,Roberts:2016lzn}. Why is it not used in merger simulations? One issue with an energy-dependent scheme is of course that the moments of $f_\nu$ within each energy bin have to be evolved, and thus the cost of the neutrino evolution scales at least linearly with the number of energy groups evolved. This would be a steep price to pay, even for simulations with relatively coarse energy resolution. More importantly, however, neutron star mergers and their post-merger remnants include regions where the fluid is moving at relativistic speed with respect to an observer at rest in the simulation frame, as well as steep velocity gradients. This creates two important issues. First, a neutrino propagating through the remnant may rapidly change energy group, if the energy discretization is done in the fluid frame. If the discretization is done in the simulation frame, on the other hand, transforming to fluid frame energies to calculate the source terms is non trivial, as that transformation depends on the unknown direction of propagation of the neutrinos (and neutrinos with the same energy in one frame may have very different energies in the other). Rapid variations of the neutrino energies can additionally lead to significant numerical diffusion in energy space and to a smoothing of the neutrino energy spectrum. Second, the flux in energy space $F_\nu^\alpha=(\nu M^{\alpha\beta\gamma}\nabla_\gamma u_\beta)$ can be large enough that explicit time stepping becomes unstable, at least for the time steps otherwise used for the evolution of the equations of general relativistic hydrodynamics. This means that either the time step has to be decreased (potentially drastically), or the energy flux has to be treated implicitly, thus coupling in an implicit time step the evolution of all energy groups. Either choice introduces a significant additional computational cost. This is not to say that an energy-dependent scheme in merger simulations is impossible. However, the development of a cost-effective and stable evolution scheme for an energy-dependent moment algorithm applicable to general relativistic merger simulations remains an unsolved problem, and thus any discussion of which scheme would be pratical is currently based on conjecture only. 

\subsubsection{Number density evolution and lepton number conservation}
\label{sec:M1Neq}

If we want more information about neutrino energies without going all the way to an energy-dependent scheme, a potentially useful extension of the standard grey two-moment approach described in the previous section is to define number-weighted moments in addition to the energy-weighted moments discussed so far, i.e., moments
\beq
N^{\alpha_1...\alpha_n}(t,x^i) = \int_0^\infty d\epsilon \epsilon^{2} \int d\Omega f_\nu(t,x^i,\epsilon,\Omega) (\hat t^{\alpha_1}+l^{\alpha_1})...(\hat t^{\alpha_n}+l^{\alpha_n}).
\eeq
In this case, the first moment is independent of the choice of time vector $\hat t^\mu$, as
\beq
N^\alpha = \int_0^\infty d\epsilon \epsilon \int d\Omega f_\nu(t,x^i,\epsilon,\Omega) p^\alpha.
\eeq
From the definition of $N^\alpha$, we see that in any given orthonormal frame, $N^{\hat t}$ is the number density of neutrinos, while $N^{\hat i}$ is the number flux of neutrinos. Taking the covariant divergence of $N^\alpha$ and combining with Boltzmann's equation, we get
\beq
\nabla_\alpha N^\alpha = S_N;\,\, S_N = \int_0^\infty d\nu \nu^2 \int d\Omega S(t,x^i,\nu,\Omega,f_\nu).
\eeq
This is simply expressing the conservation of neutrino number up to the contribution of the source term $S_N$. If we write $N^\alpha = N n^\alpha + F_N^\alpha$ with $F_N^\alpha n_\alpha=0$, this is equivalent to
\beq
\partial_t \tilde N + \partial_i \left(\alpha \tilde F_N^i - \beta^i \tilde N\right) = \alpha \tilde S_N.
\eeq
Adding this evolution equation to the evolution of $\tilde E$ and, possibly, $\tilde F_i$ has two important advantages. The first is that one can get some information about the average energy of neutrinos from the local values of $\tilde E, \tilde F_i, \tilde N$. The other is that this equation allows for explicit conservation of all lepton numbers. For example, for the electron lepton number, 
\beq
\nabla_\alpha N^\alpha_l = \nabla_\alpha \left( N^\alpha_{e^-} -N^\alpha_{e^+}+ N^\alpha_{\nu_e} - N^\alpha_{\bar \nu_e}\right).
\eeq
Transforming to the variables typically used in fluid simulations, the electron fraction $Y_e$ and weighted energy density $\rho_*$
\beq
Y_e = \frac{n_{e^-} - n_{e^+}}{n_p + n_n};\,\, \rho_* = m_b (n_p+n_n) W \sqrt{\gamma}
\eeq
with $n_i$ the number density of species $i$ in the fluid frame and $m_b$ an arbitrary reference baryon mass, we get
\beq
\partial_t( \rho_* Y_e) + \partial_i \left(\rho_* Y_e v_T^i\right) =m_b\alpha \left( \tilde S_{N,(\bar \nu_e)} - \tilde S_{N,(\nu_e)}\right)
\label{eq:Ye}
\eeq
with $v_T^i = \alpha^{-1} v^i - \beta^i$ the transport velocity. This couples the evolution of the composition of the fluid to the evolution of electron type neutrinos in a way that guarantees conservation of electron lepton number. The main disadvantage, besides the cost of evolving an additional variable (which is in practice minimal), is that we now need a closure relation for the number flux $\tilde F_N^i$, and still have to make semi-arbitrary assumption for the \emph{shape} of the neutrino spectrum \citep{Foucart:2016rxm}. 

We note that whether the number density $\tilde N$ is evolved or not, we need to calculate $\tilde S_N$. Indeed, capturing the evolution of $Y_e$ due to neutrino-matter interactions is one of the main objective of merger simulations including neutrino transport. Evaluating the source terms in Eq.~\eqref{eq:Ye} is thus a necessity in any transport scheme. 

\subsubsection{Source terms}
\label{sec:M1s}

As already mentioned, existing general relativistic simulations of neutron star mergers typically assume a source term of the form
\beq
S^\alpha(t,x^i,\nu) = \eta u^\alpha - \kappa_a J u^\alpha - (\kappa_a + \kappa_s) H^\alpha
\eeq
for neutrinos of energy $\nu$ in the fluid frame, with $\eta$ and $\kappa_a$ calculated so that
\beq
\frac{\eta_{(\nu_i)}}{\kappa_{a,(\nu_i)}} = \int d\Omega \nu^3 f_{(\nu),{\rm eq}} = 4\pi \frac{\nu^3}{1+\exp{\left(\frac{\nu - \mu_{(\nu_i)}}{k_B T}\right)}}.
\label{eq:BBs}
\eeq
This assumes that emission is isotropic in the fluid frame, and guarantees that  neutrinos reach their equilibrium energy density in optically thick regions.

We note that this is not the most general form that the source terms can take, and in fact implicitly makes a number of important assumptions. Besides isotropic emission, this choice assumes isotropic elastic scattering, and it is not an accurate model for pair processes. We will discuss alternative methods to account for pair processes in optically thin regions in Sect.~\ref{sec:M1pairs}. In optically thick regions, we tend to rely on the fact that as long as neutrinos reach their equilbrium density on a time scale short compared to the dynamical timescale of the system, imposing Eq.~\eqref{eq:BBs} will be sufficient to approximately recover the correct physical behavior. We discuss the limits of this approach in Sect.~\ref{sec:M1pairs} as well.

The coefficients $\eta,\kappa_a,\kappa_s$ can be tabulated as functions of the fluid properties $(\rho_0,T,Y_e)$ and the neutrino energy $\nu$ (as in e.g. the NuLib library \citep{OConnor:2009iuz}). For grey schemes, we need instead the integrated emissivity and average opacities, defined such that
\beq
S^\alpha_{\rm tot} = \int_0^\infty d\nu S^\alpha = \eta_{\rm tot} - \langle \kappa_a \rangle J_{\rm tot}
-( \langle \kappa_a \rangle+ \langle \kappa_s \rangle) H^\alpha_{\rm tot}
\eeq
with $J_{\rm tot}$ and $H_{\rm tot}$ the energy integrated energy density and flux.
The calculation of the total emissivity is trivial:
\beq
\eta_{\rm tot} = \int_0^\infty d\nu \eta(\nu).
\eeq
The calculation of the average opacities is however more complex. We would ideally want
\beq
\langle \kappa_a \rangle J_{\rm tot} = \int_0^\infty d\nu \kappa_a(\nu) J(\nu).
\eeq
However, while we know $\kappa_a(\nu)$, we only evolve $J_{\rm tot}$, not $J(\nu)$. Without information about the neutrino energy spectrum, the simplest choice aims to guarantee that the equilibrium neutrino energy density takes the expected value
\beq
J_{\rm tot}^{\rm eq} = \frac{\eta_{\rm tot}}{c\langle \kappa_a^{\rm eq} \rangle} 
\eeq
Unfortunately, this is only correct for the energy spectrum of neutrinos in equilibrium with the fluid. As it is quite common for neutrinos in low-density regions to have much higher energy than neutrinos in equilibrium with the fluid, and as $\kappa_a \propto \nu^2$ for many reactions, this can lead to severe underestimates of $\langle \kappa_a \rangle$.

Alternatively, given an average neutrino energy $\langle\nu\rangle$, we may take advantage of the approximate scaling of opacities with $\nu$ to write
\beq
\langle \kappa_a \rangle = \langle \kappa_a^{\rm eq} \rangle \frac{\langle \nu\rangle^2}{\langle \nu_{\rm eq}\rangle^2},
\eeq
with $\langle \nu_{\rm eq}\rangle$ the average energy of neutrinos in equilibrium with the fluid,
\beq
\langle\nu_{\rm eq}\rangle = k_B T \frac{F_3\left(\frac{\mu}{k_B T}\right)}{F_2\left(\frac{\mu}{k_B T}\right)}.
\label{eq:avnu}
\eeq
Estimates for the average neutrino energy have been taken from leakage predictions \citep{FoucartM1:2015}, or from the evolution of $N$. For example, in \cite{Foucart:2016rxm} we used
\beq
\langle\nu\rangle = W \frac{E_{\rm tot} - F_{i,\rm tot} v^i}{N_{\rm tot}}.
\label{eq:nuavg}
\eeq
This last equation is an approximation (it ignores differences between the energy-weighted average energy of neutrinos and their flux-weighted average energy), yet it should provide a significantly better estimate of $\langle \nu \rangle$ than the assumption of equilibrium with the fluid. The scattering opacity can use the same rescaling:
\beq
\langle \kappa_s \rangle= \langle \kappa^{\rm eq}_s\rangle  \frac{\langle \nu\rangle^2}{\langle \nu_{\rm eq}\rangle^2}
\eeq
with $\langle \kappa^{\rm eq}_s \rangle$ calculated assuming an equilibrium spectrum of neutrinos. We note however that in using the same scaling for $\kappa_s$ and $\kappa_a$, we again implicitly assume the same energy spectrum for $J$ and $H^\alpha$. This is generally not true in the diffusion regime (see Sec.\ref{sec:M1scat}).

We additionally need the source terms entering the evolution of $\tilde N$ and/or $Y_e$. For neutrinos of a given energy $\nu$, that source term is simply
\beq
S_N = \frac{-S^\alpha u_\alpha}{\nu} = \frac{\eta-\kappa_a J}{\nu}
\eeq
and, if we know $\eta(\nu)$, we can define an integrated number emissivity
\beq
\eta_{N,\rm tot} = \int_0^\infty d\nu \frac{\eta}{\nu}.
\eeq
As before, we would like to define $\langle \kappa_N \rangle$ such that
\beq
S_{N,\rm tot} = \int_0^\infty d\nu S_N(\nu) = \eta_{N,{\rm tot}} - \langle \kappa_N \rangle N_{\rm tot}.
\eeq
Computing $\langle \kappa_N \rangle$ requires us to once more guess at the neutrino energy spectrum. In \cite{Foucart:2016rxm} we chose $\langle \kappa_N \rangle$ so that neutrinos properly thermalize to an equilibrium spectrum at the fluid temperature if the optical depth is large enough, i.e.
\beq
\langle \kappa_N \rangle = \langle \nu_{\rm eq}\rangle \langle \kappa_a \rangle  \frac{\eta_{N,\rm tot}}{\eta_{\rm tot}}\frac{J_{\rm tot}}{\langle \nu \rangle N_{\rm tot}}\frac{\langle \nu\rangle^2}{\langle \nu_{\rm eq}\rangle^2} =  \langle \kappa_a \rangle  \frac{\eta_{N,\rm tot}}{\eta_{\rm tot}}\frac{J_{\rm tot}}{N_{\rm tot}}\frac{\langle \nu\rangle}{\langle \nu_{\rm eq}\rangle},
\eeq
with $\langle \nu\rangle$ calculated as before, and the last term included to take into account the energy dependence of the cross-sections. This choice is however far from unique. One can e.g. also choose
\beq
S_{N,\rm tot} = \frac{\eta_{\rm tot}-\langle \kappa_a \rangle J_{\rm tot}}{\langle\nu\rangle}
\eeq
with $\langle\nu\rangle$ being estimated either assuming equilibrium between the neutrinos in the fluid (at the risk of strongly underestimating its value) or from a separate estimate of neutrino energies (e.g., from the energy of neutrinos according to a leakage scheme \citep{FoucartM1:2015}).

We see that getting an accurate estimate of the source terms in a grey moment scheme is thus quite difficult, particularly when estimating the source terms entering into the evolution of $Y_e$. As that is also one of the most important parameters to be impacted by neutrino-matter interactions, this may certainly be a significant issue limiting the accuracy of current merger simulations (see \citealt{Foucart:2016rxm} and the discussion of merger simulations here for the impact of inaccurate energy estimates).

\subsubsection{Closures}
\label{sec:M1clos}

The main approximations made by moment schemes are their choice of analytical closure. Grey schemes need an energy closure specifying the neutrino spectrum. Simulations evolving only $\tilde E$ (1-moment schemes) additionally need a closure for $\tilde F_i$ and $\tilde P^{ij}$, while simulations evolving $\tilde E$ and $\tilde F_i$ (2-moment schemes) need a closure for $\tilde P^{ij}$. Energy-dependent schemes also have to specify $\tilde M^{\alpha\beta\gamma}$.

There is no unique prescriptions for these closures that work in all possible regions of a simulation, but there are regimes in which they can be fairly easily calculated -- mainly regions where neutrinos are in thermal equilibrium with the fluid (optically thick regime), as well as regions far away from a localized source of neutrinos, where all neutrinos approximately propagate away from that source. 

In the first case, we can assume that for an orthonormal tetrad in the fluid rest frame
\beq
L^{\hat i \hat j} \approx \frac{1}{3} \delta^{\hat i\hat j} J,
\label{eq:M1cthick}
\eeq
as appropriate for a relativistic gas of neutrinos. In covariant form, this is
\beq
L^{\mu\nu} \approx \frac{1}{3} h^{\mu\nu} J.
\eeq
Going back to an orthonormal tetrad in the fluid frame, we can write the moments equations in the optically thick regime as
\beqn
\partial_{\hat t} J + \partial_{\hat i} H^{\hat i} &=& \eta - \kappa_a J\\
\partial_{\hat t} H_{\hat i} + \frac{1}{3} \partial_{\hat i} J &=& -(\kappa_a + \kappa_s) H_{\hat i}
\eeqn
In this regime, neutrinos reach a quasi-equilibrium state on a time scale short compared to the time step of the simulation, and thus
\beq
H_{\hat i} \approx -\frac{1}{3(\kappa_a+\kappa_s)} \partial_{\hat i} J \rightarrow H^\mu \approx -\frac{h^{\mu\nu}}{3(\kappa_a+\kappa_s)}\nabla_\nu J.
\label{eq:M0cthick}
\eeq
This gives us closure relations appropriate for a 1-moment scheme (Eq.~\eqref{eq:M0cthick}) and for a 2-moments scheme (Eq.~\eqref{eq:M1cthick}). We note however that these equations are provided in the fluid frame. The more useful relations providing $F_i$ or $P^{ij}$ as functions of lower order moments can be recovered using the relationship between fluid-frame and simulation-frame moments (see e.g., \citealt{shibata:11,PhysRevD.87.103004,FoucartM1:2015}). Applying this optically thick closure for all values of the opacity corresponds to the {\it diffusion approximation} (Flick's law). Such a choice is however clearly problematic for low value of the opacities, as the flux becomes infinite when $(\kappa_a+\kappa_s)\rightarrow 0$.

For free-streaming neutrinos that all propagate in the same direction (as expected far from a post-merger remnant), we expect instead
\beq
F^i = E \hat f^i
\label{eq:M0cthin}
\eeq
with $\hat f^i$ the unit vector pointing away from the remnant, and
\beq
P^{ij} = \frac{F^i F^j}{E} = \frac{F^i F^j}{F^k F_k} E = \frac{F^i F^j}{\sqrt{F^kF_k}}.
\label{eq:M1cthin}
\eeq
Any form of Eq.~\eqref{eq:M1cthin} is a valid closure for a 2-moment scheme. Eq.~\eqref{eq:M0cthin} could on the other hand only be used in a 1-moment scheme if we choose a direction of propagation for the neutrinos, e.g., in a ray-by-ray transport scheme. For example, the M0 scheme of \cite{Radice:2016dwd}, which is only used to evolve free-streaming neutrinos in their mixed moment-leakage scheme, takes $\hat f^i$ to be the unit vector in the $t-r$  plane orthogonal to the fluid velocity. 

For 1-moment schemes, one way to combine these two limits is flux-limited diffusion (see e.g. \cite{1981ApJ...248..321L}), which sets
\beq
F^i  = \lambda(R) J R^i;\,\, R_i = \frac{\partial_i J}{(\kappa_a+\kappa_s)J}
\eeq
for some function $\lambda(R)$ which asymptotes to $1/3$ in optically thick regions (small $R$) and $1$ in optically thin regions (large $R$). For two-moment schemes, a number of proposals have been made for the ``optimal'' choice of closure, e.g. \cite{1975NYASA.262...54W,Minerbo1978,1981ApJ...248..321L,2000A&A...356..559S}, usually under the assumption of a preferred direction for the propogation of neutrinos (spherical or planar symmetry). If we write $F^i = f E n^i$, the tensorial closure above can then be replaced by a prescription for the scalar Eddington factor $p=\|P^{ij}n_j/E\|$. Existing general relativistic simulations use the maximum entropy closure for $p$ derived by \cite{Minerbo1978} for photon transport, and updated by \cite{1989JQSRT..42..603C} for neutrino transport. A closed-form expression for this closure was derived by \cite{1994ApJ...433..250C}. The full pressure tensor can be recovered from $p$ using the prescription of \cite{1984JQSRT..31..149L,1999CRASM.329..915D}. Overall, this results in the closure relation
\beq
P^{ij} = d_{\rm thin} P^{ij}_{\rm thin} + d_{\rm thick} P^{ij}_{\rm thick}.
\eeq
using the optically thick limit described earlier in this section (which is expressed in the fluid frame, as discussed below), as well as
\beq
P^{ij}_{\rm thin} =  \frac{F^i F^j}{F^k F_k} E
\eeq
and
\beq
d_{\rm thin} = \frac{3\chi-1}{2};\,\, d_{\rm thick} = \frac{3}{2}(1-\chi);\,\, \chi=\frac{1}{3}+\xi^2\frac{6-2\xi+6\xi^2}{15};\,\,\xi=\frac{H^\alpha H_\alpha}{J^2}.\nonumber
\eeq
We see that in optically thick regions ($\xi=0$) we recover the optically thick closure $P^{ij}_{\rm thick}$, while in free-streaming regions ($\xi=1$), we get $P^{ij}_{\rm thin}$. It is however worth noting that while the former is asymptotically correct for large opacities, the latter is not correct for free-streaming neutrinos: neutrinos in vacuum are generally not all propagating in the same direction, and thus do not satisfy Eq.~\eqref{eq:M1cthin}. When making this choice, the interpolation between two well-defined asymptotic regime is thus not the only source of error: we have to come to terms with the fact that the optically thin regime uses a closure that is inaccurate in most regions where neutrino-matter interactions are important. A well known consequence of that choice is the creation of artificial radiation shocks whenever neutrino beams cross (see Fig.~\ref{fig:beams}, and similar results in the two-beam test performed by \cite{Sdowski:2012cx}). In neutron star mergers, this also leads to an overdensity of neutrinos in the polar regions (see discussion of simulations).

\begin{figure}[ht]
\centering
\includegraphics[width=0.7\textwidth]{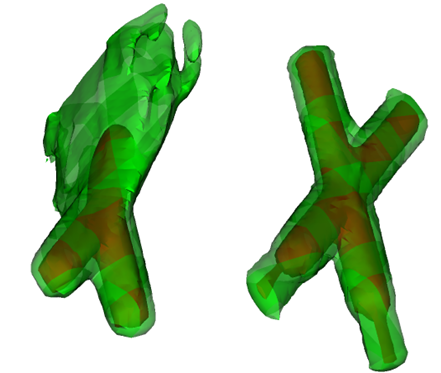}
\caption{Two neutrino beams crossing in a two-moment simulation using the Minerbo closure (Left), and in a two-moment simulation in which the pressure tensor is taken from the result of a Monte-Carlo evolution of the transport equations (Right). We see that the Minerbo closure leads to an artificial collision between the beams. Figure reproduced from \cite{Foucart:2017mbt}.}
\label{fig:beams}
\end{figure}

We also note that in the optically thin limit, we listed a number of closures that were theoretically equivalent as long as $F^k F_k = E^2$, a condition satisfied if all neutrinos propagate in the same direction. These expressions do however differ in regimes when $F^kF_k \neq E^2$. \cite{shibata:11} show that the choice
\beq
P^{ij}_{\rm thin} = E \frac{F^i F^j}{F^k F_k}
\label{eq:M1cthin2}
\eeq
has the advantage to guarantee $P^k_k = E$, but can lead to superluminal characteristic speeds for the moment equations if used directly as a closure, i.e. for $d_{\rm thin}=1$, $d_{\rm thick}=0$. On the other hand
\beq
P^{ij}_{\rm thin} =  \frac{F^i F^j}{\sqrt{F^k F_k}}
\eeq
leads to $P^k_k \neq E$ when $F^k F_k \neq E^2$. Their recommendation to prefer Eq.~\eqref{eq:M1cthin2}, consistent with \cite{1984JQSRT..31..149L,1999CRASM.329..915D}, has been followed in merger simulations so far, and appears to lead to a system of equations that is both hyperbolic and causal \citep{1999CRASM.329..915D,shibata:11}.

As $\xi$ is a function of the fluid-frame moments $(J,H_\alpha)$, which are themselves functions of the evolved simulation frame moments $(E,F_i,P^{ij})$; and as $P^{ij}$ depends on $\xi$, we see that our prescription for $\xi$ is an implicit equation.
We can solve for $\xi$ by searching for roots of the function
\beq
R(\xi) = \frac{J^2-H^\alpha H_\alpha}{E^2}
\eeq
with the known physical bounds $\xi \in [0,1]$. As $\xi$ changes relatively slowly, the value of $\xi$ at the end of a recent time step is usually a good initial guess for the current value of $\xi$, allowing for rapid convergence of many root finding algorithms.

For grey schemes evolving $(\tilde E,\tilde F_i)$, the above closure combined with a choice of neutrino energy spectrum is sufficient to close the system of equations. When evolving $\tilde N$ as well, the situation is a little more complex, as we also need to estimate the number flux $\tilde F^i_N$. For monoenergetic neutrinos, by definition,
\beq
N^\alpha = Nn^\alpha + F_N^\alpha = \frac{Ju^\alpha + H^\alpha}{\nu} \rightarrow F^i_{N} 
= \frac{JWv^i + \gamma^i_\mu H^\mu}{\nu}.
\eeq
However, after integrating over neutrino energies, we should get
\beq
F^i_{N,{\rm tot}} 
= \frac{J_{\rm tot} Wv^i}{\langle \nu \rangle_J} + \frac{\gamma^i_\mu H^\mu_{\rm tot}}{\langle \nu\rangle_H}.
\eeq
The first term in this flux is important to capture advection of neutrinos with the fluid, while the second is important to capture neutrino propagation in the fluid frame. The factor $\langle \nu \rangle_J$ is an energy-weighted average neutrino energy that can reasonably be estimated using Eq.~\eqref{eq:nuavg}. On the other hand, $\langle \nu\rangle_H$ is a flux-weighted average energy that, in dense regions, is likely smaller than $\langle \nu \rangle_J$. Indeed, low-energy neutrinos diffuse faster than high-energy neutrinos, and thus $H^\alpha(\nu)$ has a softer energy spectrum than $J(\nu)$. In fact, there is no particular reason to expect $\langle \nu\rangle_H$ to be the same for every component of $H^\alpha$! \cite{Foucart:2016rxm} developed a rather complex procedure to estimate $\langle \nu\rangle_H$ during a simulation, which involves a new evolution equation for the spectral index of each neutrino species. While that procedure was calibrated to match the result of energy-dependent radiation transport in specific test cases, and aims to capture the transition from a soft spectrum for $H^\alpha$ in the diffusion regime to a thermal spectrum close to the neutrinosphere, its accuracy for more complex physical configurations is unknown. \cite{Radice:2021jtw}, who also evolve $\tilde N$, make the simpler approximation $\langle \nu\rangle_H=\langle \nu\rangle_J=\langle \nu\rangle$. Regardless of the choice made, the difficulty of properly capturing the evolution of $\tilde N$ in the diffusion regime, and thus the diffusion of $Y_e$ through dense regions, is a limitation of existing moment schemes.

Finally, for completeness on the topic of closures in the context of general relativistic simulations, it is worth mentionning that \cite{shibata:11} and \cite{PhysRevD.87.103004} have provided closures for the third moment $M^{\alpha\beta\gamma}$. One method assumes that the expansion of the distribution function in $l^\alpha$ is truncated at second order, in which case the third moment is an explicit function of the first moment. An alternative is to use the Minerbo closure to again interpolate between optically thick and free-streaming estimates of $L^{\alpha\beta\gamma}$. These closures are not needed in existing grey moment schemes, but would be necessary for energy-dependent schemes.

\subsubsection{Numerical implementation}
\label{sec:M1num}

In the previous sections, we covered most of the ingredients needed to evolve moments of $f_\nu$: the moment equations in the simulation frame, their coupling to the fluid equations, and the required analytical closures for higher-order moments of $f_\nu$ (and for the neutrino energy spectrum in grey schemes). These equations are very similar to the equations of hydrodynamics in conservative form, i.e., they can be expressed as
\beq
\partial_t U + \partial_i \mathcal{F}^i(U) = \mathcal{S}(U)
\eeq
with $U$ the vector of evolved variables (e.g. $U=(\tilde E, \tilde F_i, \tilde N)$ in a grey M1 scheme that evolves the number density), $\mathcal{F}^i$ the fluxes, and $\mathcal{S}$ the local source terms. These equations can be evolved using the same high-order shock-capturing methods developed for the equations of fluid dynamics, guaranteeing that conservation laws are satisfied to round-off accuracy in our evolutions. The main complications will come from the source terms, which can be extremely large for radiation transport, thus requiring the right-hand side of this equation to be treated implicitly. 
The flux terms, on the other hand, can be treated explicitly as long as the timestep satisfies the usual Courant condition $\Delta t \lesssim \alpha_{CFL} \Delta x$, with $\Delta x$ the grid spacing and $\alpha_{CFL}$ a constant of order unity that depends on the exact numerical methods used to evolve these equations. In practice, it is thus useful to consider split implicit-explicit time evolutions
\beq
\partial_t U = \left( \mathcal{S}_{\rm exp}(U)-\partial_i \mathcal{F}^i(U)\right) + \mathcal{S}_{\rm imp}(U)
\eeq
where $\mathcal{S}_{\rm imp}+\mathcal{S}_{\rm exp}=\mathcal{S}$, and $\mathcal{S}_{\rm imp}$ contains all terms that we choose to treat implicitly (typically, neutrino-matter interactions). A first-order in time discretization would then be the implicit equation
\beq
U^{n+1} = U^n + \Delta t \left( \mathcal{S}_{\rm exp}(U^n)-\partial_i \mathcal{F}^i(U^n)\right) + (\Delta t) \mathcal{S}_{\rm imp}(U^{n+1}).
\eeq
where upper indices $(n,n+1)$ refer to the beginning/end of the time step. In a conservative scheme, the spatial discretization requires us to consider values of the fields at grid points, and halfway between grid points. For example, in 1D,
\beq
U^{n+1}_i = U^n_i +  \Delta t \left( \mathcal{S}_{\rm exp}(U^n_i)-\frac{\mathcal{F}^*(U^n_{i+1/2})-\mathcal{F}^*(U^n_{i-1/2})}{\Delta x}\right) + (\Delta t) \mathcal{S}_{\rm imp}(U^{n+1}_i)\nonumber
\eeq
with the subscripts referring to grid points / cell centers (integer values) and half grid points / cell faces (half-integer values). Many possible choices can be made for the calculation of the numerical fluxes $\mathcal{F}^*$, which will offer different orders of convergence and shock-capturing capabilities. A simple, very dissipative choice would for example be the Lax-Friedrich flux
\beq
\mathcal{F}^*_{i+1/2} = \frac{1}{2} \left(\mathcal{F}_i + \mathcal{F}_{i+1}\right) - \frac{c}{2}\left(U_{i+1}-U_i\right).
\label{eq:numF}
\eeq
Modern numerical simulations often use higher-order, less dissipative methods. In those advanced methods, values of $\mathcal{F},U$ at $(i,i+1)$ are replaced with left-biased and right-biased stencils used to interpolate $(\mathcal{F},U)$ on cell faces \citep{MC,Jiang1996202,Borges}, while the speed of light $c$ is replaced by the maximum characteristic speed of the system of equations. The general idea, however, remains: the numerical fluxes include a first term estimating the flux at the mid-point, and a second term providing numerical dissipation at shocks to smooth the solution and avoid instabilities. Alternatively, the evolution equations can be projected onto their characteristic fields, allowing for the use of even less dissipative methods (see, e.g., \citealt{Radice:2012cu} for general relativistic fluid dynamics). We will not review here the extensive literature discussing choices for these numerical fluxes, but note that the characteristic speeds for the two-moments algorithm can for example be found in \cite{shibata:11}, and details of the calculation of $\mathcal{F}^*$ for two-moment algorithms used in merger simulations are available in \cite{shibata:11,FoucartM1:2015,Weih:2020wpo,Radice:2021jtw,Sun:2022vri}. The only difficulty specific to the radiation transport equations is the treatment of high-density regions, discussed in more detail in Sect.~\ref{sec:M1scat}.

The treatment of the implicit terms has steadily improved over the years, starting from the hybrid leakage-moment schemes that do not require any implicit treatment of the source terms \citep{shibata:11}, to first approximate \citep{FoucartM1:2015},  and then full implicit-explicit time-stepping \citep{Weih:2020wpo} linearizing the source terms around the zero-state $(E=F^i=N=0)$, to most recently a linearization of the problem around an arbitrary state for the neutrino radiation field \citep{Radice:2021jtw}. We review here the latest methods of \cite{Radice:2021jtw}. In that work, the Jacobian matrix
\beq
J(U) = \frac{\partial S_{\rm imp}(U)}{\partial U}
\eeq
is calculated explicitly for the Minerbo closure around an arbitrary state $U$. The implicit equations for $U^{n+1}$ can then be solved using standard iterative methods for the determination of the roots of a multi-dimensional function $f(U)$, given $f$, $\partial f/\partial U$, and a reasonable initial guess $U_g$ for the solution (e.g., the value of $U$ at the beginning of a time step, or the equilibrium value of $U$). We note that in this case, $U=(\tilde E, \tilde F_x, \tilde F_y, \tilde F_z$). Each species of neutrinos is treated separately, and the evolution of the number equation (if included) can be performed after the evolution of $U$ [$J(U)$ is independent of $\tilde N$ if $U=(\tilde E, \tilde F_i)$]. As $J(U)$ is quite complex, we refer the reader to \cite{Radice:2021jtw} for its exact form. Many older two-moment schemes however use more approximate values of the Jacobian matrix in order to simplify the calculation of $J$ at the cost of some accuracy in the implicit solve.

Finally, we note that most moment algorithms use a split operator method to couple the fluid and neutrino evolution. Assuming that $U_{\rm fl}$ and $U_{\rm rad}$ are the vectors of evolved variables for the fluid and neutrinos respectively, the coupled problem is evolved using equations of the form
\beqn
U^{n+1,*}_{\rm fl} &=& f_{\rm fl}(U^n_{\rm fl},\Delta t)\\
U^{n+1}_{\rm rad} &=& f_{\rm rad}(U^{n+1,*}_{\rm fl},U^n_{\rm rad},U^{n+1}_{\rm rad},\Delta t)\\
U^{n+1}_{\rm fl} &=& U^{n+1,*}_{\rm fl} + S_{\rm fl/rad} \Delta t.
\eeqn
The first line represents the usual evolution of the fluid equations without coupling to the neutrinos. The second line represents the mixed implicit-explicit evolution of the radiation equations on a given fluid background. Finally, the third line represents backreaction of the radiation onto the fluid. It is also possible to improve on this scheme by using a guess $U^{n+1,g}_{\rm fl}$ for the fluid variables on the second line. This is particularly useful in dense regions, where we might use the expected state of the fluid once neutrinos and matter equilibrate (see e.g., \citealt{Foucart:2016rxm}). Using such a guess is sometimes necessary to avoid numerical instabilities in regions where the coupling between neutrinos and matter leads to stiff source terms in the evolution of the fluid equations (i.e., in the $S_{\rm fl/rad}$ term), in addition to the stiff source terms that nearly always exist in the evolution of the neutrino moments.

A more self-consistent method would be of course to use an implicit-explicit solver to evolve jointly the fluid equations and the moment equations for all species of neutrinos, but this would make the implicit part of the solver significantly more costly. Indeed, the standard scheme solves, for each neutrino species, a system of 4 coupled non-linear implicit equations for $(\tilde E,\tilde F_i)$ and a single linear implicit equation for $\tilde N$. A full implicit solve of the fluid and radiation equations, on the other hand, would require solving a system of $6+5N_\nu$ non-linear implicit equations for the $N_\nu$ neutrino species and the fluid (assuming $6$ fluid variables and 5 components of the moments for a grey two-moment scheme). As long as the simpler method does not lead to numerical instabilities, its significantly lower computational cost makes it much more appealing.

\subsubsection{Diffusion regime}
\label{sec:M1scat}

In regions where $(\kappa_a+\kappa_s)L\gg1$, with $L$ a typical lengthscale of our system, we expect the neutrino momentum density in the fluid frame to be $\propto -(\kappa_a+\kappa_s)^{-1} \nabla J$. This leads to an evolution equation for the energy density
in the fluid frame
\beq
\partial_{\hat t} J - \delta^{\hat i\hat j} \partial_{\hat i} \left(\frac{1}{3(\kappa_a+\kappa_s)}\partial_{\hat j} J\right)
= \kappa_a (J^{\rm eq} - J).
\eeq
i.e., a diffusion equation with diffusion coefficient $D=\left[3\left(\kappa_a+\kappa_s\right)\right]^{-1}$. However, when evolving the two-moment equations with shock-capturing methods, the dissipative terms in the numerical fluxes modify this optically thick solution.
For example, with a low-order numerical flux like Eq.~\eqref{eq:numF}, and ignoring spatial variations in the opacities, the discretized equation becomes \citep{Audit:2002hr}
\beq
\partial_{\hat t} J - \left(\frac{1}{3(\kappa_a+\kappa_s)} + \frac{c \Delta x}{2}\right) \delta^{\hat i \hat j} \partial^2_{\hat i \hat j} J
= \kappa_a (J^{\rm eq} - J).
\eeq
We clearly see that in any region where $(\kappa_a+\kappa_s)\Delta x \gtrsim 1$, numerical diffusion will be larger than physical diffusion. Higher-order fluxes will be more forgiving, but even high-order shock capturing methods behave similarly to their low-order counterparts near shocks or in underresolved regions. As $J$ may vary rapidly in the hot, dense regions of a merger remnant, the diffusion rate of neutrinos through that remnant could plausibly be impacted by numerical dissipation. 

To avoid this issue, two methods have been proposed so far. \cite{Audit:2002hr} suggest to effectively use a one-moment scheme in regions where $(\kappa_a+\kappa_s)\Delta x \gtrsim 1$, i.e., to replace the numerical flux in the evolution of $\tilde E$ by its value assuming that
\beq
H^\alpha = -\frac{h^{\alpha\beta}}{3(\kappa_a+\kappa_s)} \nabla_\beta J;\,\, L^{\alpha\beta} = \frac{h^{\alpha\beta}}{3} J.
\eeq
This leads to neutrinos being advected with the fluid, with an additional slow diffusion provided by $H^\alpha$. A more detailed discussion of this method in the merger context can be found in \cite{FoucartM1:2015}. As pointed out in \cite{Radice:2021jtw}, this method does however transform our evolution equations into a single diffusion equation, which is known to be acausal and potentially unstable in general relativity. \cite{Radice:2021jtw} thus propose an alternative, numerically simpler method: they use a high-order flux without numerical dissipation (i.e., using finite difference methods) in regions where $(\kappa_a+\kappa_s)\Delta x \gtrsim 1$. In both algorithms, one transitions smoothly between the standard shock-capturing fluxes and the modified fluxes around $(\kappa_a+\kappa_s)\Delta x \sim 1$. For example, \cite{Radice:2021jtw} use
\beq
\mathcal{F}^* = (1-a) \mathcal{F}^{\rm HO} + a \mathcal{F}^{\rm LO}
\eeq
with $\mathcal{F}^{\rm LO}$ given by Eq.~\eqref{eq:numF} (replacing $c$ by the value of the speed of light in the simulation coordinates) and $\mathcal{F}^{\rm HO}$ using the simple second-order accurate prescription
\beq
\mathcal{F}^{\rm HO}_{i+1/2} = \frac{\mathcal{F}_i + \mathcal{F}_{i+1}}{2}.
\eeq
The transition coefficient is
\beq
a = \min{\left(1,\frac{1}{\Delta x (\kappa_a+\kappa_s)}\right)}
\eeq
and the opacities are estimated using the average values of neighboring grid points. Outside of the merger context, a conceptually similar correction limiting the use of the dissipative fluxes in high-opacity regions had been used by \cite{Sdowski:2012cx}.

\begin{figure}[ht]
\centering
\includegraphics[width=0.6\textwidth]{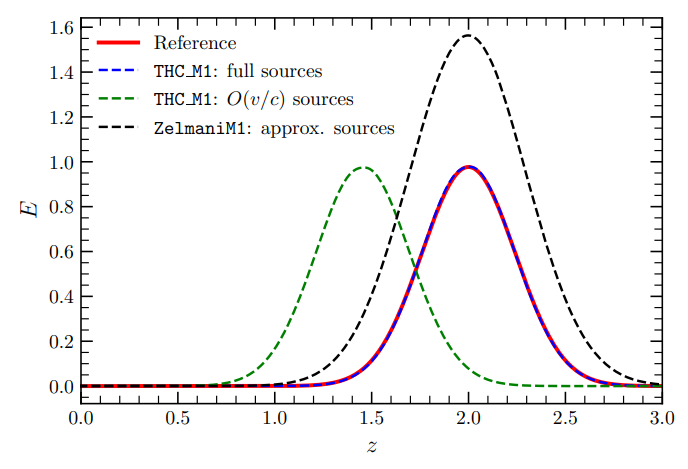}
\caption{Advection of trapped neutrinos in a rapidly moving fluid ($v=0.5c$) with high scattering opacity, and no absorption or emission. The different simulations use (i) the latest implementation of the Jacobian matrix and high-order numerical fluxes from \cite{Radice:2021jtw} (blue), (ii) approximate sources and a diffusion equation (ZelmaniM1 code, black), or (iii) only first order in $v/c$ terms (green). Figure reproduced from \cite{Radice:2021jtw}.}
\label{fig:beams}
\end{figure}

\cite{Radice:2021jtw} show that the approximate linearization of the source terms used in many existing two-moment schemes and their reliance on the solution of a diffusion equation leads to inaccuracies in the evolution of the transport equations at the very least in regions where $\kappa_s \Delta x \gg 1$, $v/c\sim 1$, and $\kappa_a \Delta x \ll 1$ (i.e., where advection and diffusion are important, and the evolution is not driven to the equilibrium value of the moments by a high absorption rate). This regime is likely to be relevant in the outer regions of a rotating post-merger neutron star remnant for heavy-lepton neutrinos. Additionally, these approximations may lead to other sources of errors in yet untested regimes. On the other hand, the use of non-dissipative fluxes could, in theory at least, lead to issues at shocks in dense region -- though this does not seem to have been observed in simulations using these newer methods so far. 

\subsubsection{Pair annihilation}
\label{sec:M1pairs}

Pair processes pose a particular challenge for moment-based transport algorithms -- and, in fact, for many other types of radiation transport schemes -- because they naturally lead to non-linear source terms in the transport equations that couple the distribution function of neutrinos and antineutrinos (see Sect.~\ref{sec:thpairs}). On the other hand, fully ignoring pair production and annihilation is not an option for muon and tau neutrinos. In grey moment schemes, we thus typically calculate an emission rate of $\nu_\mu\bar\nu_\mu$ pairs and $\nu_\tau \bar\nu_\tau$ pairs with blocking factors computed assuming an equilibrium density of neutrinos, and model the inverse reaction (pair annihilation) through an absorption opacity calculated using Kirchhoff's law. 

For a typical merger profile, we expect heavy lepton neutrinos to experience higher scattering opacities than absorption opacities. 
Deep into the neutron star, heavy lepton neutrinos are nonetheless in thermal equilibrium with the fluid, at least as long as $\kappa_a t_{\rm diff} \sim 1$. Due to the slow diffusion rate through the dense matter, this remains true even in regions where their absorption optical depth $\tau_a< 1$, as their scattering optical depth $\tau_s \gg 1$. In those regions, our assumptions for pair production and annihilation may be inaccurate, but they should nonetheless practically capture the physics of neutrino diffusion quite well. Closer to the surface of the remnant, however, we eventually reach regions where $\kappa_a t_{\rm diff} < 1$ and $\tau_s > 1$. There, heavy lepton neutrinos slowly diffuse out while out of thermal equilibrium with the fluid (typically, neutrinos will be hotter than if they were in equilibrium at the fluid temperature). In these regions, our assumptions likely underestimate the rate of pair annihilation. The impact of this approximation has not been studied so far.

For electron type neutrinos, $\nu_e$ and $\bar \nu_e$ decouple from the fluid in different regions, and the above assumptions would be even more problematic. However, in optically thick regions, pair production and pair annihilation are expected to be subdominant -- except maybe in the hottest regions, where neutrinos will be in equilibrium with the fluid regardless of the reactions included in a calculation. Accordingly, ignoring pair production for electron-type neutrinos has often been considered safer than including it in a very approximate manner.

This leaves us with one important issue, however: $\nu\bar\nu$ annihilation is expected to be an important process for energy deposition in low-density regions around the rotation axis of a post-merger remnant, and may also deposit energy and momentum in low-density matter outflows elsewhere. From Eq.~\eqref{eq:Qpairs}, we get that the appropriate energy-integrated source term for the moment equations is, for annihilation into $e^+e^-$ pairs and for the energy deposited by the neutrinos only,
\beqn
S^\alpha_{(\nu)} &=& -\int\int \frac{d^3p_{(\nu)}}{\sqrt{-g}p^t_{(\nu)}h^3}\frac{d^3p_{(\bar\nu)}}{\sqrt{-g}p^t_{(\bar\nu)}h^3} f_{(\nu)} f_{(\bar \nu)} (p^\alpha_{(\nu)})
\frac{DG_F^2}{3\pi}\left(-p_{(\nu)}^\beta p_{\beta(\bar\nu)}c^2\right)^2\nonumber\\
&=& -\frac{c^4DG_F^2}{3\pi}M_{\beta\gamma,(\bar \nu)} \int \frac{d^3p_{(\nu)}}{\sqrt{-g}p^t_{(\nu)}h^3} f_{(\nu)} p^\alpha_{(\nu)}
p^\beta_{(\nu)}p^\gamma_{(\nu)}.
\eeqn
The integral over $p_{(\nu)}$, however, does not match the moments used in our evolution equations (it includes an extra power of the neutrino energy). A similar term should also be included for the antineutrinos. \cite{Fujibayashi:2017xsz} propose the approximation
\beq
S^\alpha_{(\nu)} = -\langle \epsilon_{(\nu)} \rangle u^\alpha \frac{c^4DG_F^2}{3\pi}M_{\beta\gamma,(\bar \nu)} M^{\beta\gamma}_{(\nu)} 
\eeq
with $\langle \epsilon_{(\nu)} \rangle$ some appropriate average of the energy of annihilated neutrinos. Given the scaling in the source term, the choice
\beq
\langle \epsilon_{(\nu)} \rangle = \frac{F_4(\eta_{(\nu)})}{F_3(\eta_{(\nu)})} k_B T_{(\nu)}
\eeq
would be reasonable for a quasi-thermal spectrum at an estimated temperature $T_{(\nu)}$. 

\begin{figure}[ht]
\centering
\includegraphics[width=0.99\textwidth]{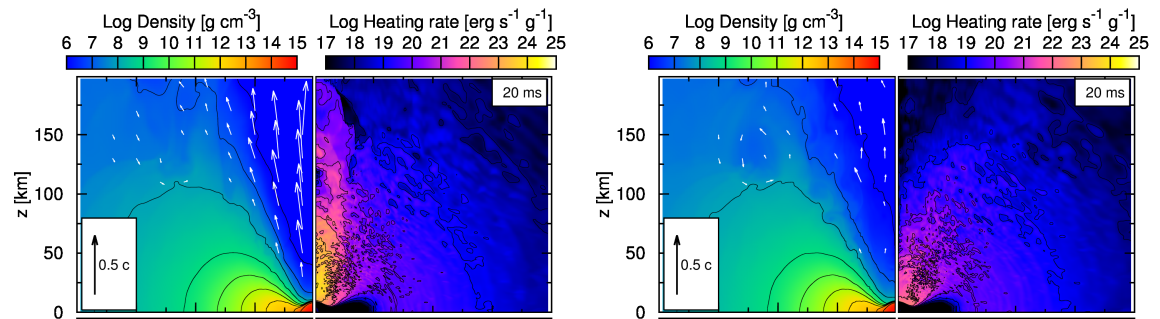}
\caption{Density, velocity, and heating rate from neutrino absorption $20\,{\rm ms}$ after a neutron star merger in a simulation using pair annihilation (Left), and a similar simulation ignoring that effect (Right). Note the significant change in the velocity of the outflows, which persists up to the end of these simulations ($300\,{\rm ms}$ post-merger). Figure reproduced from \cite{Fujibayashi:2017xsz}.}
\label{fig:pairs}
\end{figure}

This remains a significant approximation. Indeed, by using $S^\alpha \propto u^\alpha$, the above formula ignores momentum deposition in the fluid frame, even though neutrinos leaving the remnant definitely have a preferred direction of propagation. Additionally, the source term depends on the neutrino pressure tensor, which is estimated through the chosen analytical closure. Polar regions where pair annihilations are expected to be important (see Fig.~\ref{fig:pairs}) are also where that closure creates the largest errors. Despite these limitations, comparisons with Monte-Carlo transport results \citep{Foucart:2018gis} indicate that $S^\alpha$ is at least order-of-magnitude accurate when using this approximation -- which is about the best that one can hope for in the context of a grey moment scheme.

\subsubsection{Discussion}

As the main algorithms used so far for radiation transport in general relativistic merger simulations, moment schemes have allowed us to greatly increase our understanding of the role of neutrinos in mergers. In particular, moment simulations showed that neutrino-matter interactions have a fairly dramatic impact on the composition of hot matter outflows, with important consequences for r-process nucleosynthesis \citep{Wanajo:2014}. The cost and complexity of moment schemes is however significantly larger than those of the simplest leakage schemes: the moment equations without source terms are comparable in complexity to the evolution of the fluid equations, and if large regions of the computational domain require the use of implicit timestepping (as in most merger simulations that do not result in the formation of a black hole), radiation transport can easily become the main computational cost of a simulation. Mixed leakage-moment schemes can do away with that last issue, but at the cost of a more approximate treatment of neutrino diffusion in dense regions.

The most up-to-date moment schemes can likely capture very well the diffusion of single-energy neutrinos in dense regions, yet as moment schemes used in general relativistic merger simulations do not keep detailed information about the neutrino energy spectrum, and the diffusion of neutrinos through a dense medium is strongly energy dependent, their ability to properly capture the diffusion of both energy and lepton number consistently is more difficult to assess. We will see that simulations with different methods for estimating neutrino energies provide wildly different answers for the emission of heavy-lepton neutrinos \citep{Foucart:2020qjb} -- which may be in part because those neutrinos spend a significant amount of time in regions where they are out of thermal equilibrium with the fluid, yet still experiencing high scattering opacities. Approximate treatment of the source terms and the use of a solution to the acausal diffusion equations in many moment schemes may also be a source of uncertainty in that regime \citep{Radice:2021jtw}. In the semi-transparent and optically thin regimes, moment algorithms offer a reasonable approximation to the qualitative evolution of the composition and temperature of the outflows. However, simulations have shown that the assumed energy spectrum of neutrinos in those regions has a non-negligible impact on the final composition of the ejecta \citep{Foucart:2016rxm}. Energy-dependent moment schemes could do away with most of those issues, but would be computationally expensive, even if difficulties related to the choice of reference frame in which the neutrino energy is discretized were to be solved.

In optically thin regimes, the pressure closure chosen in existing simulations is also a source of concern. Factor of $\sim 2$ errors in the energy density of neutrinos in the polar regions observed in comparisons with Monte-Carlo simulations \citep{Foucart:2018gis} are very likely due to this approximate closure. Combined with the lack of information about the exact distribution function of neutrinos, this limits the ability of moment schemes to take into account the role of pair annihilation in the development of a baryon-free region and of a relativistic jet along the remnant's rotation axis -- even though existing simulations that are expected to capture the importance of pair annihilation at least at the qualitative level indicate that pair annihilation may significantly impact the properties of polar regions \citep{Fujibayashi:2017xsz,Foucart:2018gis} (see also Fig.~\ref{fig:pairs}).

Despite these limitations, comparisons of moment schemes with simulations using Monte-Carlo transport offer some reassurances regarding the validity of the moment results \citep{Foucart:2020qjb}, with disagreements at the $10\%-20\%$ level in most global quantities, including the $\nu_e$ and $\bar \nu_e$ luminosities, the average energy of neutrinos, and the mass and composition of the outflows. Given the difficulty of estimating errors directly in moment simulations, further studies of the uncertainties associated with the many different moment schemes currenlty used in the merger community are certainly required, but the overall understanding of the most important neutrino processes in merger simulations gathered from these simulations is likely accurate (except for the potential role of neutrino oscillations).

\subsection{Monte-Carlo radiation transport}

Grey moment schemes are at this point the most commonly used algorithms to approximately evolve Boltzmann's equation in merger simulations. We have seen however that one of their limitations is the strong assumptions that need to be made about the energy spectrum of the neutrinos, and the impact of these assumptions on neutrino-matter interaction rates. The use of an approximate closure for the pressure also introduces an important approximation in simulations. As a result, even in the limit of infinite resolution, a two-moment scheme does not converge to the correct solution to Boltzmann's equations.

Direct discretizations of Boltzmann's equation in both position space and momentum space using finite difference, finite volume, or spectral methods have not yet been used in general relativistic merger simulations, and are likely beyond our current computational capabilities. While we wait for these methods to become practical in the merger context, however, an interesting alternative is the use of Monte-Carlo radiation transport. Monte-Carlo methods are particularly efficient for low-cost evolutions of high-dimensional, highly inhomogeneous problems. Their scaling with increasing computational resources is significantly worse than the other methods mentioned so far, and they will thus nearly certainly become less efficient than those other algorithms at some point in the future -- but at the moment, they are the only algorithms going beyond the moment formalism to have been implemented in general relativistic merger \citep{Foucart:2020qjb} and post-merger \citep{Miller:2019dpt} simulations.

As for moment schemes, the use of Monte-Carlo methods for the evolution of radiation coupled to a fluid has a long history outside of the context of neutron star mergers and general relativistic radiation hydrodynamics. Monte Carlo methods have been developed to, among other applications, study stellar profiles \citep{1999A&A...344..282L}, stellar formation \citep{2003MNRAS.340.1136E,2012MNRAS.420..562H}, black hole accretion \citep{Ryan2015}, post-merger accretion disks \citep{richers:15}, supernova ejecta \citep{Kasen:2006ce,2012MNRAS.425.1430N,Wollaeger:2013sta}, or core-collapse supernovae \citep{10.1007/978-1-4612-2988-9_57,Abdikamalov:2012zi}.
In this section, we will first review the formalism of Monte-Carlo radiation transport in general relativistic simulations (Sect.~\ref{sec:MCform}). We will then discuss technical issues related to the coupling of the neutrinos to the fluid (Sect.~\ref{sec:MCfl}), the treatment of optically thick regions (Sect.~\ref{sec:MCthick}), and the inclusion of non-linear source terms such as pair processes (Sec~\ref{sec:MCpairs}). As the use of Monte-Carlo methods in neutron star merger simulations is significantly less mature than the use of leakage or moment schemes, and the exact methods are likely to change rapidly in the next few years, we will keep our discussion more general than in the previous section, focusing on the important components of an algorithm and the main difficulties that have been encoutered so far. We base our discussion of general relativistic Monte-Carlo transport in large part on the methods developed for photon transport in black hole accretion disk by \cite{Ryan2015}, with modifications made for applications to the merger problem in \cite{Foucart:2021mcb}. We also comment on the recent development of another Monte-Carlo code aimed at axisymmetric post-merger simulations by \cite{Kawaguchi:2022tae}. That code generally makes less simplifying assumptions and aims for higher order methods than \cite{Foucart:2021mcb}, taking advantage of the expected lower cost of axisymmetric simulations.

As in the previous section, we assume $h=c=1$ unless noted otherwise.

\subsubsection{Formalism}
\label{sec:MCform}

The general idea behind Monte-Carlo methods for radiation transport is to discretize the distribution function of neutrinos using Monte-Carlo packets (or superparticles) each representing a large number of neutrinos, i.e.
\beq
f_{(\nu)} = \sum_{k\in P} N_k \delta^3\left(x^i-x^i_k(t)\right)\delta^3\left(p_i-p_i^k(t)\right)
\eeq
with $P$ the ensemble of all packets, $x^i_k(t)$ the position of packet $k$ as a function of time, and $p_i^k(t)$ the spatial components of the momentum one-form of the neutrinos in that packet. A Monte-Carlo algorithm needs prescriptions to create packets (emission), propagate them on the grid, destroy them (absorption), and handle non-destructive interactions with the fluid or other neutrinos (e.g. scattering events). The propagation of neutrinos can simply be performed by moving packets along geodesics. All other processes have to be treated probabilistically, in such a way that the ensemble of packets remains a good sampling of the true distribution function $f_{(\nu)}$. 

The finite number of packets in a Monte-Carlo simulation will inevitably lead to sampling noise in $f_{(\nu)}$, and in any variable derived from $f_{(\nu)}$. We typically expect relative errors in quantities obtained by summing over $N$ packets to be $\sim N^{-1/2}$. If one wanted $\sim 1\%$ errors for the energy density of neutrinos within each cell of a computational simulation at any given time, one would thus need $\sim 10^4$ packets per cell... and scaling to smaller error bars is extremely expensive. A saving grace for Monte-Carlo simulations in the merger context, however, is that neutrino-matter interactions are generally not dynamically important, and tend to change the properties of the fluid over relatively long time scales (with a few exceptions). In practice, this means that we will be able to perform reasonably accurate Monte-Carlo simulations by relying on averaging neutrino-matter interactions over time scales significantly longer than our numerical time step in the vast majority of the computational domain. Practically, many regions can in fact be evolved stably and reliably with less than a packet per grid cell. This should be contrasted with astrophysical systems where radiation is dynamically important (e.g. radiation-dominated accretion disks). In those systems, sampling noise in e.g. the radiation pressure can be problematic. The price to pay for the use of a small number of packets, however, is that we lose the ability to get instantaneous estimates of the neutrino distribution function. 

\textbf{Emission}: The creation of Monte-Carlo packets is theoretically simple to handle, as long as we know the emission rate of neutrinos as a function of momentum, and are provided with a prescription to choose the number of neutrinos represented by each packet. For example, if we assume that each individual packet in a given grid cell of coordinate volume $\Delta V$ represents a total neutrino energy $E_p$, and that the energy-integrated emissivity is $\eta_{\rm tot}$, we create
\beq
N = \frac{\eta_{\rm tot} \sqrt{-g} \Delta V \Delta t}{E_p}
\eeq
packets over a time step $\Delta t$ in that cell. Non-integer values of $N$ can be treated statistically (e.g. $N=0.2$ implies a $20\%$ chance of creating a single packet). The direction of propagation of the packets is drawn from an isotropic distribution in the fluid frame, and their energy from the distribution $f(\nu) = \eta(\nu)/\eta_{\rm tot}$. The initial time and position of a packet can be drawn from an homogeneous distribution within the 4-volume $(\Delta V \Delta t)$, or possibly from a spatial distribution more adapted to the problem (e.g. in axisymmetry, it is beneficial to assume that the volume element is $\propto rdr$, making it more likely that packets are created in the outer region of a cell than in its inner region). 

In many simulations, we have at our disposal tabulated values of $\eta(\nu)$ within specific energy bins, rather than a known continuous function $\eta(\nu)$. In that case, a common strategy is to calculate within each bin the number of emitted packets $N_b$. Packets can then be created in the fluid frame with the energy of the center of their energy bin. Giving all packets the energy of the center of the bin guarantees that if we have tabulated values of $\kappa_a$ at the center of the same energy bins, the equilibrium energy density of neutrinos will match its expeceted physical value to numerical roundoff \citep{richers:15}. 

We see that the emission step is thus conceptually simple. However, it involves a choice crucial to the efficiency of Monte-Carlo methods: that of the energy of a packet $E_p$. That energy can vary from cell to cell, over time, and from energy bin to energy bin, and is one of the main tool available to distribute computational resources efficiently through a simulation. In neutron star mergers, for example, it is convenient to choose $E_p$ in optically thick regions in order to get a desired number of packets per cell (and thus a desired statistical noise), while $E_p$ in optically thin regions can be chosen to obtain a desired number of packets over the entire simulation (setting the overall cost of the simulation). Such an algorithm is detailed in \cite{Foucart:2021mcb}. We note however that many other choices of $E_p$ are possible, and that the optimal choice is problem dependent. \cite{Kawaguchi:2022tae} instead propose to choose $E_p$ as a fraction of the energy of neutrinos in thermal equilibrium with the fluid in a cell, with a prescription to destroy and resample low energy packets at the end of each step to avoid the continuous evolution of a large number of low-energy packets in optically thin regions. Simulation of accretion disks around black holes can rely on simpler prescriptions for $E_p$ (e.g. constant packet energy), as they do not need to deal with the dense, hot regions observed in neutron star remnants.

\textbf{Propagation:} Neglecting the finite mass of neutrinos, we expect packets to propagate along null geodesics in between interactions with the fluid and other neutrinos. \cite{Hughes1994} showed that to do so, it is convenient to evolve the position $x^i$ and spatial components of the momentum one-form $p_i$. The corresponding geodesic equations were initially developed for ray-tracing and photon transport, but are also appropriate for neutrinos:
\beqn
\frac{dx^i}{dt} &=& \gamma^{ij} \frac{p_j}{p^t} -\beta^i\\
\frac{dp_i}{dt} &=& -\alpha p^t \partial_i \alpha +p_j \partial_i \beta^j -\frac{1}{2}p_j p_k \partial_i \gamma^{jk}\\
p^t &=&\frac{\sqrt{\gamma^{ij}p_ip_j}}{\alpha}.
\eeqn
The first two lines are evolution equations for $x^i$, $p_i$, while the third comes from the constraint that $p^\mu p_\mu=0$. Numerically, the main choices to make here are the time stepping algorithm, and a method for the computation of the metric and its derivatives at the location of the packet. We note that as packets are either slowly diffusing through the system (in which case we do not directly use these equations; see below) or rapidly free-streaming out of the computational domain, even low-order methods perform well enough that the propagation step is not a leading source of error in simulations (see e.g., \citealt{Foucart:2021mcb}). 

\textbf{Additional interactions:} Finally, we have to allow packets to interact with the fluid and/or other neutrinos. If we imagine that we have a set of processes with known mean free path $\lambda_i$, or opacity $\kappa_i= \lambda_i^{-1}$, then for each process we can randomly draw the time to the next interaction from a Poisson distribution, e.g.
\beq
\Delta \tau_i = -\lambda_i \ln r_i = -\kappa^{-1}_i \ln r_i
\eeq
with $r_i$ a random number drawn from an homogeneous distribution in $[0,1)$ (the repeated index $i$ does not imply summation here). We note that $\Delta \tau$ is in the reference frame in which we provide $\kappa_i$. In simulations, this is usually the fluid frame. Transforming to the coordinate time, we get \citep{Ryan2015}
\beq
\Delta t_i =  -\kappa^{-1}_i \frac{p^t}{\nu} \ln r_i.
\eeq
The time to the next event is then $\Delta t_{\rm next} = \min{(\Delta t_i,\Delta t_{\rm step})}$ with $\Delta t_{\rm step}$ the time step, and with the minimum taken over all possible processes. If the minimum is $\Delta t_{\rm step}$, a packet is simply propagated for that time. Otherwise, whichever process provided the minimum $\Delta t$ occurs after that time interval. Existing simulations have considered an absorption opacity $\kappa_a$ and an isotropic elastic scattering opacity $\kappa_s$, as in moment schemes, but this method can be generalized to a larger number of interactions. Absorption simply results in the packet being removed from the simulation, while isotropic elastic scattering results in redrawing the direction of propagation of the packet from an isotropic distribution in the fluid frame, under the constraint that the fluid frame energy is conserved. 

As for the propagation of packets, an important step here is how to interpolate the opacities to the position of the packets and, if opacities are tabulated, in energy space. Additionally, our estimate for $\Delta t_i$ is only valid for a constant $\kappa_i$. If opacities are changing rapidly, taking too large of a time step can introduce significant errors. The existing simulation of a neutron star merger with Monte-Carlo transport used constant $\kappa_i$ within each grid cell \citep{Foucart:2020qjb}, which is numerically convenient but may need to be improved. In addition to being low-order, this choice leads to an algorithm that is  sensitive to the time at which we determine a packet leaves a cell, and thus recompute the opacities. As discussed in \cite{Foucart:2021mcb}, recovering accurate diffusion rates through high-opacity regions is then only possible if the packets take fairly small time steps when close to a cell boundary. Higher-order methods are desirable, but complicate calculations of the times-to-interaction $\Delta t_i$ and, to be consistent, also require the use of higher-order estimates of the emissivities, and thus inhomogeneous production of packets within a single grid cell. A second-order accurate scheme has recently been developed by \cite{Kawaguchi:2022tae}, for applications to axisymmetric systems, where the additional cost of a higher-order scheme is more manageable than in full 3D merger simulations. 

\textbf{Potential Issues}: We see that the main steps of a Monte-Carlo algorithm are fairly straightforward, at least when using low-order methods. They also naturally follow the expected behavior of individual neutrinos. Unfortunately, this simple algorithm runs into a number of important roadblocks in practice. The simplest one to solve is that the evolution still needs to be coupled to the fluid. Ideally, this would be done in a way that satisfies conservation laws and avoids unnecessary shot noise in the fluid evolution. We will see that in practice, there is a trade-off between these two objectives (Sect.~\ref{sec:MCfl}). A more significant issue is that the algorithm as proposed would be extremely inefficient and potentially unstable in optically thick regions. This is because we then expect rapid creation, annihilation, and scattering of individual packets, requiring expensive calculations and potentially creating stiff source terms in the evolution of the fluid equations. This is despite the fact that the outcome of the evolution of neutrinos in that regime is known: they simply get into statistical equilibrium with the fluid (if $\kappa_a \Delta t_{\rm step} \gg 1$), and/or slowly diffuse through the fluid (if $\kappa_s \Delta t_{\rm step}\gg 1$). We discuss how one may deal with these issues in Sect.~\ref{sec:MCthick}. Finally, neutrino packets are very inhomogeneously distributed through the computational grid. This is a desirable effect for optimal use of our computational resources, but it means that we cannot simply divide our computational domain into regions with roughly the same number of grid cells, and then distribute those regions onto different processors. Parallelization is not a fundamental roadblock to the use of Monte-Carlo transport, as packets can be distributed to processors in such a way that the number of packets on each processor is well balanced. Doing so does however require significant reorganization in merger codes that were not built with radiation transport in mind, and thus tend to assume that the computational cost of evolving a given grid cell is roughly the same regardless of the chosen cell.

\textbf{Moments of $f_\nu$}: Before getting into these issues, a useful additional piece of formalism to discuss is the computation of moments of $f_{(\nu)}$ in a Monte-Carlo code. Moments can be useful to compute neutrino-matter interactions, and of course to compare Monte-Carlo results with moment transport algorithms. The Dirac $\delta$ in the definition of $f_{(\nu)}$ can be practically problematic when considering moments of $f_{(\nu)}$, and it is thus often preferable to consider the average moments within a volume $V$ (e.g. a grid cell), defined as
\beq
\bar M^{\alpha_1...\alpha_n}(t,x^i) = \int_V \frac{dV\epsilon^{3}}{V} \int d\Omega f_\nu(t,x^i,\epsilon,\Omega) (\hat t^{\alpha_1}+l^{\alpha_1})...(\hat t^{\alpha_n}+l^{\alpha_n}).
\eeq
For the second moment, this is
\beq
\bar M^{\alpha\beta} = \sum_{k\in V} N_k \frac{p^\alpha_k p^\beta_k}{\sqrt{-g} V p^t_k}
\eeq
and similary for the number flux
\beq
\bar N^\alpha = \sum_{k\in V} N_k \frac{p^\alpha_k}{\sqrt{-g} V p^t_k}.
\eeq
In both cases, the sum is over all packets within the volume $V$. From these expressions, calculating moments only requires tensor projections. For example, the energy density in the fluid frame is
\beq
\bar J  =  \sum_{k\in V} N_k \frac{\nu_k^2}{\sqrt{-g} V p^t_k}
\eeq 
with $\nu_k = -p^\mu_k u_\mu$. We note that while this expression will turn out to be convenient for numerical simulations, its theoretical interpretation is a little convoluted, as it represents a moment in the fluid frame, averaged over a volume element in the simulation frame at a constant simulation time.

\subsubsection{Coupling to the fluid}
\label{sec:MCfl}

In the moment formalism, the back-reaction of the neutrinos onto the fluid was relatively easy to compute, because the moment equations are directly equivalent to the equations of conservation of energy, momentum, and, if evolving the number density, lepton number. A coupling scheme that explicitly conserves these quantities can also be designed for Monte-Carlo methods, with a few additional calculations; but we'll see that it comes with some pitfalls.

To get a conservative scheme, we can keep track of changes in the momentum of neutrinos due to emission, absorption, and scattering. If the $N_k$ neutrinos in a packet undergo a change of 4-momentum $\Delta p_k^\mu = p^\mu_{\rm after} - p^\mu_{\rm before}$, then the fluid should undergo a change of 4-momentum $-N_k \Delta p_k^\mu $ as a reaction. While this change is generally instantaneous, we can distribute it over a 4-volume element $\Delta V\Delta t$ to get the 4-force density \citep{Ryan2015}
\beq
G^\mu = -\sum_{\rm events} \frac{N_k \Delta p_k^\mu}{\sqrt{-g} \Delta V \Delta t},
\eeq
with the sum being over all events changing a packet's 4-momentum within the given 4-volume. We note that a single packet may be subject to no event, or to many events, and that a packet may interact with the fluid in different volume elements over the course of a time step; thus, properly computing that sum requires careful bookkeeping. The fluid equations become
\beq
\nabla_\mu T_{\rm fl}^{\mu\nu} = G^\nu.
\eeq
When using operator splitting, one might instead want to apply the total change of momentum within the 4-volume to the evolved fluids variables as a postprocessing step, after evolving the fluid and neutrinos. This requires a time integration of this equation. The relevant changes for the evolved fluid variables within a volume $V$ after a time step $\Delta t$ are
\beq
\Delta (\tilde T^{\mu\nu}n_\nu) =  \sum_{\rm events} \frac{N_k \Delta p_k^\mu}{\Delta V}.
\eeq
The most commonly evolved fluid variables in merger simulations are the internal energy density $\tilde \tau=\tilde T^{\mu\nu}n_\mu n_\nu-\rho_*$ and momentum density $\tilde S_i = -\tilde T^\mu_i n_\mu$. For these variables, we get
\beqn
\Delta \tilde \tau &=&  \sum_{\rm events} \frac{N_k \Delta p_k^\mu n_\mu}{\Delta V}\\
\Delta \tilde S_i &=& - \sum_{\rm events} \frac{N_k \Delta p^k_i}{\Delta V}.
\eeqn
Similarly, for the evolution of the electron fraction, 
\beq
\Delta (\rho_* Y_e) = - m_b \sum_{\rm events} \frac{N_k s_k}{\Delta V}
\eeq
with $s_k=0$ for muon and tau neutrinos and for any scattering event, $s_k=1$ for emission of $\nu_e$ and absorption of $\bar \nu_e$, and $s_k=-1$ for emission of $\bar\nu_e$ and absorption of $\nu_e$. The mass $m_b$ is again the reference baryon mass entering the definition of the rest mass density. This method has the advantage of imposing exact conservation laws: whatever the neutrinos gain, the fluid looses, and vice-versa. Its disadvantage is to be fairly sensitive to shot noise. If a packet is emitted, absorbed, or scattered in a very low density regions, that event would lead to extremely large changes in the temperature, momentum and composition of the fluid. These changes may even lead to unphysical values of the fluid variables after coupling to the neutrinos. It is best suited to simulations with fairly homegeneous packet energies, or a sufficiently large number of packets to avoid shot noise issues. This is the method adopted e.g. in \cite{Miller:2019gig}.

One alternative is to give up on exact conservation laws, counting instead on conservation laws being statistically satisfied over many interactions. In such an algorithm, described for example in \cite{Foucart:2021mcb}, one may write the source terms for the fluid equations as in the moment equations, i.e., for neutrinos of a given energy and considering only emission, absorption, and elastic scattering,
\beq
G^\mu = -(\eta - \kappa_a J) u^\mu + (\kappa_a +\kappa_s) H^\mu.
\eeq
As when using exact conservation laws, we then need to integrate this source term over a time step $\Delta t$ and a small volume $\Delta V$, as well as over the entire neutrino energy spectrum. We then get
\beq
\int d\nu \int dV \int dt G^\mu = -\eta_{\rm tot} \Delta V \Delta t u^\mu +
\sum_{k,j} \frac{\Delta t_{k,j}N_k \nu_k}{\sqrt{-g}p^t_k}
 \left(\kappa_{a,k} p^\mu
+ \kappa_{s,k} p_k^\alpha h^\mu_\alpha\right).\nonumber
\eeq
In this expression, the sum is taken over all packets $k$, and all time intervals $\Delta t_{k,j}$ during which packet $k$ is propagating along a geodesic while inside of the volume $\Delta V$ and time interval $\Delta t$ (due to scattering events, a single packet may appear multiple times in this sum). Converting to the change per unit volume in the evolved variables over a time $\Delta t$, we get
\beqn
\Delta \tilde \tau &=& -\tilde\eta_{\rm tot} \alpha W \Delta t +
\sum_{k,j} \frac{\Delta t_{k,j}N_k \nu_k}{\Delta V p^t_k}
 \left( (\kappa_{a,k}+\kappa_{s,k}) \epsilon_k
- \kappa_{s,k} W \nu \right)\\
\Delta \tilde S_i &=& -\tilde \eta_{\rm tot} \alpha u_i \Delta t +
\sum_{k,j} \frac{\Delta t_{k,j}N_k \nu_k}{\Delta V p^t_k}
 \left( (\kappa_{a,k}+\kappa_{s,k}) p_i^k
- \kappa_{s,k} \nu u_i \right).
\eeqn
For the electron fraction, we get instead
\beq
\Delta(\rho_* Y_e) = m_b \alpha \left(\tilde \eta^{(\bar\nu_e)}_{N,\rm tot}-\tilde \eta^{(\nu_e)}_{N,\rm tot}\right) 
 + m_b \sum_{k,j} s_k \kappa_{a,k}\Delta t_{k,j} N_k \frac{\nu_k}{\Delta V p^t_k}
\eeq
with $s_k=1$ for $\nu_e$, $s_k=-1$ for $\bar\nu_e$, and $s_k=0$ otherwise. With these choices, the source terms in low-density regions ($\kappa\Delta t \ll 1$) are a lot less noisy. Indeed, every packet passing through such a region contributes to the source terms exactly the expectation value of these source terms, instead of creating large source terms when a packet actually interacts with the fluid and none otherwise. Importantly, the time-averaged value of the source terms is the same as in the previous method \citep{Foucart:2021mcb}. We decrease the shot noise in the source terms at the cost of loosing exact conservation laws. We note that in this latter technique, an important decision is when to switch a packet from contributing to one grid cell to contributing to its neighbor. The obvious answer would be to do so at the exact time a packet crosses a cell boundary, but this is not necessarily trivial to predict in high-opacity and/or high-curvature regimes. Any algorithm using this method thus has to consider a balance between the cost of accurately determining the time at which a packet crosses a cell boundary, and the cost of using approximate values for the $\Delta t_{k,j}$.

An intermediate scheme, used e.g. by \cite{Kawaguchi:2022tae}, is to absorb full packets in optically thick regions, but to damp $N_k$ in optically thin regions, following the method of \cite{2009ApJS..184..387D}. The number of neutrinos in a packet then evolves as
\beq
\frac{dN_k}{d\tau} = -\kappa_a N_k
\eeq
with $\tau$ the time in the fluid frame. In this case, the energy / lepton number absorbed over a time step can easily be added to the fluid while guaranteeing that we exactly satisfy the relevant conservation laws, but one may end up with a larger number of packets as individual packets are no longer destroyed by default (an issue solved in \cite{Kawaguchi:2022tae} by resampling packets in optically thin regions at the end of a step anyway).

As more Monte-Carlo simulations are performed in the future, other methods are likely to be developed. At the moment, exact conservation appears preferable if the neutrinos contain a significant fraction of the total energy /momentum of the system, or are dynamically important to the evolution of the system, and if a lot of packets are used in the simulations. In mergers and post-merger remnants, where neutrinos are not dynamically important and we tend to use few packets in low-density regions, the second method may be preferable.

\subsubsection{Optically thick regime}
\label{sec:MCthick}

The optically thick regime poses a more fundamental problem for Monte-Carlo methods. One issue is that if $(\kappa_a+\kappa_s)\Delta t = N$, we expect $\sim N$ neutrino-matter interactions per time-step. In dense or hot regions of mergers, we can easily get $N\gg 1$, making evolutions costly. A related issue is that if we calculate $\eta,\kappa_a,\kappa_s$ using the properties of the fluid at the beginning of a time step, and the source terms lead to large changes in the fluid variables, the evolution can become numerically unstable. To avoid taking extremely small time steps, we would then need to at the very least obtain a good guess for the fluid variables at the end of the time step, or even iteratively solve for these values. The latter would be costly for Monte-Carlo simulations, as all packets have to be reevolved for every new guess of the fluid variables. We note that in the merger context, these issues are mostly due to the presence of a dense, hot neutron star remnant. For their simulations of post-merger accretion disks, \cite{Miller:2019gig} simply require a time step smaller than the cooling timescale of any given cell, and find that condition to be generally less constraining than the Courant condition. Accordingly, the problem of very optically thick regions has only really been encountered in the merger simulations of \cite{Foucart:2020qjb}, and handled with approximate methods inspired by the implicit Monte-Carlo method of \cite{1971JCoPh...8..313F} for regions of high absorption optical depth, and the random walk method of \cite{1984JCoPh..54..508F} for regions of high scattering optical depth. We briefly summarize their methods here, as well as the proposed algorithm of \cite{Kawaguchi:2022tae}, but note that given the relative novelty of the problem, these methods remain poorly tested and understood.

A first change that can be made to the standard Monte-Carlo algorithm in dense region, appropriate when $\kappa_s \Delta t \gg 1$, is to once more rely on the idea that, in that regime, neutrinos are slowly diffusing in the reference frame of the fluid. \cite{Richers:2017awc} and \cite{Foucart:2017mbt} propose similar methods to treat this regime, based on the work of \cite{1984JCoPh..54..508F}. Both assume that in regions of sufficiently high $\kappa_s \Delta t$, neutrinos are advected with the fluid, while undergoing a slow random walk away from the fluid's motion. When determining the outcome of that random walk, \cite{Richers:2017awc} draw from the distribution of times needed for a neutrino to diffuse a certain distance in the fluid frame. \cite{Foucart:2017mbt} draws instead from the distribution of distances that the neutrinos move in the fluid frame after a fixed time, and additionally draws the final momentum of the diffusing neutrinos from a distribution function calibrated to a solution of the full Boltzmann equation. Both algorithms also correct the solution of the diffusion equation so that neutrinos cannot move faster than the speed of light (i.e., effectively assume that if the diffusion equation predicts superluminal motion, no scattering occurs and neutrinos just propagate at the speed of light along a geodesic). We refer the reader to \cite{Richers:2017awc,Foucart:2017mbt} for the exact choices of distribution functions. When using such an approximation, calculating the source terms for coupling to the fluid can be slightly more involved. If using the actual value of the momentum transfer, we can still compute $\Delta p^\mu$ of each packet between the beginning and end of a step, but should in theory correct that result for the natural change in $p^\mu$ due to the evolution of a packet along a geodesic in curved spacetime. If using expectation values of the interaction rates instead, \cite{Foucart:2017mbt} uses the approximation that $\langle p^\mu\rangle = \nu u^\mu$ for a fraction $f_{\rm adv}$ of the time step, and $p^\mu = \nu(u^\mu + l^\mu)$ for the remaining of the time step, with $f_{\rm adv}$ and $l^\mu$ chosen so that the packet ends at the desired location. The source terms can then be computed as in the previous section, using these approximate values of the neutrino 4-momentum. Overall, this process has been demonstrated to work well even down to $\kappa_s \Delta t \sim 3$ \citep{Foucart:2017mbt}.

Regions with $\kappa_a \Delta t \gg 1$ create a more difficult challenge. In these regions, we expect neutrinos to reach statistical equilibrium with the fluid on a time scale $\kappa_a^{-1} \ll \Delta t$, and the temperature and composition of the fluid themselves may quickly change as more neutrinos or antineutrinos are emitted, absorbed, or diffuse away. In these regions, we expect the energy density of neutrinos of a given energy to be set by $\eta/\kappa_a$, and their diffusion timescale to be set by $(\kappa_a+\kappa_s)^{-1}$. Then, the evolution of the system can be approximately captured if we accept that we cannot resolve what happens on timescales much shorter than $\Delta t$ \citep{Foucart:2021mcb}, and rely instead on changes to the source terms inspired by Implicit Monte-Carlo methods \citep{1971JCoPh...8..313F}. In \cite{Foucart:2021mcb}, we propose the transformation
\beqn
\eta' &=& a \eta\\
\kappa_a' &=& a \kappa_a\\
\kappa_s' &=& \kappa_s + (1-a) \kappa_a
\eeqn
which guarantees that $\eta'/\kappa_a' = \eta/\kappa_a$ and $\kappa_a+\kappa_s =\kappa_a' + \kappa_s'$, but modifies the equilibration time scale of neutrinos from $\kappa_a^{-1}$ to $(a\kappa_a)^{-1}$. We then choose $a$ such that $\kappa_a' \Delta t \lesssim 1$, making all relevant time scales longer than the time step. We note that a different value of $a$ may be used for each energy bin and each neutrino species. This method has the advantage of limiting the number of emissions and absorptions of neutrinos to roughly what is needed to get to statistical equilibrium in a few time steps, while avoiding stiff source terms in the fluid evolution equations. For neutron star merger remnants, we find sub-percent errors in the neutrino luminosities when comparing this method to the solution of Boltzmann's equation \citep{Foucart:2021mcb}. We note however that this method could easily impact the diffusion rate of neutrinos within the densest regions if the neutrino spectrum changes on scales smaller than the grid scale, and that its accuracy in the presence of significant inelastic scattering is also uncertain. As opposed to the more complex method used by \cite{1971JCoPh...8..313F}, the method proposed here is not a direct discretization of the coupled fluid-radiation equations, though deviations from those equations are all related to sub grid scale and sub time step features of the solution. There is no doubt that further improvements and/or tests of the method would be benefitial to properly understand and potentially reduce the errors that it creates.

In their proposed Monte-Carlo algorithm, \cite{Kawaguchi:2022tae} adopt an algorithm much closer to that of \cite{1971JCoPh...8..313F}. A single parameter $a$ is chosen for all energy bins of a given neutrino species, and the additional scattering opacity induced by the algorithm is assumed to be inelastic: the total energy of scattered packets does not change, but the energy of individual neutrinos within the packet samples the equilibrium energy distribution of neutrinos. With such an algorithm, Implicit Monte-Carlo can be seen as just a specific discretization of the original transport equations \citep{1971JCoPh...8..313F}, a significant theoretical advantage. This comes at the cost of slightly less flexibility in our ability to reduce absorption and reemission of packets, and more complexity in the treatment of scattering events.

\subsubsection{Pair processes and other non-linear source terms}
\label{sec:MCpairs}

In this last section discussing Monte-Carlo methods, we turn to pair processes as an example of issues that may arise when non-linear terms in the neutrino distribution functions come into play. In theory, Monte-Carlo methods provide us with a direct discretization of the neutrino distribution functions in 6D. One could thus evaluate blocking factors and non-linear source terms explicitly. However, in the presence of a large number of packets, this is an expensive computation. If on ther other hand only a small number of packets are present, sampling noise in the distribution function may lead to large errors in the resulting reaction rates. 

Consider for example the $\nu_i\bar\nu_i \rightarrow e^+ e^-$ reaction. The absorption cross section for neutrinos of 4-momentum $p^\alpha$ could be written, under the same assumptions as Eq.~\eqref{eq:Qpairs} and assuming that each packet of antineutrinos represent a \emph{spatially uniform} distribution of neutrinos within a grid cell of volume $V$, as
\beq
\kappa_{\nu\bar\nu} = \frac{DG_F^2}{3\pi} \frac{1}{\sqrt{-g}Vp^t}  \sum_{k,\bar\nu_i \in V} N_k\frac{(-p^\alpha \bar p_{\alpha,k})^2}{\bar p^t_k}
\eeq
with $\bar p_k^\alpha$ the 4-momentum of antineutrinos in packet $k$, and the sum covering all antineutrino packets of the correct species in the chosen grid cell. We note that as opposed to previous opacities, which were computed in the fluid frame, this opacity is computed in the simulation frame, i.e., if we draw the time to the next annihilation event, we should use $\Delta t_{\nu\bar \nu} = \kappa^{-1}_{\nu\bar\nu} \ln{r}$ with $r\in [0,1)$. While the mathematical expression is relatively simple, this requires for each packet iteration over all packets of antineutrinos in the same cell. This is costly if the number of packets is large, and sensitive to shot noise if the number of packets is small. We note however that according to our earlier expression for the stress-energy tensor in Monte-Carlo, this is equivalent to
\beq
\kappa_{\nu\bar\nu} = \frac{DG_F^2}{3\pi} \frac{p^\alpha p^\beta}{p^t}  \bar T_{\alpha\beta}
\eeq
with $\bar T_{\alpha\beta}$ the stress-energy tensor of the antineutrinos. This allows for faster calculations for large numbers of antineutrinos if we precompute $\bar T_{\alpha\beta}$ (the algorithm will scale linearly with the number of packets, instead of scaling with the number of packets squared). Additionally, this method allows us to approximate $\bar T_{\alpha\beta}$ in other ways in regions where few packets are available, e.g., using an average over a larger spatial region, or a longer time interval, in order to gather information from a larger number of packets at the cost of a smoothing of the neutrino distribution function. Quite importantly however, as the opacity depends explicitly on the 4-momentum of the neutrinos, we have to recompute it for each packet, at each time. To avoid costly contractions with the metric, a code that stores $p^t$ and $p_i$ would ideally express this opacity as
\beq
\kappa_{\nu\bar\nu} = \frac{DG_F^2}{3\pi p^t} \left([p^t\alpha]^2 \bar E - 2 [p^t \alpha] [p_i \bar F^i] + p_i p_j \bar P^{ij}\right)
\eeq
and thus specifically precompute $\bar E, \bar F^i, \bar P^{ij}$ to minimize computations. These expressions have however not been used in general relativistic merger simulations so far. 

A potential complication is that if pair processes are included in the emission rate of neutrinos (as is typically done for muon and tau neutrinos), and the absorption opacity is calculated according to Kirchoff's law, then the opacity for pair annihilation is already included in the simulations under the assumption that neutrinos and antineutrinos are both in equilibrium with the fluid. So simply adding the opacity calculated here to a simulation would double count pair annihilation in optically thick regions. On the other hand, we know that when assuming equilibrium we underestimate the rate of pair annihilations by many orders of magnitude in low-density regions above the remnant. Without a fully self-consistent calculation of pairs everywhere (which would require the inclusion of blocking factors in the emissivity, and thus make the emission step dependent on the current distribution of neutrinos), the method to calculate pair annihilation outlined here should only be used in optically thin regions.


\subsubsection{Discussion}

The use of Monte-Carlo algorithms in merger and post-merger simulations is a relatively novel development, with few simulations published so far. Early results however indicate that Monte-Carlo methods can be used at a surprisingly low cost, comparable to or lower than that of the most complex moment schemes, while automatically taking into account the energy dependence of the distribution function $f_\nu$. Monte-Carlo simulations have already allowed important tests of the accuracy of moment schemes, and will certainly continue to remain a valuable tool until the evolution of Boltzmann's equations with higher-order methods becomes a realistic prospect. Nevertheless, these methods have their own drawbacks. For monoenergetic neutrinos, they are certainly less accurate than moment schemes in high-density regimes, were approximations are currently made to avoid the use of implicite time stepping and the calculations of a large number of interactions. For more realistic neutrino spectra, existing tests of these approximations indicate that they are probably subdominant sources of error in current merger simulations, but only a few of these tests have been performed so far. Further improvements to the behavior of Monte-Carlo algorithms in dense regions, possibly combined to the use of implicit methods for their coupling to the fluid, may be desirable in the future.

Additionally, one of the main supposed advantage of an evolution of the full Boltzmann equations is the availability of $f_\nu$. For existing Monte-Carlo scheme, this availability is doubtful. Current simulations use a very small number of packets in low-density regions, so that $f_\nu$ can only be recovered with reasonably low statistical noise if one averages over long time scales and/or spatial scales. We have noted the importance of this problem for pair annihilation, but this could also be an issue for calculations of oscillations due to fast flavor instabilities. The FFI is typically triggered due to changes in the sign of the net lepton flux (flux of $\nu_e$ minus flux of $\bar \nu_e$), and calculations of that net flux from existing Monte-Carlo simulations would be entirely dominated by statistical noise. While Monte-Carlo simulations are certainly an important step forward in our modeling of radiation transport, they are thus far from a one-size-fit-all solution to the problem of radiation transport in merger simulations.

\section{General relativistic merger simulations}
\label{sec:results}

In the previous sections, we provided a detailed discussion of the three broad classes of algorithms used in general relativistic simulations of neutron star binary mergers so far. These algorithms have been used in a large number of simulations, and discussing each of these results goes beyond the scope of this review. However, in order to put these algorithms in context, and provide examples of their limitations and known sources of errors, it is useful to discuss at least two aspects of these simulations: what they broadly tell us about neutrino physics in neutron star mergers, and how different algorithms compare when used to simulate the same physical configuration.  

\subsection{Neutrinos in neutron star merger simulations}

Neutrinos play two major roles in the evolution of the properties of post-merger remnants: cooling the remnant disk and, if present, the remnant neutron star, and modifying the composition of the fluid. There is broad agreement that, for systems that rapidly collapse to a black hole, or for black hole-neutron star systems, shock heating during the merger and the circularization of the accretion disk will lead to bright neutrino emission peaking at $10^{53-54}\,{\rm erg/s}$. That emission however rapidly decays on $\sim 10\,{\rm ms}$ timescale \citep{Deaton:2013sla,Sekiguchi:2016bjd,Radice:2021jtw}. Late time emission depends on the relatively poorly constrained heating due to magnetically-driven turbulence in the post-merger disk, as well as on the mass of the disk itself \citep{Fernandez:2020oow,Shibata:2021xmo,Fujibayashi:2022ftg}. It may remain as high as $\sim 10^{52-53}\,{\rm erg/s}$ for $O(100\,{\rm ms})$. This emission is not sufficient to create a thin disk (the disk aspect ratio remains $\sim 0.2$), but simulations that do include cooling find remnant disks that are significantly more compact than simulations without cooling, indicating a loss of gravitational binding energy. Once the mass accretion rate drops sufficiently (to roughly $\dot M\lesssim 10^{-3}\,M_\odot\,{\rm /s}$ \citep{De:2020jdt}), neutrino emission is no longer sufficient to cool the disk, which becomes an advection dominated accretion flow. 

\begin{figure}[ht]
\centering
\includegraphics[width=0.7\textwidth]{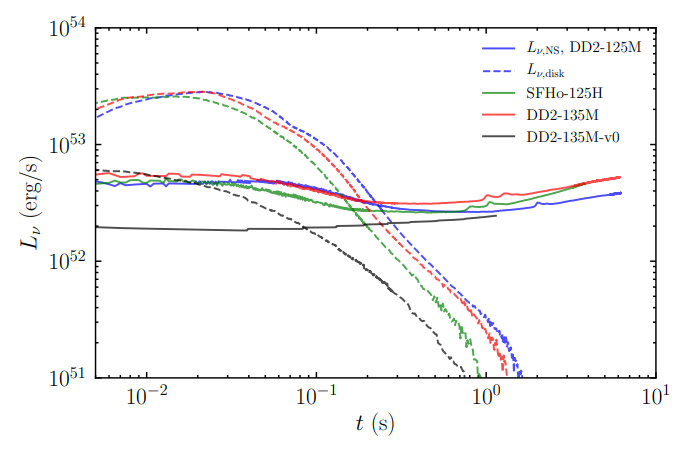}
\caption{Total neutrino luminosity from the neutron star (solid curves) and accretion disks (dashed curve) in three long simulations of NSNS merger remnants in which the central object remains a neutron star. We observe both the decay of the disk emission on $O(100\,{\rm ms})$ timescales and the sustained emission from the neutron star. Figure reproduced from \cite{Fujibayashi:2020dvr}.}
\label{fig:Lnu}
\end{figure}

For systems with a massive neutron star remnant, the peak emission is at a level comparable to the black hole-disk remnant, but emission from the neutron star can continue over much longer timescales. For example, the long axisymmetric simulations of \cite{Fujibayashi:2020dvr} find neutrino luminosities of $\sim 10^{52-53}\,{\rm erg/s}$ more than $5\,{\rm ms}$ post-merger, with no sign of the luminosity decreasing on those timescales. The neutron star remnant will thus be the dominant source of neutrinos after $O(100\,{\rm ms})$ (see Fig.~\ref{fig:Lnu}).

In both cases, simulations generally agree on an energy hierarchy $\epsilon_{\nu_x}>\epsilon_{\bar\nu_e}>\epsilon_{\nu_e}$, with average energy of $\sim$\,10\,MeV for $\nu_e$ but potentially above 20\,MeV for $\nu_x$. This is simply due to the higher absoption opacity of the fluid to $\nu_e$, which puts the $\nu_e$ neutrinosphere farther out in the remnant than the $\bar\nu_e$ neutrinosphere (and even more the $\nu_x$ neutrinosphere). As the temperature of the remnant close to the neutrinosphere increases with density, heavy-lepton neutrinos have higher energies.

\begin{figure}[ht]
\centering
\includegraphics[width=0.42\textwidth]{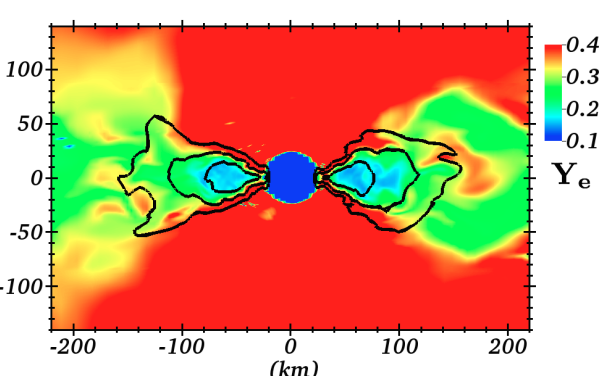}
\includegraphics[width=0.56\textwidth]{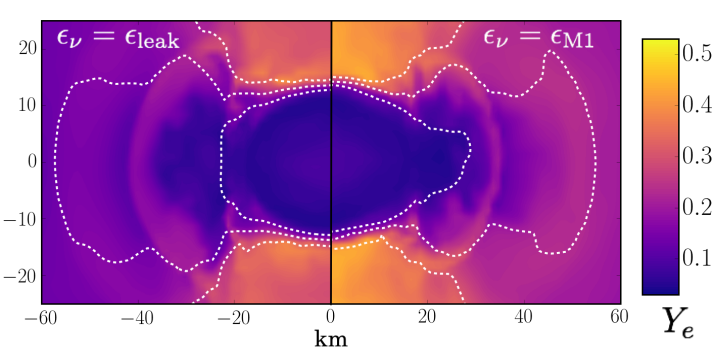}
\caption{Electron fraction in post-merger remnants. We show a vertical slice through a NSBH merger (Left) and a BNS merger (Right). The BNS merger is evolved using two different choices for the energy closure in a two-moment scheme. Figures reproduced from \cite{FoucartM1:2015} and \cite{Foucart:2016rxm}.}
\label{fig:ye}
\end{figure}

Weak reactions during the merger and early post-merger evolution are typically well out of equilibrium, and existing simulations find an initial overabundance of $\bar\nu_e$ emission over $\nu_e$ emission as the remnant's electron fraction rapidly increases - for both NSBH and BNS mergers, and regardless of whether a remnant black hole is formed or not. After formation of an accretion disk, however, different regimes can be found. Long simulations of accretion disks with mild electron degeneracy using a leakage scheme have found that weak interactions can lead to self-regulation of the composition of the disk midplane to $Y_e\sim 0.1$ \citep{2018ApJ...858...52S}, with a lower density outer disk and hotter at significantly higher $Y_e$ (Fig.~\ref{fig:ye}). Simulations of accretion disks with moment transport find that neutrino absorption in the disk can lead to higher electron fractions $Y_e\sim 0.15- 0.2$ \cite{FoucartM1:2015,Just:2021cls}. These values can also be significantly impacted by irradiation of the disk by a central neutron star (Fig.~\ref{fig:ye}). Farther out in the disk, or once neutrino emission becomes inefficient and the disk becomes advection dominated, the electron fraction largely freezes out. The initial density and mass accretion rate of the disk thus plays a role in determining the composition of post-merger outflows \citep{Fernandez:2016sbf}, and that composition is also affected by the location from which the matter is ejected (midplane vs corona). In that respect, it is worth noting that outflows produced from self-consistent MHD simulations and outflows produced by simulations using $\alpha$ viscosity models as a subgrid model to capture angular momentum transport and heating from magnetically-driven turbulence \citep{1973A&A....24..337S}, even when they agree on the amount of matter ejected, make very different predictions for the history of the fluid elements ejected from an accretion disk -- which may impact predictions of the final $Y_e$ of the outflows. MHD simulations starting with different magnetic field configurations can also produce very different amounts of outflow,indicating that predictions for matter outflows in post-merger remnants are likely to depend on the unknown large scale properties of magnetic fields in the post-merger remnant \citep{Christie:2019lim,2022arXiv221107158H}.

All of the above features are found in simulations using leakage, moments, or Monte-Carlo transport, although disagreements between methods can be found on the exact neutrino luminosity and composition of the remnant (see below). Simulations that include reabsorption of neutrinos in the matter outflows also find that hot outflows originating from the colliding cores of two neutron stars or the hot corona of an accretion disk rapidly evolve to electron fractions $Y_e \sim (0.2-0.4)$\cite{Wanajo:2014,FoucartM1:2015,Sekiguchi:2016bjd,Radice:2018pdn,Foucart:2020qjb,Radice:2021jtw,Camilletti:2022jms}, though the exact value of $Y_e$ can vary significantly depending on the chosen numerical algorithm (see below). Faster and colder outflows associated with the tidal disruption of a low-mass neutron star by a more massive companion (in either NSBH or BNS mergers) do not capture enough neutrinos to undergo sigificant changes of composition, and thus remain very neutron-rich ($Y_e\lesssim 0.05$), even in simulations that include neutrino absorption \citep{Sekiguchi:2016bjd,Foucart:2016vxd,Radice:2018pdn,Camilletti:2022jms}. There is higher uncertainty on the composition of the post-merger outflows. Neutrino-driven outflows observed in post-merger simulations that do not include magnetic fields tend to have very high $Y_e$, but fairly low mass $M_{\rm outflow}\approx 10^{-(3-4)}M_\odot$ \citep{2015MNRAS.448..541J,FoucartM1:2015} (see e.g. polar regions of Fig.~\ref{fig:ye}). They are likely to be dominated by magnetically-driven winds \citep{2018ApJ...858...52S,Fernandez:2018kax}, if those winds are significant (see above), and by viscous outflows in the advection-dominated phase \citep{Fernandez:2013tya}. In the absence of strong magnetically-driven winds, neutrino-driven winds could however impact kilonova signals, as they might be geometrically separated from both the dynamical ejecta and the viscously-driven winds. Magnetically-driven outflows are faster ($\sim 0.1c$), and have less time to absorb neutrinos, while viscous outflows are slower ($\sim 0.03c$) and have a composition set by the electron fraction of the fluid at the time at which weak interactions freeze-out. Long 3D simulations including magnetic fields and approximtely accounting for neutrino absorption in post-processing find outflows with $Y_e\sim 0.15-0.25$, starting from very neutron-rich initial conditions \citep{2018ApJ...858...52S}. Shorter simulations using an advanced Monte-Carlo transport scheme and including magnetic fields \citep{Miller:2019dpt}, with similar initial conditions, find a broader distribution of $Y_e$ peaking just below $Y_e\sim 0.2$ and extending up to $Y_e\sim 0.4$. Similarly, long simulations of a NSBH merger remnant with a moment scheme and magnetic fields \citep{Hayashi:2021oxy}, initialized from the outcome of a merger simulation, find a broad $Y_e\sim 0.15-0.4$ distribution peaking just above $Y_e\sim 0.2$. Very few long 3D simulations including both magnetic fields and neutrinos are however available. Parameter space exploration with axisymmetric simulations (using artificial viscosity instead of evolving the magnetic field) find a significant sensitivity of the results on the compactness of the disk \citep{Fernandez:2020oow}, the lifetime of the massive neutron star remnant (if present) \citep{Metzger:2014ila}, and the initial composition used in the simulation \citep{Fernandez:2016sbf}. 3D simulations have also demonstrated the importance of the large scale structure of the post-merger magnetic field both with a black hole remnant \citep{Christie:2019lim} and a neutron star remnant \citep{deHaas:2022ytm}. Given our limited understanding of the properties of post-merger remnant as a function of the initial binary configuration, it is thus fairly difficult at this point to build reliable models of post-merger outflows.

Finally, in the previous sections we already emphasized the difficulties of properly implementing $\nu\bar\nu$ pair annihilation in relativistic merger simulations. \cite{Fujibayashi:2017xsz}, using the approximate moment method described in Sect.~\ref{sec:M1pairs}, finds that pair annihilation can accelerate the matter in the polar regions to mildly relativistic speeds (Lorentz factor $\Gamma\sim 2$) -- sufficient to be important to the dynamics of the fluid in those low-density regions, but not enough to power SGRBs (see Fig.~\ref{fig:pairs}). In NSBH remnants with very little matter in the polar regions, \cite{just:16} even find relativistic outflows powered by pair annihilation, but not with enough energy to power the brightest SGRBs. \cite{Foucart:2018gis} find that the energy deposition predicted by the approximate moment scheme is likely accurate within a factor of $\sim$\,2--3. More advanced transport calculations had earlier been performed on a prescribed fluid background \citep{1999ApJ...518..356P,2006JPhG...32..443K,Zalamea:2008dq,Dessart:2008zd,Perego:2017fho}. These simulations find that at most a few percents of the neutrino luminosity is deposited in the polar region through neutrino annihilation, with a rapid drop in efficiency as the accretion rate decreases. These two sets of results are qualitatively consistent, indicating that while neglecting pair annihilation may be acceptable when considering only the low-velocity outflows powering kilonovae, it may be important to take into account when attempting to resolve the evolution of the polar regions, and in particular the formation of jets in that region.

\subsection{Comparisons of numerical algorithms}

To finish our discussion of general relativistic transport methods for neutron star mergers, let us look into a few direct comparisons of numerical algorithms to gauge their accuracy.

A direct comparison of a leakage scheme (whithout absorption) and a two-moment scheme (with Minerbo closure, evolving $\tilde E$ and $\tilde F_i$) was performed in the context of a low-mass neutron star merger in \cite{Foucart:2015gaa}. In that simulation, leakage and moment schemes agreed quite well on the $\bar\nu_e$ luminosity, while other luminosities varied by factors of 2--3 in the first 10\,ms following the merger. The inclusion of neutrino absorption led to the production of a neutrino-driven wind that did not exist in the leakage simulation, but the mass outflow rate was only $\dot M \sim 10^{-2}\,M_\odot/s$, i.e., less than what one might expect at that time from magnetically-driven winds. Outflows in the moment simulation were also significantly hotter and less neutron-rich than in the leakage simulation ($\langle Y_e\rangle =0.2$ vs $\langle Y_e\rangle=0.1$, and $\langle s\rangle=20k_B$ vs $\langle s\rangle=10k_B$ per baryon). This confirms that neutrino luminosities are only order-of-magnitude accurate in simple leakage schemes, and the crucial impact of neutrino absorption on the properties of matter outflows.

\begin{figure}[ht]
\centering
\includegraphics[width=\textwidth]{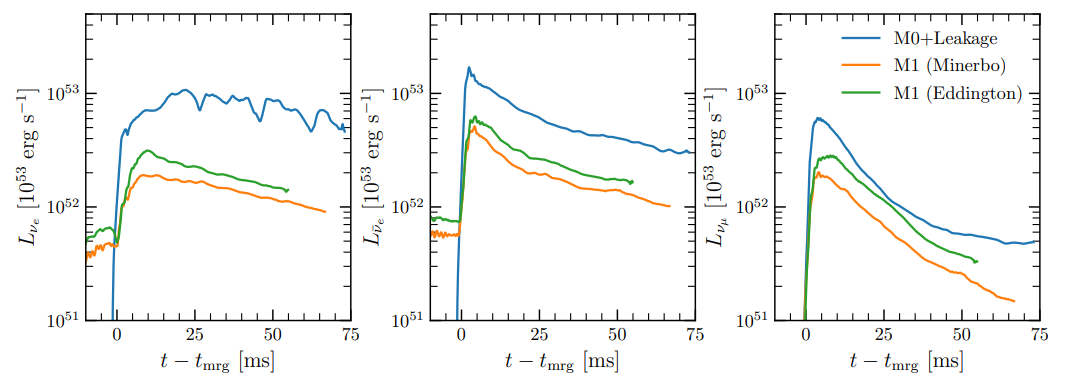}
\caption{Neutrino luminosities for $\nu_e$, $\bar\nu_e$ and $\nu_x$ in a BNS merger with a long-lived remnant. Results are shown for three algorithms: a hybrid leakage-moment scheme (blue), a two-moment scheme with Minerbo closure (orange), and a two-moment scheme with Eddington closure (green). Figures reproduced from \cite{Radice:2021jtw}.}
\label{fig:M1M0}
\end{figure}

In \cite{Radice:2021jtw}, the authors consider two neutron star mergers, one for which the remnant collapses to a black hole a few milliseconds after contact, and one forming a long-lived neutron star remnant. They compare results using a hybrid moment-leakage scheme, the standard two-moment scheme with Minerbo closure, as well as a two-moment scheme using the Eddington closure (i.e., the optically thick closure everywhere). The hybrid schemes overestimates neutrino luminosities by a factor of $\sim 2$ with respect to the moment simulations, while the two-moment simulations with different closures are in much better agreement (10\%--30\% differences, see Fig.~\ref{fig:M1M0}). All three schemes agree very well on the average energy of escaping neutrinos for $\nu_e\bar\nu_e$, while early emission of higher energy heavy-lepton neutrinos is predicted by the two-moment scheme but not the hybrid scheme. The two-moment scheme also predicts $\sim 30\%$ less mass ejected than the hybrid scheme, and significantly higher electron fractions ($\Delta Y_e\sim 0.1$). The use of the Eddington or Minerbo closure had again a much smaller effect ($\sim 10\%$ relative error in mass, shifts of a few percents in $Y_e$). Differences between the hybrid and two-moment scheme are thus slightly lower than between a pure leakage and a two-moment scheme, but nonetheless significant. The fact that the Minerbo and Eddington closure are in much closer agreement is however quite encouraging, as a reasonable proxy for the error due to the use of an approximate closure in semi-transparent regions.

The impact of the chosen energy closure in grey two-moment schemes was investigated in \cite{Foucart:2016rxm}, for the same physical configuration as in \cite{Foucart:2015gaa}. In that manuscript, two energy closures are considered. In the first, the average energy of neutrinos is taken from a black-body distribution at temperature $T_{\nu} = \max{(T_{\rm leak},T)}$, with $T_{\rm leak}$ the temperature predicted by a leakage scheme (globally) and $T$ the temperature of the fluid. In the second, the neutrino number density is evolved to obtain a local estimate of $T_{\nu}$, and the neutrino spectral index is evolved to attempt to accounting for softer spectra in the regions optically thick to scattering but optically thin to absorption. The luminosity of neutrinos initially is in reasonable agreement in both schemes (20\%--30\% differences), but differences increase over time to $\sim$\,50\% at the end of the evolution (8\,ms post-merger). This is less than the difference between leakage and two-moment schemes, but still quite significant. One likely source of error here is divergence in the evolution of $Y_e$ in the remnant, due in part to the fact that the scheme that does not evolve the number density cannot explicitly conserve the total lepton number of the system. As in \cite{Radice:2021jtw}, the average energy of electron-type neutrinos is reasonably close in both schemes, but calculations based on the leakage scheme underestimate the initial energy of heavy-lepton neutrinos. Both schemes have similar outflow masses, but difference in electron fraction $\Delta Y_e\sim (0.05-0.1)$ (see Fig.~\ref{fig:ye}). We note that this is despite the fact that the estimated average neutrino energies are similar in both simulations, and mostly due to the fact that polar neutrinos (which interact with hot outflows) have higher energy than equatorial neutrinos (which do not), and are thus more strongly absorbed than when opacities are computed using a global estimate of the average energy. Using the local fluid temperature to estimate the neutrino energies would lead to significantly larger errors: factors of a few changes in the neutrino energies instead of tens of percent.

One can also take a broader view of these comparisons between transport schemes. Instead of directly comparing numerical simulations, \cite{Nedora:2020qtd} compare datasets of simulations using different microphysical inputs, and provide numerical fits for the outcome of these simulations. Their results are consistent with the direct comparisons discussed above, and show that the choice of neutrino transport algorithm in merger simulations remain an important source of error in outflow modeling.

\begin{figure}[ht]
\centering
\includegraphics[width=\textwidth]{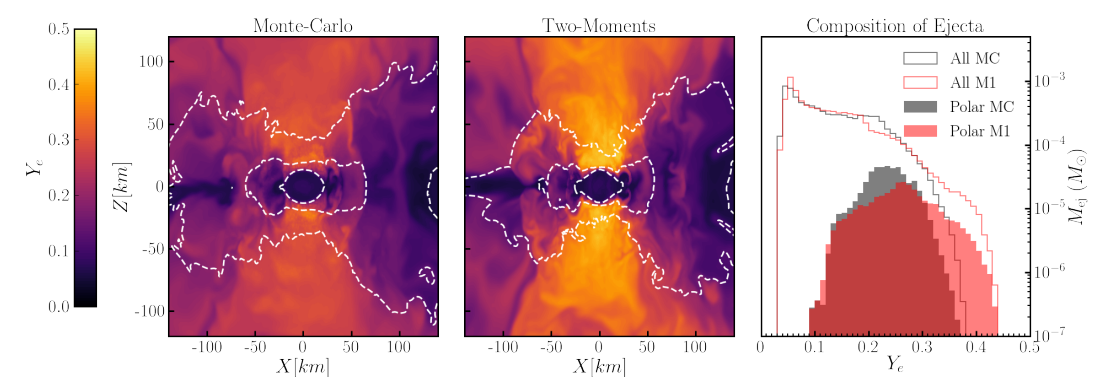}
\caption{Comparison of the electron fraction $Y_e$ in a vertical slice of a post-merger remnant for simulations using Monte-Carlo methods (Left) and a two-moment scheme (Middle). The right panel shows the $Y_e$ distribution of matter outflows in both simulations. Figures reproduced from \cite{Foucart:2020qjb}.}
\label{fig:MCM1ye}
\end{figure}

Finally, comparisons of a Monte-Carlo transport scheme with a two-moment scheme evolving the number density were performed without back-reaction of the Monte-Carlo code to the fluid (i.e., using the same fluid evolution as in the moment evolution) \citep{Foucart:2018gis} as well as with coupling to the fluid \citep{Foucart:2020qjb} in a neutron star merger simulation. In the fully coupled simulation, the luminosity of $\nu_e$ and $\bar\nu_e$ and the energy of neutrinos was in better agreement (10\%--20\% differences) than in other comparisons between transport methods discussed so far, with the exception of the comparison between different choices of closure relations performed by \cite{Radice:2021jtw}. The heavy-lepton luminosity showed $\sim50\%$ changes, showing once more the difficulty of properly capturing the evolution of those neutrinos. Similarly, the composition of the outflows was in much better agreement than in other comparisons discussed in this section, with the average $Y_e$ changing by $\Delta Y_e\sim 0.03$, and the maximum $Y_e$ by $\Delta Y_e\sim 0.05$ -- not necessarily negligible changes, but a significantly better agreement than in other comparisons (see Fig.~\ref{fig:MCM1ye}). 

\begin{figure}[ht]
\centering
\includegraphics[width=\textwidth]{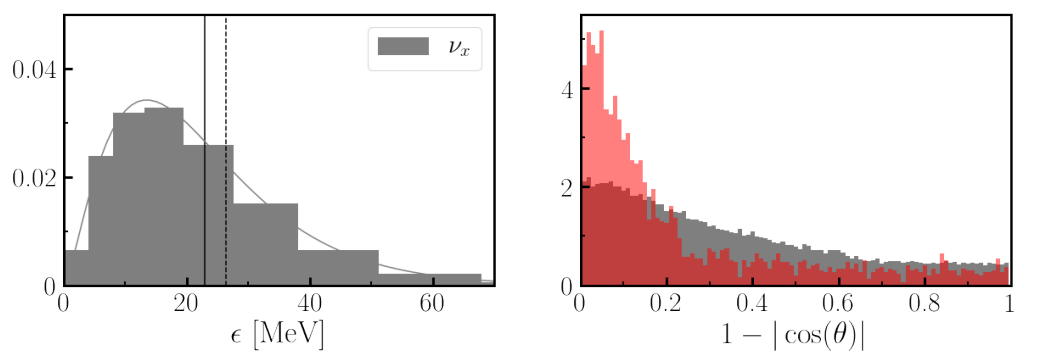}
\caption{{\it Left}: Energy spectrum of the $\nu_x$ neutrinos 14\,ms after a BNS merger, according to a Monte-Carlo simulation. Vertical lines show the average energy esimated using Monte-Carlo (Solid line) or a two-moment scheme (dashed line). {\it Right}: Angular distribution of the neutrinos in the same simulations, using a Monte-Carlo evolution (black) or a two-moment evolution (red). Figures reproduced from \cite{Foucart:2018gis}.}
\label{fig:MCM1Lum}
\end{figure}

The simulation that does not fully couple the Monte-Carlo algorithm to the fluid evolution, thus avoiding any drift of the fluid variables over time due to small differences between the algorithms, shows similar differences in the neutrino energies, but better agreement for the $\nu_x$ luminosity. A more detailed study of the spatial distribution of neutrinos however indicates that the moment scheme greatly overestimates the density of neutrinos close to the pole (by $\sim$\,50\%--100\%), and underestimates their density farther out -- an important consequence of the choice of closure made for the pressure tensor. That simulation also computed the rate of neutrino pair annihilation using the moment scheme and the Monte-Carlo methods. We note that, as shown in Sect.~\ref{sec:M1pairs}, the moment calculation requires the use of both an approximate average energy for the pairs and of the approximate pressure closure in a regime in which it is inaccurate. Additionally, it is naturally impacted by errors in the estimated energy density of neutrinos close to the poles. Interestingly, for the specific numerical algorithms studied here, those errors partially cancelled out, leaving us with factors of 2--3 errors in the actual annihilation rate for the moment scheme. There is however now guarantee that this would also be true for a different binary configuration, or with different estimates for the neutrino energy, as individually some of these errors can modify the annihilation rate by close to an order of magnitude (see Fig.~\ref{fig:MCM1Lum}).

Overall, we thus note that there has been a significant improvement in the magnitude of those errors, with the more modern moment \citep{Foucart:2016rxm,Radice:2021jtw} and Monte-Carlo \citep{Foucart:2020qjb} schemes getting estimated relative errors in the $\sim$\,10\%--20\% range for the variables that most significantly impact astrophysical observables -- by which point neutrino transport is certainly a subdominant source of error when compared to the underresolved evolution of magnetic fields, the nuclear physics uncetainties in kilonova modeling, or maybe even the impact of neglected / poorly modeled neutrino physics (oscillations, inelastic scattering, pair process).

\section{Conclusions}

The inclusion of radiation transport algorithm in neutron star merger simulations has taken significant step forward over the last decade. The development of improved two-moment schemes and Monte-Carlo algorithms, in particular, allows for reasonably accurate evolution of the transport equations \emph{for relatively simple neutrino physics}. 

This leaves us however with a few important challenges. First, very few simulations have made use of these new methods, and thus efforts to model the observable counterparts to neutron star mergers still heavily rely on results obtained with simpler microphysics. As a result, current model are often unreliable \citep{Henkel:2022naw}, and dependent on the algorithms used for neutrino evolution in the simulations used to calibrate them \citep{Nedora:2020qtd}. Second, we know that a number of potentially important processes are \emph{not} included in existing simulations, or are poorly modeled in those simulations. This include at least neutrino oscillations, pair annihilation, inelastic scattering, and potentially direct and modified URCA processes. Finally, neutrino transport is only one part of the problem when modeling neutron star mergers. Magnetic fields are crucial to the evolution of neutron star mergers and their post-merger remnants. The growth of large scale magnetic fields due to MHD instabilities is not sufficiently resolved even in simulations that do not include neutrinos, or that include them very approximately. Combining high-resolution MHD simulations with our most advanced neutrino transport schemes over the seconds time scales needed to follow the evolution of a post-merger remnants remains an extremely difficult problem that will likely remain an important source of uncertainty in our modeling of electromagnetic signals from neutron star mergers for years to come.

\backmatter

\bmhead{Acknowledgments}

This article has been accepted for publication, after peer review, but is not the Version of Record and does not reflect post-acceptance improvements or any corrections. The version of record is available online at https://link.springer.com/article/10.1007/s41115-023-00016-y.

The author is grateful to Adam Burrows, Gail McLaughlin, Philipp Moesta, Elias Most and Sherwood Richers for valuable discussions on various aspects of this review during its preparation, and to Sho Fujibayashi, Kota Hayashi, and David Radice for allowing reproduction of their results in this manuscript. The author also gratefully acknowledges the many improvements suggested by an anonymous reviewer after first submission of the manuscript. The author acknowledges support from the DOE Office of Science, Office of Nuclear Physics, under contract number DE-AC02-05CH11231, from NASA through grant 80NSSC18K0565, and from the NSF through grant AST-2107932. Part of this work was performed at the Aspen Center for Physics, which is supported by National Science Foundation grant PHY-1607611.

\section*{Declarations}

\bmhead{Conflict of interest}
The author declares no conflicts of interest.


\newpage
\phantomsection
\addcontentsline{toc}{section}{References}
\bibliography{sn-bibliography}


\end{document}